\documentclass[fleqn,usenatbib]{mnras}

\usepackage{newtxtext,newtxmath}

\usepackage[T1]{fontenc}

\usepackage{ae,aecompl}
\usepackage{graphicx}	
\usepackage{mathtools}
\usepackage{amsmath}
\usepackage{csquotes}

\usepackage{listings}
\usepackage{fancyvrb}
\usepackage{bbding}
\usepackage{alltt, xcolor}
\usepackage[export]{adjustbox}

\title[MeerKAT observations of XCS J2215]{MeerKAT observations of starburst galaxies and AGNs within the core of XMMXCS J2215.9 -- 1738 at $z$ = 1.46}
\author[D.Y.  Klutse]{D.Y. Klutse$^{1,2}$
\thanks{E-mail: dianasmoke1@gmail.com}, 
	M. Hilton$^{3,2}$
\thanks{Email: matt.hilton@wits.ac.za},
	I. Heywood$^{4,5,6}$,
	I. Smail$^{7}$,
	A.M. Swinbank$^{7}$, 
	K. Knowles$^{5,6}$,
	\newauthor S.P. Sikhosana$^{1,2}$ 
	\\
	$^{1}$Astrophysics Research Centre, University of KwaZulu-Natal, Westville Campus, Durban 4041, South Africa \\
	$^{2}$School of Mathematics, Statistics \& Computer Science, University of KwaZulu-Natal, Westville Campus, Durban 4041, South Africa\\
 	$^{3}$Wits Centre for Astrophysics, School of Physics, University of the Witwatersrand, Private Bag 3, 2050, Johannesburg, South Africa\\
	$^{4}$Astrophysics, Department of Physics, University of Oxford, Keble Road, Oxford OX1 3RH UK\\
	$^{5}$Centre for Radio Astronomy Techniques and Technologies,  Department of Physics and Electronics, \\ Rhodes University, P.O. Box 94, Makhanda 6140, South Africa\\
	$^{6}$South African Radio Astronomy Observatory, 2 Fir Street, Black River Park, Observatory, Cape Town 7925, South Africa\\
	$^{7}$Centre for Extragalactic Astronomy, Department of Physics, Durham University, South Road, Durham DH1 3LE, UK}

\date{Accepted XXX. Received YYY; in original form ZZZ}

\pubyear{2023}

\begin{document}
	\label{firstpage}
	\pagerange{\pageref{firstpage}--\pageref{lastpage}}
	\maketitle
	
\begin{abstract}
	We present the first detailed radio study of the  galaxy cluster XMMXCS J2215.9-1738 at $z$ = 1.46 using MeerKAT $L$-band (1.3\,GHz) observations. We combine our radio observation with archival optical and infrared data to investigate the star formation and AGN population within $R_{200}$ ($R = $ 0.8~Mpc) of the cluster centre.~Using three selection criteria; the radio luminosity, the far-infrared radio ratio ($q_{\rm{IR}}$) and the mid-infrared colour, we distinguish galaxies with radio emission predominantly powered by star formation from that powered by AGNs.~We selected 24 cluster members within $R_{\rm{200}}$ in the MeerKAT image based on either their photometric or spectroscopic redshift.~We classified 12/24 ($50\%$) as galaxies whose radio emission is dominated by star-formation activity, 6/24 ($25\%$) as intermediate star-forming galaxies and  6/24 ($25\%$) as AGN-dominated galaxies.~Using the radio continuum luminosities of the star-forming cluster galaxies, we estimated an integrated star formation rate (SFR) value of 1700 $\pm$ 330 M$_{\odot}$yr$^{-1}$ within $R_{200}$.~We derived a mass-normalized integrated SFR value of $(570 \pm 110) \times 10^{-14}$~yr$^{-1}$.~This supports previous observational  and theoretical studies that indicated a rapid increase in star formation activity within the core of high-redshift clusters. We also show that the high AGN fraction within the cluster core is consistent with previous cluster observations at $z >$ 1.5.
\end{abstract}	


\section{INTRODUCTION}\label{sec:intro}
Galaxy clusters provide a powerful means to study the formation of large-scale structures, the evolution of galaxies and the thermodynamics of the intergalactic medium.~They also complement other existing probes of structural growth over cosmological time due to their high masses \citep{2012KravtsovBorgani}. 

Environment plays a major role in galaxy evolution and this is evident via the  relationship of star formation  rate to local galaxy density.~Star formation rate (SFR) in galaxies at $z \lesssim$ 1.0 increases with  increasing galaxy density up to galaxy group scales, and  then declines in the denser regions of galaxy clusters
 \citep{2001MNRASKodama,2002MNLewis,2003ApJGomez,2009Blanton,2020ApJCluver,2021Pearson}.~Denser regions within galaxy clusters at $z \lesssim$ 1.0 are mostly populated by elliptical and lenticular galaxies (early-type galaxies) which show minimal star formation activity in contrast to low-density regions mostly occupied by star-forming spirals 
\citep[late-type disk galaxies,][]{1997ApJDress,2005ApJSmith}.

Several studies have suggested a reverse in this order at higher redshifts (i.e., in some clusters at $z > 1$, the fraction of star-forming galaxies increases with galaxy density) leading to the so-called reversal in the star formation density relation \citep{2010ApJTran,2010MNHayshi,2012MNRASTadaki,2013ApJBrodwin,2015MNRASantos,2024Smail}.~For example, a study of the  $z = 1.56$  galaxy cluster XMMU J1007.4+1237 showed  strong starburst activity within the cluster core \citep{2011AFassbender}.~Whilst \citet{2016ApJWang} detected nine star-forming cluster galaxies within the central 80 kpc  region of the X-ray detected cluster CL J1001+0220 at $z = 2.5$. A SCUBA-2 (sub-millimetre observation) study of the CL 0218.3-0510 cluster at $z = 1.6$ by \citet{2014ApJSmail} showed that active star formation activities take place outside the cluster core, whilst the outcome of a 24-$\mu$m (mid-infrared) observation of the  same cluster from \citet{2010ApJTran}  showed that this high redshift cluster was actively forming stars.
In contrast, an 850~$\mu$m  SCUBA-2  continuum observations of eight  X-ray-detected massive galaxy clusters at $z \approx  0.8 - 1.6$ studied by \citet{2019Cooke} revealed that the mass-normalized SFR for clusters at $1 < z < 1.6$ is a factor of  $1.5 \pm 0.3$ lower than the field galaxies and they do not find any reversal for local star-formation-rate–density (SFRD) relation in their study.~In a study of eight submillimetre (850~$\mu$m) galaxy clusters at $z = 1.6-2.0$,  \citet{2024Smail}  found a reversal in the  local  SFRD relation  for  clusters and fields at $z = 1.8$. 	

There may be a wide range of star formation activity in galaxy clusters at $z >$ 1.~A narrow band imaging survey of the [\textsc{Oii}]  emitters of the CIG J0218.3-0510 cluster at $z = 1.62$ using the Suprime-Cam on Subaru Telescope led to the detection of some very high star formation rate galaxies  as well as some quiescent galaxies within the cluster core \citep{2012MNRASTadaki}.~Similarly, the  study of a higher redshift cluster JKCS 041 at $z = 1.8$ \citep{2009Andreon} using the Wide Field Camera 3 of the \textit{Hubble Space Telescope} showed that the cluster core was populated by quiescent galaxies.  \citep{2014ApJNewman}.

Studies have shown that galaxy mergers, interactions and active galactic nuclei (AGN) activities take place in high redshift clusters \citep{2016ApJAlberts,2013ApJLotz,2017MNRASKrishnan} but whether these processes lead to the enhancement or suppression of star formation has not been conclusively established.~The study of star formation and AGNs in 11 spectroscopically selected massive clusters at  1.0 $< z <$ 1.75 by \citet{2016ApJAlberts} showed an increase in star formation rate, excess AGN activity and an increase in galaxy merger rate within the clusters at this high redshift.~ The central 24 $\times$ 24 kpc region within the cluster XDCP J0044.0-2033 at $z \gtrsim$ 1.5 by \citet{2022Lepore} showed that  high star formation and gas-rich merger-driven nuclear activities take place in the cluster core (i.e., within $\approx$ 0.16 Mpc from the cluster centre).

One well studied cluster to investigate star-burst activity at  high-redshift is XMMXCS J2215.9-1738 at $z = 1.46$ (subsequently J2215, located at R.A. $=$ 22h 15m 58.5s and  Dec. $=17^{\circ} 38^{\prime} 02^{\prime \prime}$ ).~It was one of the distant clusters with a well-developed structure to be discovered at X-ray wavelengths  \citep{2006ApJStanford}.~The cluster's velocity dispersion $\sigma_{v}$ within a virial radius $R_{200}$ (i.e., $R$ = 0.8 Mpc) is 720 $\pm$ 110 kms$^{-1}$ \citep{2010ApJHilton}, resulting in a virial mass of M$_{\rm{cl}} = 3 \times 10 ^{14}$ M$_{\odot}$. A SCUBA-2 survey found several highly star-forming galaxies in the cluster's core with an integrated star formation rate of $\approx$ 1400 M$_{\odot}$ yr$^{-1}$ \citep[Salpeter initial mass function (IMF) assumed,][]{2015ApJMa}.
Thus J2215  appears to be an example of a distant and highly star-forming galaxy cluster. 

Twelve [\textsc{Oii}]  emitting cluster members and four star forming dust-obscured ultra luminous infrared galaxies (ULIRGs) were detected at $<$ 0.25 Mpc from the cluster centre by \citet{2010MNHayshi} and  \citet{2015ApJMa} respectively. A 24-$\mu$m Spitzer/MIPS cluster survey conducted by \citet{2010ApJHilton} resulted in the detection of three star-forming galaxies and a potential AGN all located at $R <$ 0.25 Mpc, suggesting that both star-burst and AGN activities take place in the cluster core at higher redshifts. 

There is also evidence of mergers between cluster galaxies within this high redshift cluster; a survey of the core region of J2215 cluster at $z = 1.46$ using the Atacama Large Millimetre Array (ALMA) and the MUSE spectrograph on the Very Large Telescope (VLT)  by \citet{2017ApJStach} led to the detection of 14-millimetre sources within $\approx$ 0.5 Mpc from the cluster centre. The result obtained by \citet{2017ApJStach} indicated that there was an intense star formation  within the cluster core and evidence of a likely merger event taking place.~The most recent study of the J2215 cluster from the ALMA CO $J = 2-1$  line (0.4$^{\prime \prime}$ resolution) and 870 $\mu$m continuum (0.2$^{\prime \prime}$ resolution) observation by \citet{2022ApJIkeda} has shown evidence of enhanced star formation in the central region of the cluster, with 6 out of 17 cluster members observed to be early-stage mergers. 

Observations have also shown that both quiescence galaxies and star-forming systems reside within the cluster core; \citet{2018ApJHayashi} detected 12 quiescent and 27 star-forming galaxies within the core of the J2215 cluster at $z = $ 1.46 via the study of the ALMA band 3 data. The study of the J2215 cluster using the KMOS ($K$-band Multi-Object Spectrograph) by \citet{2019Maier} led to the detection of apparently slow quenching systems and slightly lower star formation activity within the cluster core.

As noted above, there have been several studies of the J2215 galaxy cluster conducted at infrared and millimetre/submillimetre wavelengths. Here we present the first detailed radio observation of the J2215 galaxy cluster using the MeerKAT radio telescope. In this work, we combine archival optical and infrared data with new $L$-band radio observations from MeerKAT to further investigate the star formation, AGN and merger activities within the core of the J2215 cluster.

This paper is structured as follows.~We describe the MeerKAT observations of the J2215 cluster, the data reduction process and briefly describe the optical and infrared archival observations used for this work in Section \ref{sec:Observation_and_Data_Analysis}.~We describe the MeerKAT source detection, cluster membership classification scheme and the radio luminosity and continuum star formation rate estimation in Section \ref{sec:Source_Catalogue_and_membership_Classification}.
In Section \ref{subsection:MORPHOLOGY_COLOURMAGNITUDE} we show the colour-magnitude relation of the cluster galaxies and their morphologies.~In Section \ref{sec:Results} we characterize our galaxy samples into normal star-forming galaxies, intermediate star-forming galaxies and AGNs using three indicators; the radio luminosity value, the far-infrared radio ratio ($q_{\rm{IR}}$) value and the mid-infrared colours.
~In Section  \ref{subsec:AGN Activity in higher and low redshift surveys} we compare the  AGN activity in the J2215 cluster core to other higher and low redshift surveys.
We discuss the star formation activity in J2215 and the evolution of clusters with redshift in Section \ref{sec:STAR_FORMATION_WITHIN_J2215}.~We conclude this work by giving a summary of the entire work done and the prospects in Section \ref{section:summary}. 

We assume a $\Lambda$CDM cosmology with H$_0$ = 70 km s$^{-1}$~Mpc$^{-1}$, $\Omega_{\Lambda} = 0.7$ and $\Omega_{m} = 0.3$.~The AB magnitude system and the \citet{2003Chabrier} IMF were used throughout this work unless stated otherwise. 

\vspace{10pt}

\section{OBSERVATION AND DATA REDUCTION} \label{sec:Observation_and_Data_Analysis}
This work is based on the radio observation of the J2215 cluster obtained from the MeerKAT telescope in combination with other archival data.

Sections \ref{subsec:MeerKAT_observations} and  \ref{subsec:The_oxkat_pipeline}  cover the details of the MeerKAT observation and data processing respectively, while we summarize the properties of the archival multi-wavelength data used in this study in Section \ref{sec:The Archival Observations}.

\vspace{15pt}

\subsection{MeerKAT  Observations} \label{subsec:MeerKAT_observations}
The J2215 galaxy cluster was observed in May 2019 using the $L$-band receivers of the 64-dish MeerKAT telescope.~The $L$-band receiver covers a frequency range of 856 $-$ 1712~MHz  with a central frequency of 1284~MHz. ~The observation was conducted on two different days, from 11 $-$ 12 May 2019 between the hours of 02:16 $-$ 10:20 UTC each day. A total of 61 antennas were in operation during the observation.~The 4096 (4K) channel wideband-coarse mode of the  Square Kilometre Array Reconfigurable Application Board (SKARAB) processing nodes correlator was used resulting in 209~kHz channel resolution.The data was recorded in full polarization. The correlator dump time was 8 seconds, and the data were acquired for all four polarization products, labelled as XX, XY, YX and YY.~The dataset in measurement set  format for the two-day observation is about 5.2~TB with an approximate 2.6~TB measurement set obtained each day.~J1939-6342 was used as the primary calibrator (i.e., for flux, delay and bandpass calibration) while a much closer source ( $\approx{13}$ degrees of the target source position), J2225-0457 (3C446) served as the secondary calibrator for amplitude and phase calibration.
~We observed the bandpass calibrator for 10 minutes after every two hours and two minutes on the gain calibrator after every 15-minute target scan.~In total, we spent $\approx$ 12 hours on the target source and the total integration time for the entire observation is $\approx$ 16 hours i.e., 8 hours of total integration each day.

\vspace{15pt}

\subsection{Data Processing} \label{subsec:The_oxkat_pipeline}
Data reduction was performed using a semi-automated pipeline, \textsc{Oxkat} \footnote{https://github.com/IanHeywood/oxkat} \citep{2020softwareHeywood}.~\textsc{Oxkat} comprises a set of python scripts that incorporates other radio astronomical packages to process the MeerKAT data.~The first step in the \textsc{Oxkat} pipeline was to duplicate the MS and average to 1024 channels to make the continuum data processing easier i.e., faster with shorter computation time.~The primary calibrator was then re-phased to the correct position via the \textsc{CASA} \citep{2007McMullin} task \textsc{fixvis}.~The known RFI channels documented in the \enquote{MeerKAT-Cookbook} were flagged, other flags were also applied to all the fields using the \textsc{CASA} task \textsc{flagdata} (i.e., manual, clip, quack mode) while auto-flaggers were applied to only the calibrators.~We derived model data visibilities for the primary calibrator defined by Stevens-Reynolds 2016 flux density scale \citep{2016ApJPartridge}.

We derive an intrinsic model for the secondary calibrator based on the primary calibrator and further derived delay ($K$), bandpass ($B$), gain ($G$) calibrations from the primary and secondary calibrators.~The gain solutions were then applied to all the calibrators and the target.~$K$, $B$, and $G$ corrections are derived iteratively, with rounds of residual flagging in between.~The calibrated target data was split out into individual MS, with the reference calibrated data in the DATA column of the new MS.~We then imaged the radio continuum emission with \textsc{wsclean} \citep{2014Offringa} in Stokes I and then conducted one round of phase and then amplitude self-calibration. 

\begin{figure*}
	\centering
	\includegraphics[height=6.55cm]{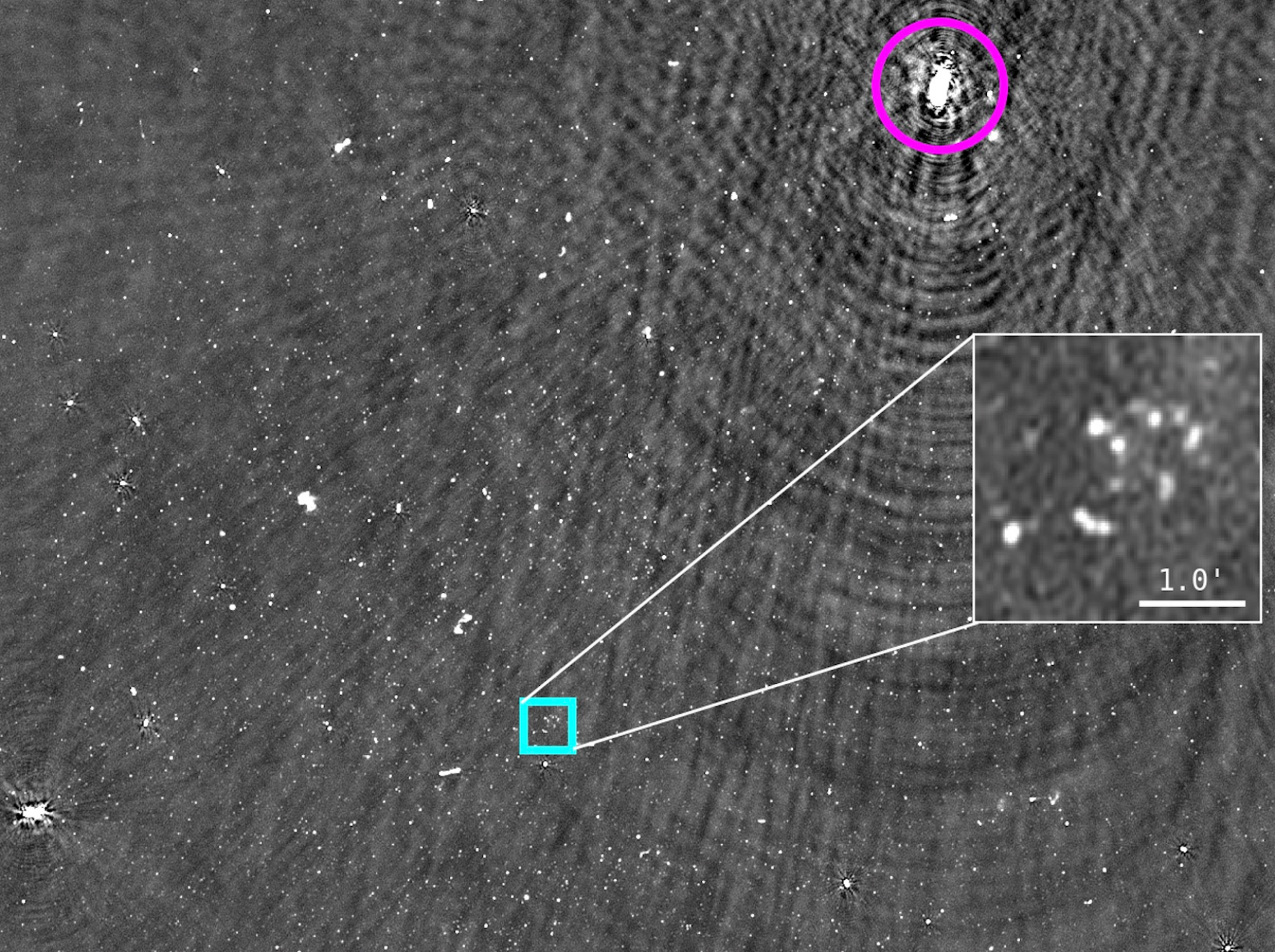}
	\includegraphics[height=6.55cm]{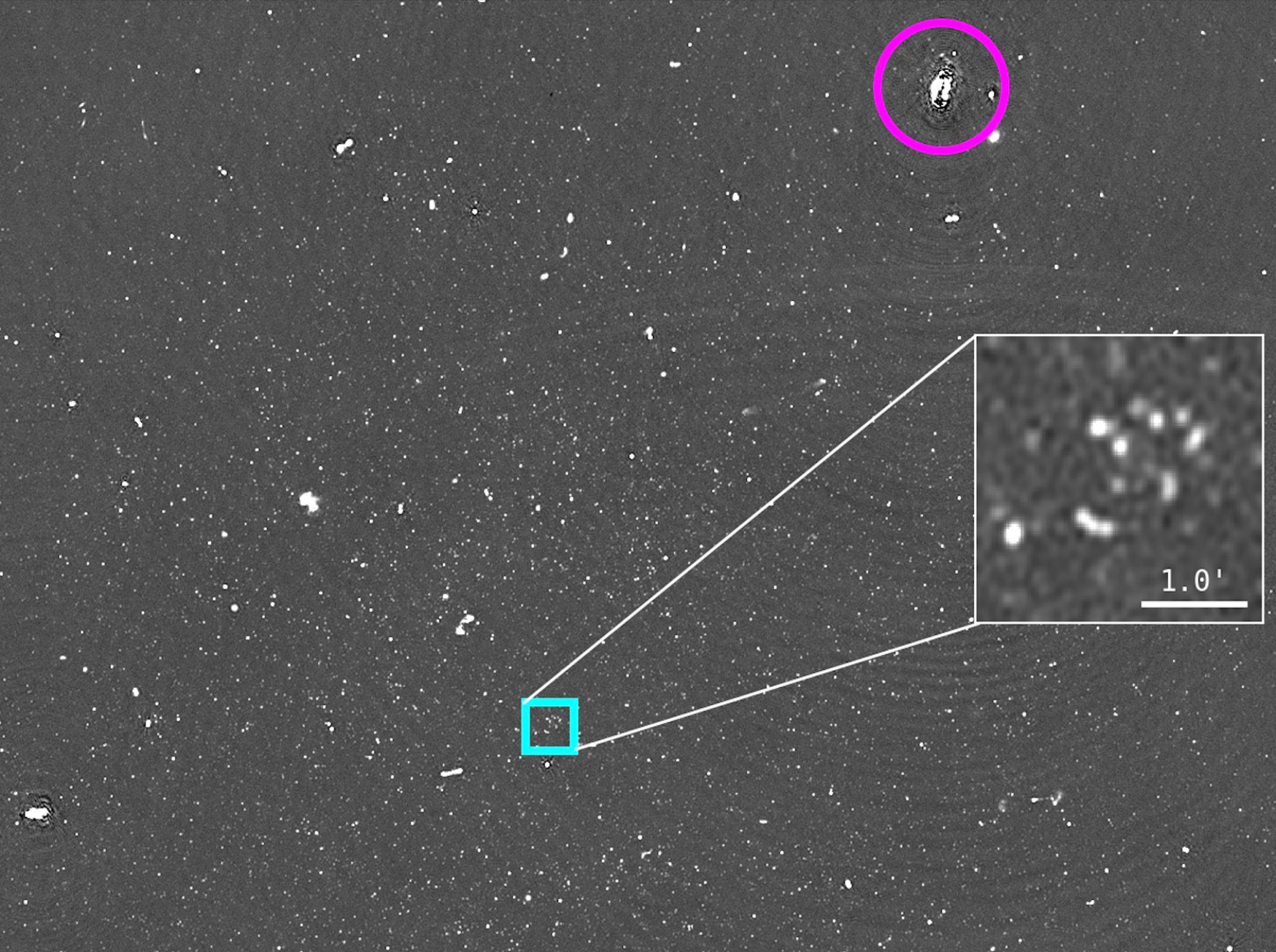}
	\caption{Left panel: A $\approx 70^{\prime} \times 50 ^{\prime}$ zoom-in image of the radio continuum image produced from MeerKAT's $L$-band observation with the image before peeling displayed (rms noise value = 3.9~$\mu$Jy\rm{beam}$^{-1}$).~The very bright $\approx$ 4.6~Jy source is visible in the image  (indicated with $\approx$ 4.0$^{\prime}$ radius magenta circle) and the target source is located at a distance of 0.7$^{\circ}$ (indicated with a 2.75$^{\prime}$ $\times$ 2.75$^{\prime}$  cyan box). Right panel: Same as the left panel but the very bright $\approx$ 4.6~Jy source has been mitigated after the peeling operation. The overall image has been well improved after the peeling process (rms noise value = 3.5~$\mu$Jy\rm{beam}$^{-1}$).}
	\label{fig:TroubleSource}
\end{figure*}

Imaging was performed using a cell size of 1.1 arcsecs, and a Briggs weighting of $ -0.3$. Due to the presence of a $\approx$ 4.6~Jy extended source 2MASX J22142575-1701362 located at $\approx$ 0.7 degrees away from our target (see Figure \ref{fig:TroubleSource}), we performed a peeling operation on the \enquote{troublesome source}  in our field.~This involved modelling  and subtracting using \textsc{CubiCal} \citep{2018Kenyon}.~We exploit \textsc{DDFacet} \citep{2018Tasse} on the residual visibility with the best available cleaning mask from the previous \textsc{wsclean} run (cropped to the appropriate \textsc{DDFacet} image size i.e.,  $10125 \times 10125$ pixels) to obtain a sky model and then defined the directions that will form the centres of the tesselated sky.~We derived the gain solutions for each of the tesselated sky models using \textsc{killMS} \citep[][]{2014KmsTasse,2015KmsSmirnov}. 

Further, we re-ran \textsc{DDFacet} to apply the gain corrections derived from the \textsc{killMS} run and also the primary beam correction.~The MeerKAT $L$-band primary beam model  was generated using the \textsc{eidos} software \citep{2019Asad}. The primary beam has a diameter of $\approx$ 1.4 $^{\circ}$. 
The resulting 1284 MHz image has a synthesized beam size of $6.01\arcsec \times 5.26\arcsec$ with a positional angle, PA of $-$10.4~degrees and an rms noise level of $\approx$ 3.5~$\mu$Jy\,beam$^{-1}$.~This is the most sensitive radio observation of the cluster to date. 

We determined the accuracy of the astrometry of our image by comparing the coordinates of the MeerKAT sources (10890, see Section \ref{subsec:MeerKATSourceDetection}) with its closest counterpart in the JVLA  (Karl G. Jansky Very Large Array) map (298 sources). The processed JVLA map with a sensitivity of $\approx$ 7.5~$\mu$Jy\,beam$^{-1}$ was obtained from \citet{2015ApJMa}. 
~The source extraction was done following the same procedure described in Section \ref{subsec:MeerKATSourceDetection}.~The mean offset of the 275 counterparts in the JVLA map was $\bigtriangleup$RA = ($-0.22^{\prime \prime} \pm 0.61^{\prime \prime}$) and $\bigtriangleup$Dec = ($-0.11 ^{\prime \prime}\pm 0.61 ^{\prime \prime})$ which is insignificant so we did not correct the astrometry.

\begin{figure*}
	\includegraphics[scale=1.0, center]{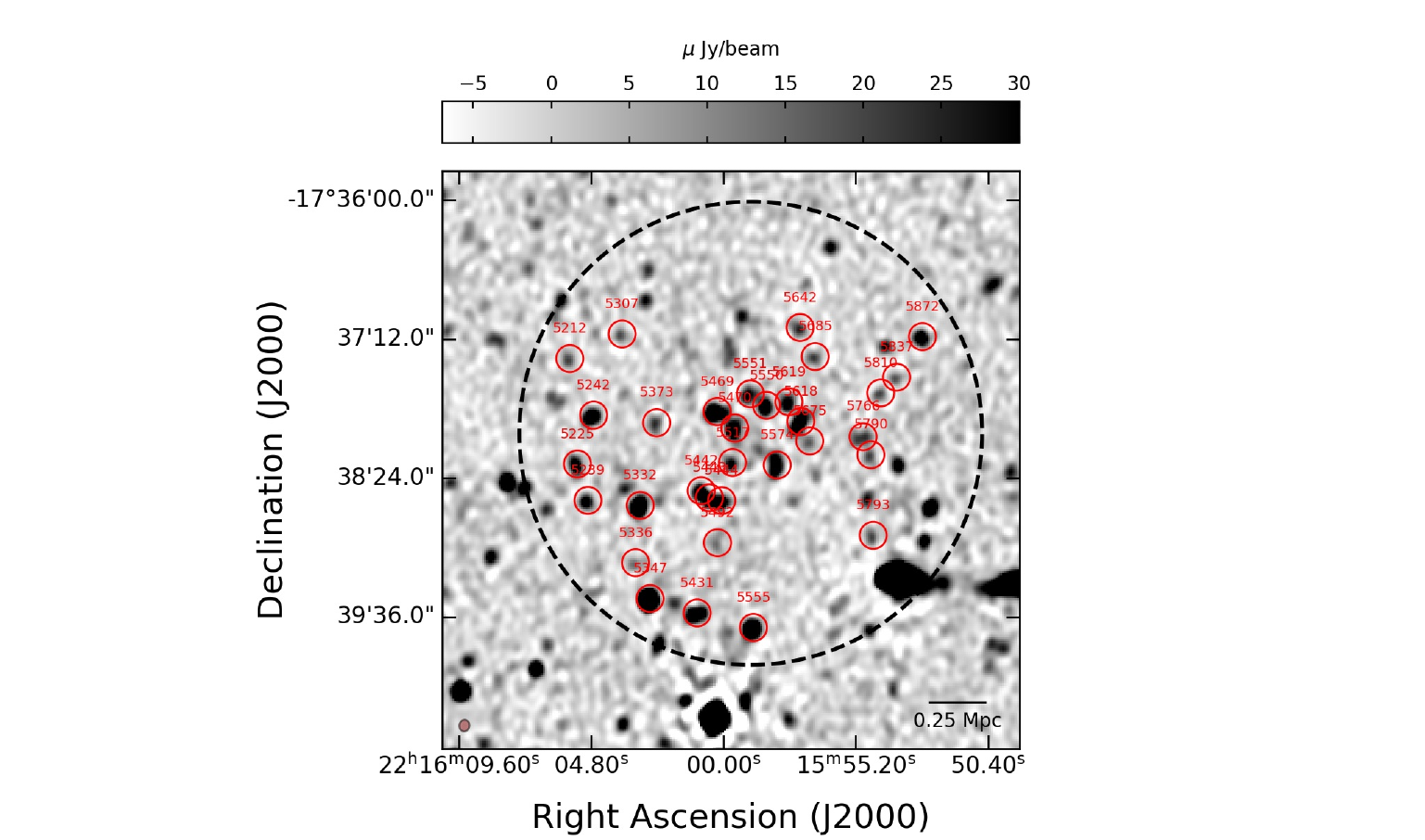}
	\caption{A zoom-in of the MeerKAT radio map centred on the cluster members.
	~The positions of the cluster members and the ID numbers of the sources in Table \ref{table:table1SFR} are shown with red circles and numbers respectively.~The $\approx$ 6 $^{\prime \prime}$ synthesized beam size is displayed in the lower-left corner of the image.}
	\label{fig:figure1 cluster members}
	\vspace{8pt}
\end{figure*}

\vspace{10pt}

\subsection{Archival Multi-Wavelength Observations} \label{sec:The Archival Observations}
To study the multi-wavelength properties of the radio sources detected in the cluster we made use of archival images, photometry or catalogues from  
sub-millimetre/far infra-red data from the Sub-millimetre Common User Bolometer Array 2, SCUBA-2 \citep[][]{2015ApJMa}, sub-millimetre/millimetre observations from Atacama Large Millimetre/Sub-millimetre Array, ALMA band 3 \citep{2010MNHayshi}, ALMA 1.25~mm (Band-6) observation \citep[][]{2017ApJStach}, optical data from the \textit{Hubble Space Telescope} (HST), Advanced Camera for Surveys, ACS and infrared data from the Multi-Object Infrared Camera and Spectrograph, MOIRCS  mounted on the 8.2~m Subaru telescope \citep{2009ApHilton}, the Infrared Array Camera (IRAC) on board the  \textit{Spitzer Space Telescope} \citep{2004ApJFazio}, the mid-infrared (24-$\mu$m) observations of the cluster obtained using the Multiband Imaging Photometer for Spitzer, MIPS \citep{2010ApJHilton} and the  PACS (Photodetecting Array Camera and Spectrometer) on board the \textit{Herschel Space Observatory}  \footnote{https://irsa.ipac.caltech.edu/Missions/herschel.html \label{foot}} .~A summary of the archival observations can be found below.



\textbf{SCUBA-2 450~$\mu$m and 850~$\mu$m}:~The J2215 cluster was observed in band 1 using the SCUBA-2 camera by \citet{2015ApJMa}  between July–August 2013.~The total observation time of the 450~$\mu$m and 850~$\mu$m surveys was 8 hours (i.e., 12 scans and each scan lasted for 40 mins).~For detailed information on  data reduction and imaging see \citet{2015ApJMa}.~The resulting noise level of the 450~$\mu$m and the 850~$\mu$m  data was 5.4 and 0.63 mJy~$\rm{beam}^{-1}$ respectively.

\textbf{ACS/HST}:~The J2215 galaxy cluster was among the 25 clusters observed during the HST type Ia Supernovae search observed with the Wide Field Channel (WFC) of the Advanced Camera for Surveys (ACS) mounted on the HST from July 2005 to December 2006.~The total integration time for the observation in the $i_{775}$ band was 3320~s (resulting in a 5$\sigma$ point source magnitude limit of $\approx{25.1}$) whilst a total of 16,935~s exposure time was recorded in the $z_{850}$ band (reaching a deeper magnitude limit of $\approx$ 26.0). ~Information on the observation is reported by \citet{2009AJDawson} also see \citet{2009ApHilton} for detailed information on the data processing. 

\textbf{MOIRCS/Subaru}:~Observation of the J2215 cluster was carried out with the Multi-Object Infrared Camera and Spectrograph  \citep[MOIRCS,][]{2006Ichikawa} mounted on the Subaru telescope.~The 5$\sigma$ point source limiting magnitude from the $J$ and $K_{s}$-band image was 24.~More detailed information on the observation can be found in \citet{2009ApHilton}.

\textbf{IRAC/Spitzer}:~The IRAC imaging of J2215 was obtained through Program 50333 using all four channels (Ch1-Ch4: 3.6, 4.5, 5.8 and 8.0$~\mu$m) on July 12 2008.~The total observation time was 1500s exposures.~The 5$\sigma$ point source magnitude limit obtained from the IRAC catalogue was $\approx$ 23 mag (Ch1 and Ch2) and $\approx$ 20 mag (Ch3 and Ch4) \citep[see][for details]{2010ApJHilton}.  

\textbf{MIPS/Spitzer observation}:~The target cluster was observed on June 21 2008 before the IRAC observation was conducted via the same proposal.~The total integration time for the observation was 27000~s.~The 50\% completeness limit of the 24~$\mu$m observation was estimated to be 70~$\mu$Jy at 5~$\sigma$.~Detailed information about data reduction and photometry can be found in \citet{2010ApJHilton}.

\textbf{ALMA (BAND 3)}:~A CO($J=$2–1) emission line survey of the cluster was by carried out by \citet{2017ApJHayashi} using  ALMA in band 3 in May 2016.~The total integration time for the observation was 3.12 hours.~Data reduction and processing were done using \textsc{CASA} with a standard pipeline.~The mosaicked 3D cubes with five different velocity resolutions 50, 100, 200, 400, and 600 km s$^{-1}$ yielded noise levels of 0.17, 0.12, 0.11, 0.12, and 0.12 mJy beam$^{-1}$ respectively   \cite [see] [for detailed information]{2017ApJHayashi}.  

\textbf{ALMA  (BAND 6)}:~The CO($J=$5–4) observations of the core of the J2215 were carried out by \citet{2017ApJStach} using forty-two 12 m antennae on June 19 2016 under the project ID: 2015.1.00575.S.~The six pointing observations yielded a synthesized beam size of $0. 66^{\prime \prime} \times 0. 47^{\prime \prime}$ (PA $=78^{\circ}$).~The rms noise recorded for the final mosaicked image was 48 $\mu$Jy beam $^{-1}$.~Detailed information on the data acquisition and reduction process can be found in \citet{2017ApJStach}. 

\textbf{PACS Herschel 100~$\mu$m (green) and  160~$\mu$m (Red)}: 
We obtained the J2215  PACS 100~$\mu$m and 160~$\mu$m data from the \textit{Herschel} Science Archive. We conducted  aperture photometry at the positions of the detected MeerKAT sources (see Section \ref{subsec:MeerKATSourceDetection} below)  using the \texttt{photoutils} package \citep{2016Bradley}. Fluxes were extracted within aperture sizes  6$^{\prime \prime}$  and  12$^{\prime \prime}$  for  the 100~$\mu$m and 160~$\mu$m  images respectively. The noise in the images was estimated using the median absolute deviation (MAD) of the data in source-free regions using apertures of  6$^{\prime \prime}$  and  12$^{\prime \prime}$. We obtained a noise level of 0.2 and 0.3~mJy  at 100~$\mu$m and 160~$\mu$m respectively. Comparing our values to that of the catalogue  from the \textit{Herschel}/PACS Point Source Catalogue (HPPSC) via the Herschel  User  Provided  Data Products $^{\ref{foot}}$, our values are consistent with that of the  HPPSC (i.e., a median ratio of  $\approx$ 1 for both bands). However, our values are higher by a factor of $\approx$  3  and  2  for the green and red bands respectively with reference to the values obtained by  \citet{2015ApJMa} even after conducting aperture photometry using similar aperture sizes used in that work i.e., 4.2$^{\prime \prime}$  and 8.5$^{\prime \prime}$  for the green and red bands respectively. Note that \citet{2015ApJMa} erroneously treat the 100\,$\mu$m PACS data as if it were 70\,$\mu$m data, although no 70\,$\mu$m PACS observations of J2215 exist in the archive.

\vspace{15pt}

\subsection{Photometric Redshifts} \label{subsec:Photometric_redshift}
The photometric redshift catalogue from \citet{2009ApHilton} was used in this work.~\citet{2009ApHilton} used the \texttt{EAZY} spectral template fitting code of \citet{2008ApJBrammer} to estimate the photometric redshift for the cluster galaxies.~A default option within \textsc{EAZY} was selected to fit a linear combination of all of the spectral energy distribution templates to the cluster galaxies available in the catalogue within a redshift range of 0 $<  z < $ 4.~The $K-$magnitude based Bayesian redshift prior used in the code resulted in a maximum likelihood redshift estimate $z_{p}$ and this estimated redshift value was adopted as the photometric redshift of the J2215 cluster galaxies, see \citet{2009ApHilton} for detailed information.

\vspace{15pt}

\section{Analysis} \label{sec:Source_Catalogue_and_membership_Classification}
\subsection{MeerKAT Source Detection} \label{subsec:MeerKATSourceDetection}
We extracted sources with $>4\sigma$ detection significance from the MeerKAT image using the Python Blob Detector and Source Finder \citep[\texttt{PyBDSF},][]{2015Mohan} resulting in an initial catalogue of 10,892 radio sources.~We performed a \enquote{negative peak analysis test} to quantify the level of noise contamination similar to the approach used by \citet{2019Patil}.~This was accomplished by inverting the MeerKAT continuum map and re-running PyBDSF  
using the same parameters we used for the original source extraction. We detected 154 negative sources as compared to the $>$ 10K sources detected in the original map which corresponds to a false positive detection rate of $\approx$ 1.4 $\%$.

We estimated a 4$\sigma$ flux detection limit of 14~$\mu$Jy from the MeerKAT image which translates to a limiting luminosity value of 1.5$\times$10$^{23}$ WHz$^{-1}$ at the cluster redshift of $z = 1.46$.~This luminosity is equivalent to SFR =  46 M$_{\odot}$yr$^{-1}$ when assuming the \citet{2003ApJBell} relation between radio continuum luminosity and SFR assuming a \citet{2003Chabrier} IMF.  

\vspace{15pt}

\subsection{XMMXCS J2215.9-1738 Cluster Membership Classification} 
We cross-matched the MeerKAT source catalogue with the archival photometric \citep{2009ApHilton} and spectroscopic \citep{2010ApJHilton} redshift catalogues of  J2215 at $z$ = 1.46 using a cross-matching radius of 6.0$^{\prime \prime}$ to determine their cluster membership. We selected galaxies as cluster members within $R_{200}$  based on their photometric or spectroscopic redshifts. For the photometric selection, we adopted a similar redshift cut-off that was used to define the cluster membership in \citet{2009ApHilton} (i.e., cluster galaxies within the redshift range 1.27  $<  z_{p}  <$  1.65 were assumed to be cluster members) whilst sources having spectroscopic redshifts = 1.46 $\pm$ 3$\sigma$ (where $\sigma$ is the line of sight velocity dispersion) were also selected as cluster members. We also cross-matched the MeerKAT source catalogues with catalogues obtained from [\textsc{Oii}] emitters via the narrow band (NB) observation of the J2215 cluster at $z$ = 1.46 using the Suprime-Cam on the Subaru Telescope. Two  sources were selected from the cross-match between the NB912 \citet[NB912 $\leftthreetimes$ = 9139\AA,~$\Delta \leftthreetimes$ = 134\AA][]{2010MNHayshi,2011Hayashi}, one source was detected using the NB921 and seven sources using the NB 912+921 catalogues \citep{2014Hayashi}.~In total, we detected 24 cluster members within 0.8~Mpc of the cluster centre (see Table \ref{table:table1SFR}).~The median angular size of the selected sources presented in Table \ref{table:table1SFR} is $\approx$ 7 $^{\prime \prime}$ which suggests that the majority of our selected radio sources are barely resolved given a synthesized beam size of $\approx$ 6$^{\prime \prime}$.  

We further cross-matched these 24 cluster members detected in the MeerKAT image with catalogues from other archival infrared observations as described in Section \ref{sec:The Archival Observations}. 
The final sample of the J2215 cluster members in the MeerKAT $L$-band and their counterparts in other archival observations are summarized in Table \ref{table:table222}.
We discuss how the far infrared and radio luminosities were derived in Sections \ref{subsection:SEDCIGALE} and   \ref{sec:Radio Luminosities} respectively.
\vspace{10pt}

\subsection{SED Fitting with CIGALE} \label{subsection:SEDCIGALE}
We performed SED fitting using  \texttt{CIGALE V2022.0} \citep{2022ApJYang}  to derive the far-IR luminosity values of the cluster members.
We used the following models; the delayed star formation history model SFH model \citep[see equation 4 of][]{2019Boquien}, the single stellar populations (SSP) model of \citet{2003MNRASBruzual}, the nebular emission template adopted based on \citet{2011MNRASInoue}, dust attenuation templates based on the \citet{2000ApJCharlot} model, dust emission template of \citet{2014ApJDraine}, AGN models from \citet{2016MNRAStalevski} and galaxy radio synchrotron emission based on the radio-infrared correlation of \citet{1985ApJelou}. {\tt{CIGALE V2022.0}} also models the AGN components of galaxies using the radio-loudness parameter \citep{2012Ballo} and assumes an AGN power-law SED within the wavelength range 0.01 - 100 cm. The fitting parameters we used in this work are listed in Table \ref{table:CigaleParameters}.~One of the output parameters derived from the SED fitting was the infrared luminosity.

~Comparing the infrared luminosity values derived after fitting the models to the observed SEDs (i.e., optical to radio bands, see Appendix \ref{Appendix:sedplots}) to that of the counterpart values in Table 2 of \citet{2015ApJMa} we observed that the infrared luminosity values derived in this work are about a factor of 2.4 higher than that estimated in \citet{2015ApJMa}.~We probed further and realized that the J2215 data was taken using the 100~$\mu$m and the 160~$\mu$m filters of the PACS camera on board the \textit{Herschel Space Telescope} and not the 70~$\mu$m filters presented in \citet{2015ApJMa} and therefore the luminosity values they obtained after the SED fit are likely to be in error (see Section  \ref{sec:The Archival Observations}).
We also fitted the SED models to only the mid-infrared to the radio data \citet[the exact bands used in][]{2015ApJMa} to ascertain whether the inclusion of the optical and near-infrared bands had any impact on the results generated, we obtained a median ratio of 1.6 which is still higher than the value obtained in \citet{2015ApJMa}. 

We also compared the MeerKAT flux densities reported in this work to those of the 9 detected JVLA sources reported in \citet{2015ApJMa}. We find a median flux ratio of 1.34, with the MeerKAT flux densities being higher on average. However, only 6/9 of the detected JVLA sources reported on in \citet{2015ApJMa} are considered to be cluster members in our work, and as indicated in Table~\ref{table:table1SFR}, all of these are either blended in the MeerKAT image, or are thought to be AGN, which may explain the higher MeerKAT flux densities for these sources. We have verified that the MeerKAT flux density measurements in our catalogue are within 5 per cent on average of sources detected in the NRAO VLA Sky Survey \citep[NVSS;][]{1998Condon}. 

We show in Appendix \ref{subsect:CIGALE Model_Parameters} examples of SED fitting using  \texttt{CIGALE V2022.0}, the goodness of the fit is measured via the reduced $\chi^{2}$ value. 
The median reduced $\chi^{2}$ value is 1.9 for all galaxies selected as cluster members.~In total, 17/24 (71\%) sources had a reduced $\chi^{2}$ value of less than 3. Whilst 20/24 (83\%) sources had reduced $\chi^{2}$ value less than 5. This implies that  the observed SED data are reasonably well-fitted with the models we selected for the SED fitting with \texttt{CIGALE V2022.0}.

\begin{table*}
	\centering
	\addtolength{\tabcolsep}{5pt}
	\begin{tabular}{rrrrrrrrrrrrr}  
		\hline   \hline 
		\multicolumn{1}{c}{(1)} & 
		\multicolumn{1}{c}{(2)} &
		\multicolumn{1}{c}{(3)} &
		\multicolumn{1}{c}{(4)} &
		\multicolumn{1}{c}{(5)} & 
		\multicolumn{1}{c}{(6)} &  
		\multicolumn{1}{c}{(7)} &  
		\multicolumn{1}{c}{(8)}  \\  
		\multicolumn{1}{c}{MKT ID} & 
		\multicolumn{1}{c}{R.A.} & 
		\multicolumn{1}{c}{Dec.} &  
		\multicolumn{1}{c}{$F_{1.3}$} &  
		\multicolumn{1}{c}{$L_{1.4}$} &  
		\multicolumn{1}{c}{$q_{IR}$} & 
            \multicolumn{1}{c}{SFR} &  
	 	\multicolumn{1}{c}{$R$}  \\  
		\multicolumn{1}{c}{} & 
		\multicolumn{1}{c}{(J2000)} & 
		\multicolumn{1}{c}{(J2000)} &  
		\multicolumn{1}{c}{($\mu$Jy)} &  
		\multicolumn{1}{c}{($10^{23}$ WHz $^{-1}$)} & 
            \multicolumn{1}{c}{} &
		\multicolumn{1}{c}{(M$_{\odot}$yr$^{-1}$)} &    
		\multicolumn{1}{c}{(Mpc)}  \\
		\hline 
5242$^{\ast}$&22:16:04.79&$-$17:37:52.5&102$\pm$8&10.6$\pm$0.8& 2.37$\pm$0.25&336$\pm$26&0.76  \\  
5307&22:16:03.75&$-$17:37:10.4&24$\pm$7&2.5$\pm$0.8&2.53$\pm$0.37 &81$\pm$24&0.77  \\  
5332$^{\ast}$&22:16:03.10&$-$17:38:39.3&197$\pm$8&20.4$\pm$0.8&1.74$\pm$0.39 &650$\pm$26&0.64  \\  
5336&22:16:03.26&$-$17:39:09.0&18$\pm$7&1.9$\pm$0.7& 1.95$\pm$1.0&60$\pm$23&0.80  \\  
5373&22:16:02.50&$-$17:37:56.5&40$\pm$9&4.2$\pm$1.0&1.76$\pm$0.35&133$\pm$31&0.49  \\  
5442$\dagger$&22:16:00.88&$-$17:38:31.7&77$\pm$7&8.0$\pm$0.7&2.48$\pm$0.22&254$\pm$22&0.38  \\  
5443$\dagger$&22:16:00.59&$-$17:38:35.3&61$\pm$6&6.3$\pm$0.6&2.55$\pm$0.22&201$\pm$20&0.38  \\  
5469$^{\ast}$&22:16:00.30&$-$17:37:50.7&174$\pm$12&18.1$\pm$1.3&1.85$\pm$0.42&576$\pm$40&0.24 \\  
5470$^{\ast}$$\dagger$&22:15:59.67&$-$17:37:59.2&125$\pm$11&13.0$\pm$1.2&2.45$\pm$0.35&412$\pm$37&0.14  \\  
5492&22:16:00.29&$-$17:38:58.7&21$\pm$8&2.2$\pm$0.8&2.19$\pm$0.26&   70$\pm$26&0.52  \\  
5517$\dagger$&22:15:59.75&$-$17:38:17.0&47$\pm$7&4.9$\pm$0.8& 2.48$\pm$0.24&154$\pm$24&0.19  \\  
5551$\dagger$&22:15:59.10&$-$17:37:41.5&69$\pm$11&7.2$\pm$1.1& 2.45 $\pm$0.34&228$\pm$35&0.19  \\  
5574$^{\ast}$&22:15:58.13&$-$17:38:18.3&108$\pm$10&11.2$\pm$1.1& 2.21$\pm$40.41& 356$\pm$34&0.14  \\  
5618$^{\ast}$$\dagger$&22:15:57.27&$-$17:37:55.8&140$\pm$11&14.5$\pm$1.1&2.01$\pm$0.33&462$\pm$35&0.16  \\  
5619$\dagger$&22:15:57.70&$-$17:37:45.6&66$\pm$8&6.9$\pm$0.8& 2.31$\pm$0.37&219$\pm$25&0.17  \\  
5642&22:15:57.30&$-$17:37:07.1&42$\pm$9&4.3$\pm$1.0&2.18$\pm$0.5&137$\pm$30&0.49  \\  
5675&22:15:56.95&$-$17:38:05.9&28$\pm$9&2.9$\pm$0.9&2.32$\pm$0.38&92$\pm$30&0.19  \\  
5685&22:15:56.75&$-$17:37:22.2&29$\pm$8&3.0$\pm$0.8&2.26$\pm$0.59&96$\pm$25&0.40  \\  
5766$\dagger$&22:15:55.01&$-$17:38:03.7&73$\pm$14&7.6$\pm$1.5&2.50$\pm$0.45&   242$\pm$47&0.42 \\  
5790&22:15:54.73&$-$17:38:13.0&30$\pm$8&3.1$\pm$0.9&...&99$\pm$27&0.46  \\  
5793&22:15:54.65&$-$17:38:54.8&29$\pm$8&3.0$\pm$0.8&2.55$\pm$0.27&96$\pm$27&0.64  \\  
5810&22:15:54.37&$-$17:37:41.1&23$\pm$7&2.4$\pm$0.7&...&76$\pm$22&0.53  \\  
5837&22:15:53.80&$-$17:37:32.9&21$\pm$7&2.2$\pm$0.7&2.49$\pm$0.34&71$\pm$22&0.62  \\  
5872&22:15:52.86&$-$17:37:11.8&74$\pm$7&7.7$\pm$0.7&2.19$\pm$0.22&244$\pm$23&0.80   \\  \hline 
	\end{tabular}
	\caption{The MeerKAT-detected cluster galaxies within 0.8 Mpc of the cluster centre. Columns: (1) The ID numbers of each source in the MeerKAT $L$-band image selected as a cluster member; (2) MeerKAT pointing coordinates of each galaxy the units of right ascension (R.A.) are in hours, minutes and seconds; (3) The units of declination (Dec) are in degrees, arcminutes, and arcseconds; (4) The observed flux measured in $\mu$Jy; (5) Radio luminosities are measured in WHz$^{-1}$; (6) The far infrared radio luminosity Ratio; (7) Star formation rates measured in M$_{\odot}$yr$^{-1}$. ~Note that these estimates will not be reliable for galaxies where the radio emission is primarily due to the presence of an AGN; (8) Radial distance of sources measured with respect to the  cluster X-ray position  measured in Mpc. The ~potential radio AGNs are marked with asterisks ($^{\ast}$) whilst objects blended in the MeerKAT radio image are denoted by the obelisk mark ($\dagger$).}
	\label{table:table1SFR}
\end{table*}

\begin{table*}
	\centering
	\addtolength{\tabcolsep}{2pt}
	\small
	\begin{tabular}{rrrrrrrrrrrrr}  
		\hline   \hline
		\multicolumn{1}{c}{(1)}  &
		\multicolumn{1}{c}{(2)}  &
		\multicolumn{1}{c}{(3)}  &  
		\multicolumn{1}{c}{(4)} &  
		\multicolumn{1}{c}{(5)} &  
		\multicolumn{1}{c}{(6)}  &  
		\multicolumn{1}{c}{(7)}  & 
		\multicolumn{1}{c}{(8)}  & 
		\multicolumn{1}{c}{(9)} \\  
		\multicolumn{1}{c}{MKT ID} & 
		\multicolumn{1}{c}{M15ID} &
		\multicolumn{1}{c}{H9ID} &  
		\multicolumn{1}{c}{H10ID} &  
		\multicolumn{1}{c}{HY17ID} &  
		\multicolumn{1}{c}{ST17ID}  &  
		\multicolumn{1}{c}{C18}  & 
		\multicolumn{1}{c}{Morph}  & 
		\multicolumn{1}{c}{log$_{10}$M$_{\ast}$[M$_{\odot}$]} & 
		\multicolumn{1}{c}{Notes} \\ 
		\hline 	
		5242$^{\ast}$&10&709&...&...&...&...&Sp+Irr&11.11$\pm$0.06&... \\  
		5307&...&460&...&...&...&...&Sp+Irr&9.58$\pm$0.09&... \\  
		5332$^{\ast}$&2&1022&...&...&...&...&Sp+Irr&9.72$\pm$0.15&... \\  
		5336&...&1201&...&...&...&...&Sp+Irr&9.61$\pm$0.08&... \\  
		5373&...&728&...&...&...&...&Sp+Irr&10.63$\pm$0.11&... \\  
		5442$\dagger$&3&983&...&ALMA.B3.16&...&...&Sp+Irr&11.02$\pm$0.12&NB921 [O II] \\  
		5443$\dagger$&...&1004&...&...&...&575&E&9.81$\pm$0.13&... \\  
		5469$^{\ast}$&...&692&53&ALMA.B3.14&5&1011&E&11.06$\pm$0.10&NB912 [O II] 
		\\ 
		5470$^{\ast}$$\dagger$&6&747&...&ALMA.B3.06&9&930&Sp+Irr&11.49$\pm$0.07&NB921 [O II] \\  
		5492&...&1130&...&ALMA.B3.17&...&...&Sp+Irr&11.29$\pm$0.05&NB912+NB921 [O II] \\  
		5517$\dagger$&...&874&...&ALMA.B3.13&12&724&...&11.01$\pm$0.09& ... \\  
		5551$\dagger$&...&614&...&ALMA.B3.11&2&1118&E&11.03$\pm$0.05&NB912+NB921 [O II] \\  
		5574$^{\ast}$&4&884&899&ALMA.B3.05&13&732&...&9.82$\pm$0.16&NB921 [O II] \\  
		5618$^{\ast}$$\dagger$&13&734&35&ALMA.B3.07&7&940&Sp+Irr&9.64$\pm$0.02&NB912+NB921 [O II] \\  
		5619$\dagger$&...&653&...&ALMA.B3.09&...&1038&Sp+Irr&11.25$\pm$0.11&NB912+NB921 [O II] \\  
	5642&...&419&...&...&...&...&E&9.48$\pm$0.13&NB912+NB921 [O II]\\  
	5675&...&786&39&ALMA.B3.12&...&829&Sp+Irr&10.65$\pm$0.25&NB912 [O II] \\  
		5685&...&529&529&...&...&1261&Sp+Irr&10.93$\pm$0.08&... \\  
		5766$\dagger$&...&777&...&...&...&798&S0&9.85$\pm$0.06&... \\  
		5790&...&863&...&...&...&...&...&...&... \\  
		5793&...&1118&1118&...&...&412&...&11.27$\pm$0.08&... \\  
		5810&...&658&...&...&...&...&...&...&... \\  
		5837&...&571&...&...&...&...&...&10.41$\pm$0.14&... \\  
		5872&...&446&...&...&...&...&...&11.52$\pm$0.09&... \\  
		\hline 
	\end{tabular}
	\caption{We compare the sources detected in the MeeKAT $L$-band image to other archival observations. 
   Columns: (1) Same as in Table \ref{table:table1SFR}; 
   (2) source ID numbers from the SCUBA-2 observation compiled by \citet[][]{2015ApJMa}; 
   (3) ACS/HST  \citet[][]{2009ApHilton}; 
   (4) 24$\mu$m  MIPS/Spitzer observation \citet[][]{2010ApJHilton}; 
   (5) ALMA (Band 3) \citet[][]{2017ApJHayashi}; 
   (6) ALMA (Band 6) \citet[][]{2017ApJStach}; 
   (7) KMOS \citet[][]{2018ApJChan}; 
   (8) Morphology E=elliptical, S0=lenticular, Sp=spiral, Irr=irregular  \citet[see Section 3.2 of][for details]{2009ApHilton};
   (9) $\log_{10} M_{\ast}$[M$_{\odot}$];  the logarithmic expression of the cluster galaxy stellar mass.}
	\label{table:table222}
	\vspace{-4pt}	
\end{table*}

\vspace{10pt}

\subsection{Radio Luminosities and Continuum Star Formation Rates}\label{sec:Radio Luminosities}
The observed 1.3~GHz radio flux $F_{1.3}$, obtained from the MeerKAT survey was scaled to the restframe 1.4~GHz  $\left(F_{1.4}\right)$ using a power law with a slope ($\alpha$) of $-$0.8  \citep[i.e., $F_{1.4} = \left(\nu_{1.4}/\nu_{1.3}\right)^{\alpha}$ $\times$ F$_{1.3}$, this value of $\alpha$ is assumed for non-thermal radio emissions,][]{2003ApJBell} and then converted  the F$_{1.4}$ into a rest-frame 1.4 GHz luminosity ($L_{1.4}$) assuming a spectral index ($\alpha$) dependent K-correction factor \citep{1992Condon,2011Karim},
\begin{equation}
   L_{1.4}  = \dfrac{9.52 \times 10^{12}  \times   F_{1.4} D_{L}^{2}   4\pi }{(1 + z)^{1 + \alpha}}, 
\end{equation}\label{eq:Qvalue2}where, $D_{L}$ is the luminosity distance in Mpc, $L_{1.4}$ in units of WHz$^{-1}$, $F_{1.4}$ in units of $\mu$Jy and  $z$ = 1.46.
We converted the $L_{1.4}$ into SFR using the calibration of the Far Infrared Radio Correlation (FIRRC)  by \citet{2003ApJBell} scaled to a Chabrier IMF  \citep{2011Karim}, i.e.,	
\begin{equation}
	\text{SFR}~\text{(M$_{\odot}$yr$^{-1}$)} =
	\begin{cases}
		3.18 \times 10 ^{-22} L_{1.4},
		\qquad L_{1.4} > L_{c}, \\ \\
		\dfrac{ 3.18 \times  10 ^{-22} L_{1.4}}{0.1 + 0.9 
			\left(\dfrac{L_{1.4}}{L_{c}}\right)^{0.3}}, 
		\qquad L_{1.4} \leqslant L_{c},       
	\end{cases}
	\label{eq:SFR}
\end{equation}
where L$_{c}$ is the radio luminosity of an L$_{\ast}$ galaxy (i.e., $ 6.4 \times 10^{21}$ ~WHz$^{-1}$).~\citet{2003ApJBell} sees the need to distinguish between low-luminous and high-luminous star-forming galaxies since non-thermal (low-luminous) galaxies may have their SFR rate underestimated.~Though the MeerKAT observation mostly exploits the non-thermal radio regime, our radio luminosities mostly fall above the L$_{\rm{c}}$ threshold i.e., L$_{1.4} \geq 10^{23}$ WHz$^{-1}$ and also fall below radio luminosity  (L$_{\rm{rad}}$) value of 10$^{25}$ WHz$^{-1}$ where radio emissions at L$_{\rm{rad}}$  $\gtrsim$  10$^{24-25}$ WHz$^{-1}$ are mostly dominated by radio loud AGNs \citep{1990Miller,2001Yun}.

\section{Morphologies and optical colours of radio sources in J2215} \label{subsection:MORPHOLOGY_COLOURMAGNITUDE}
AGN host galaxies can be identified via their colours and morphologies. Previous colour-magnitude relation studies have shown that most galaxies found on the red sequence or blue cloud are normal galaxies \citep{2004ApJBaldry}
whilst  the majority of the AGNs are located in the green valley \citep{2007ApJNandra,2010ApJSchawinski,2012Povic,
	2017Wang,2018ApJGu,2020MNRASLacerda}.
The green valley represents a transitional region between the blue cloud (mostly populated by late-type galaxies having spiral or irregular morphologies) and the red sequence (populated by early-type galaxies having elliptical or lenticular morphologies).
In this work, we investigated the position of the MeerKAT-detected cluster members on the colour-magnitude diagram and their morphologies.~We describe the morphological classification in Section \ref{subsec:Morphology} and the position of the morphologically classified  radio galaxies on the colour-magnitude diagram in Section \ref{subsec:ColourMagnitude}.  


\subsection{Morphological Classification} \label{subsec:Morphology}
\begin{figure*}
	\centering
	\includegraphics[width=2.5cm,height=2.5cm]{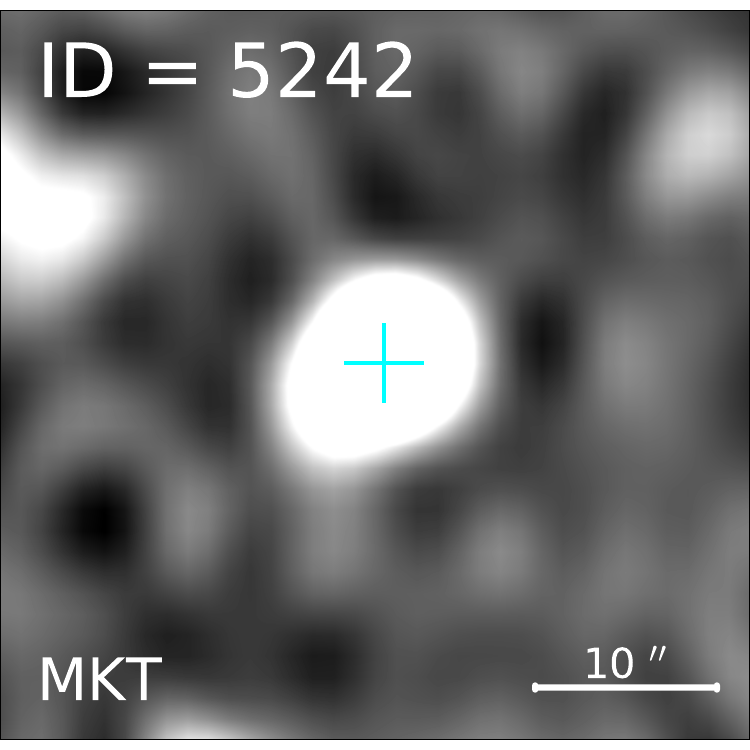}\hspace{-2.00mm}
	\includegraphics[width=2.5cm,height=2.5cm]{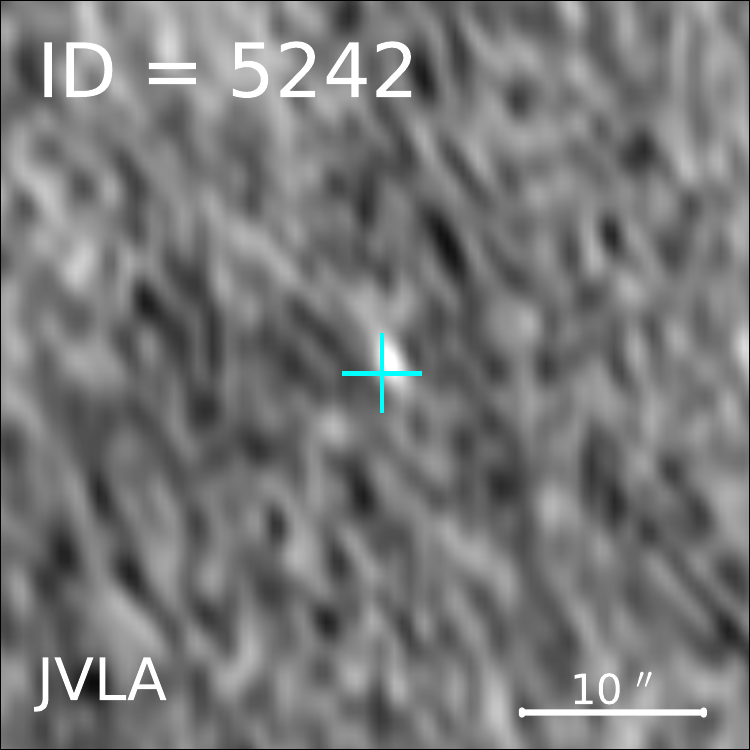}\hspace{-2.00mm}
	\includegraphics[width=2.5cm,height=2.5cm]{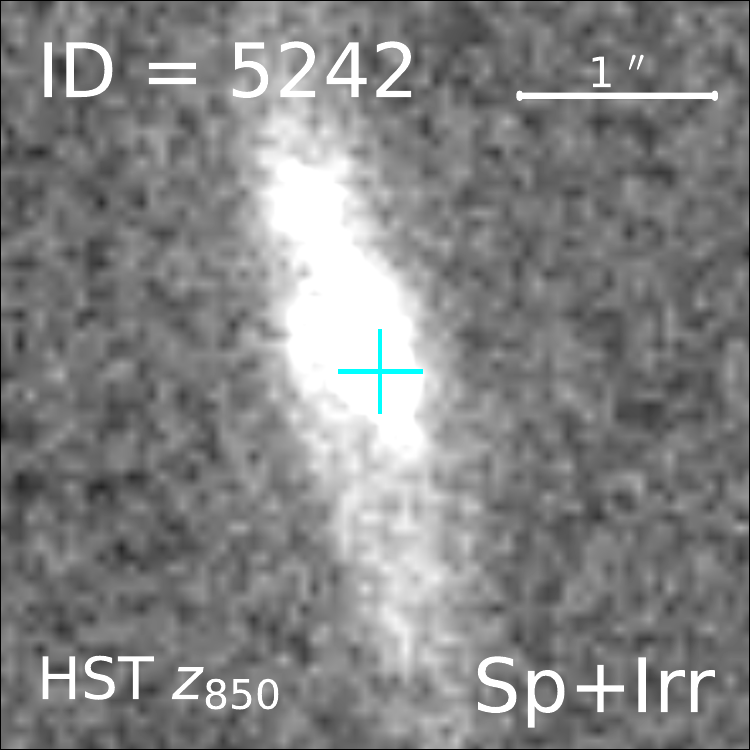}
	\includegraphics[width=2.5cm,height=2.5cm]{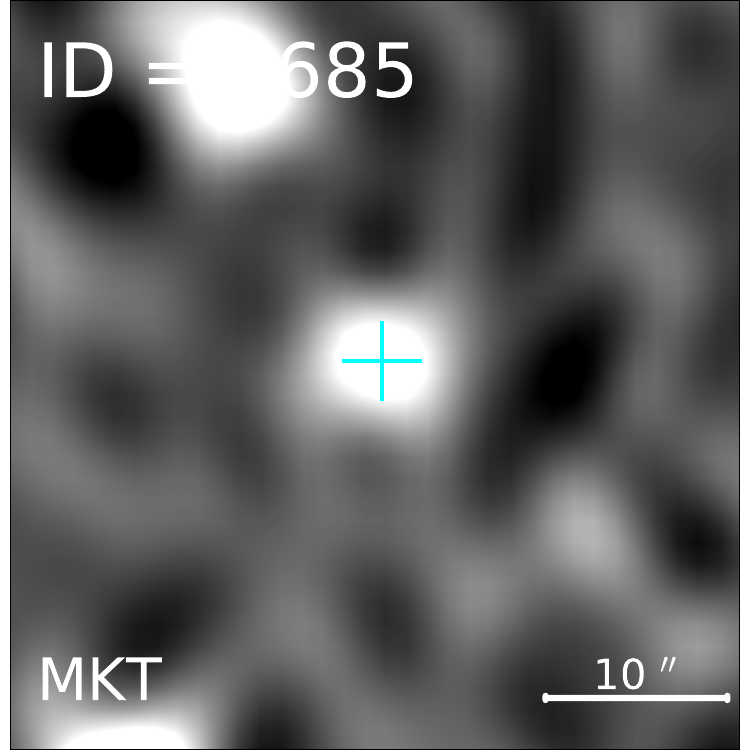}\hspace{-2.00mm}
	\includegraphics[width=2.5cm,height=2.5cm]{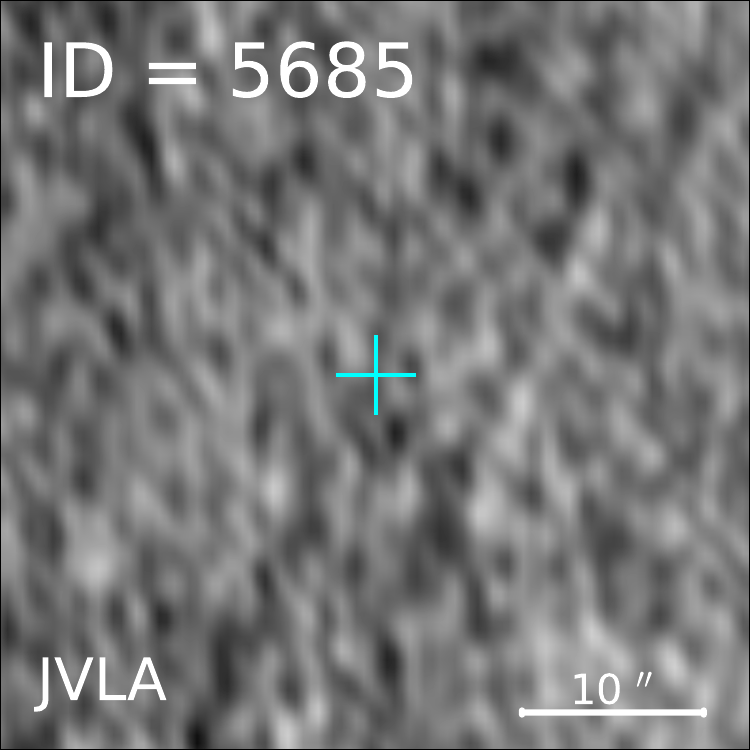}\hspace{-2.00mm}
	\includegraphics[width=2.5cm,height=2.5cm]{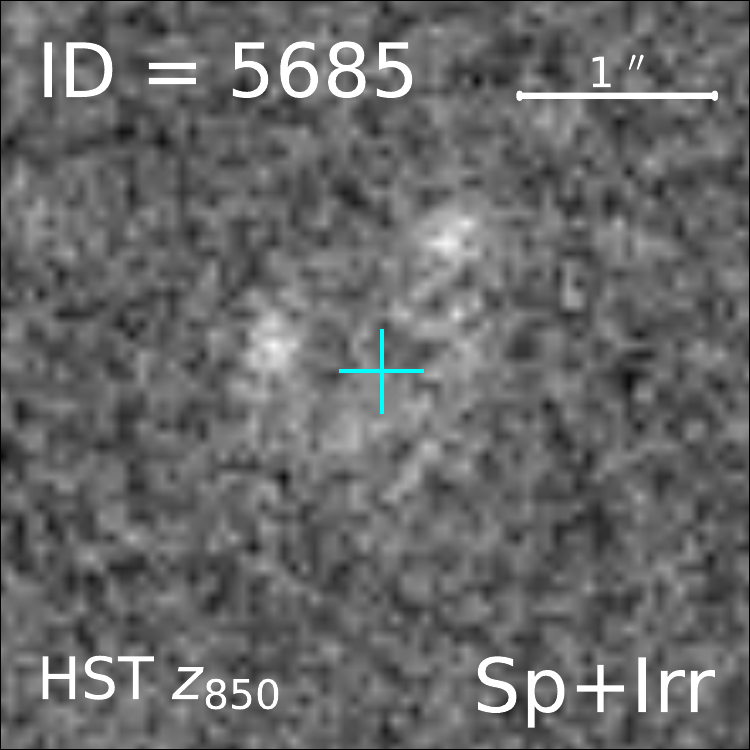}  \\
	\includegraphics[width=2.5cm,height=2.5cm]{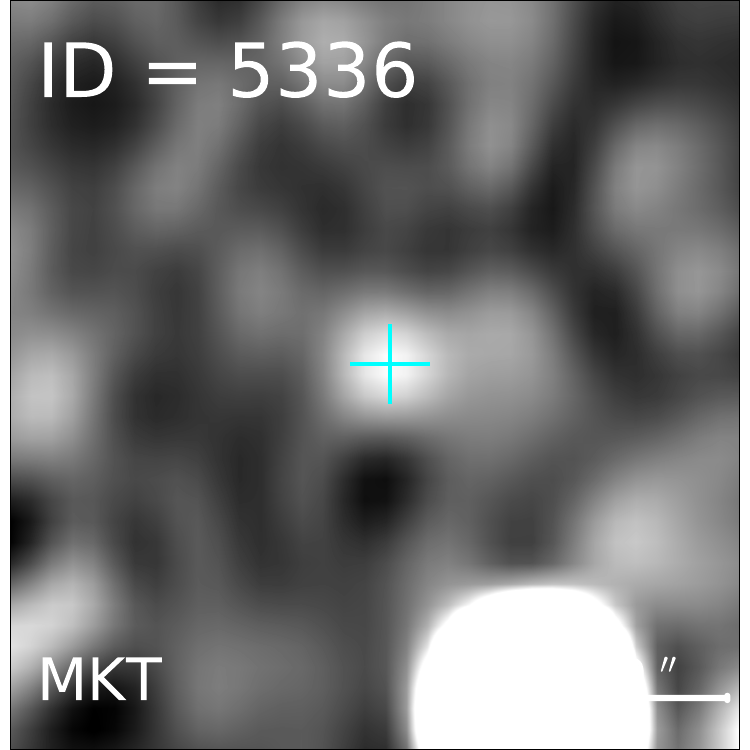}\hspace{-2.00mm}
	\includegraphics[width=2.5cm,height=2.5cm]{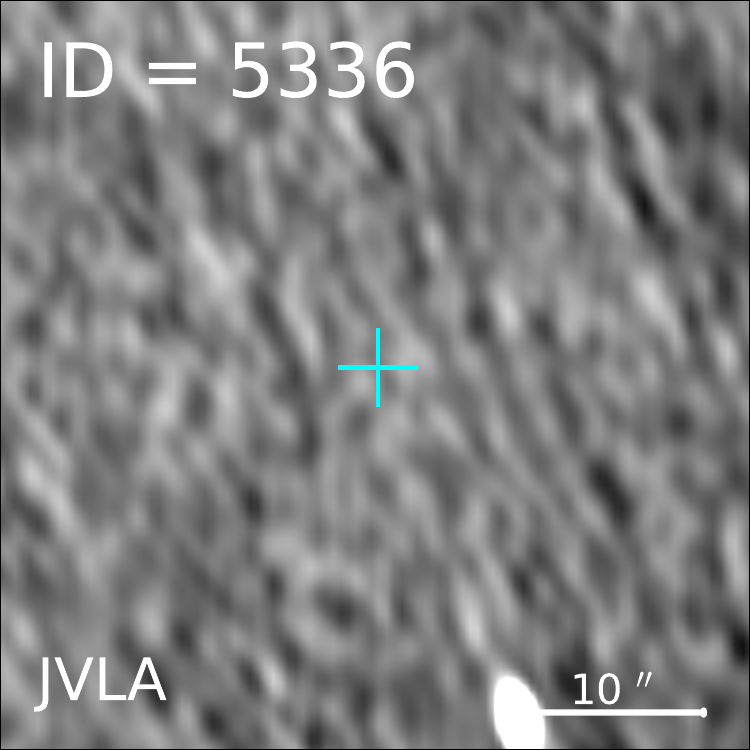}\hspace{-2.00mm}
	\includegraphics[width=2.5cm,height=2.5cm]{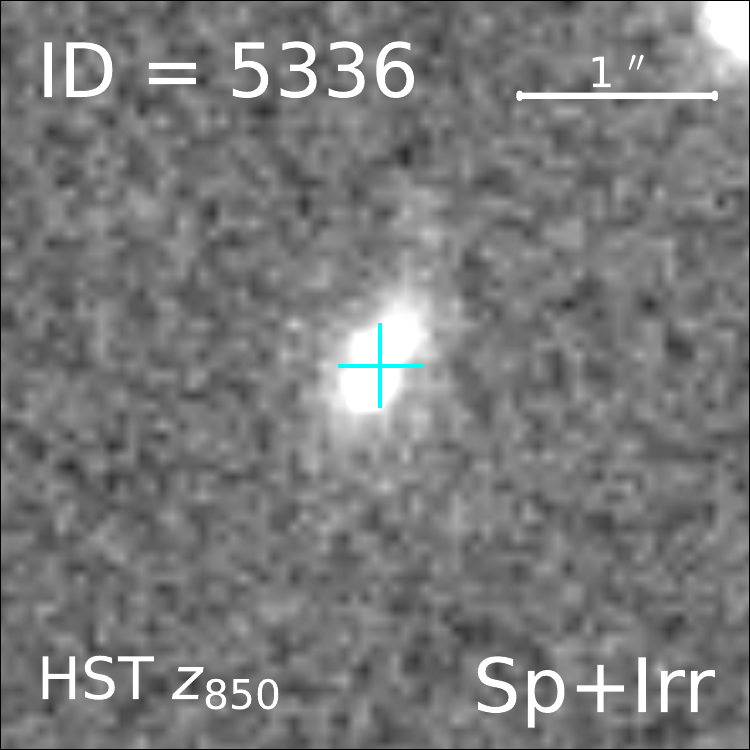}  
	\includegraphics[width=2.5cm,height=2.5cm]{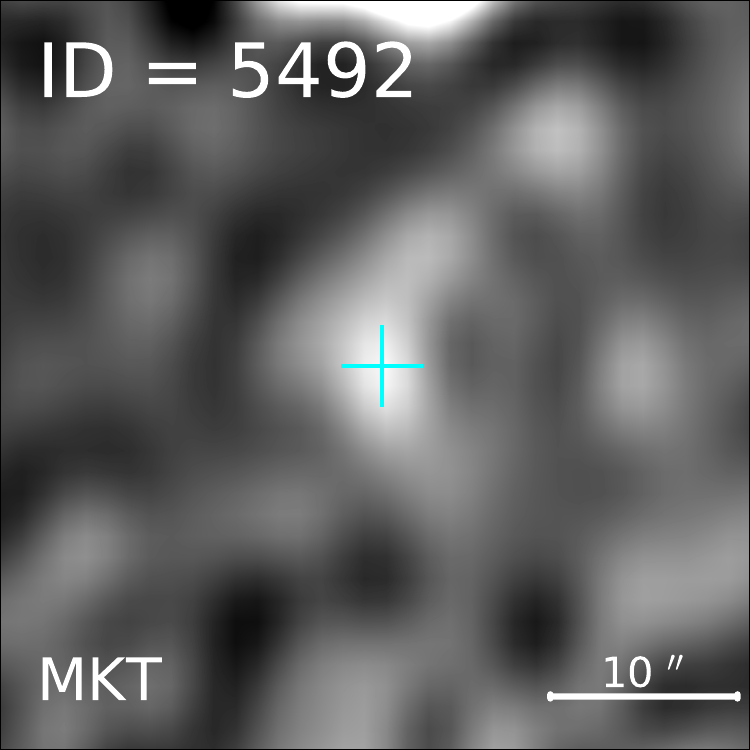}\hspace{-2.00mm}
	\includegraphics[width=2.5cm,height=2.5cm]{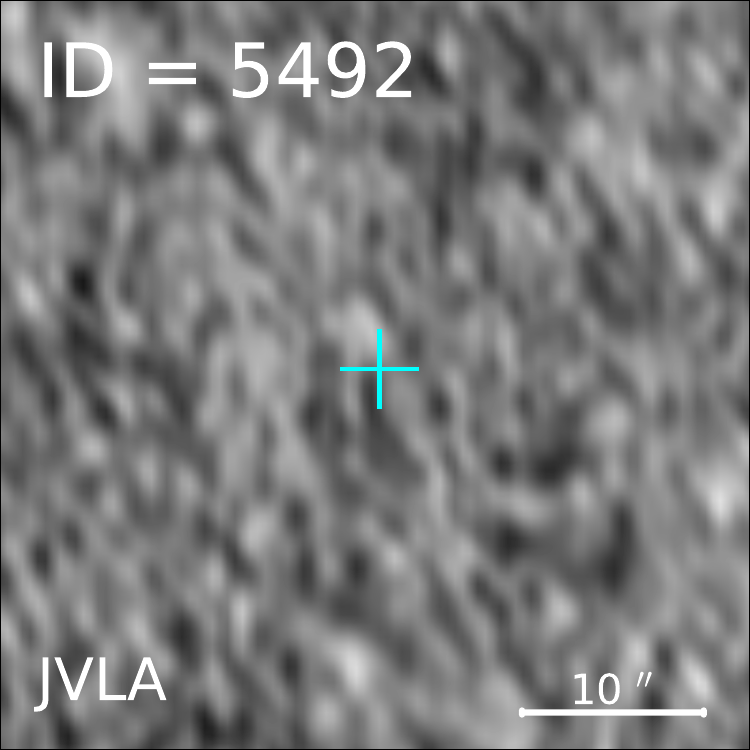}\hspace{-2.00mm}
	\includegraphics[width=2.5cm,height=2.5cm]{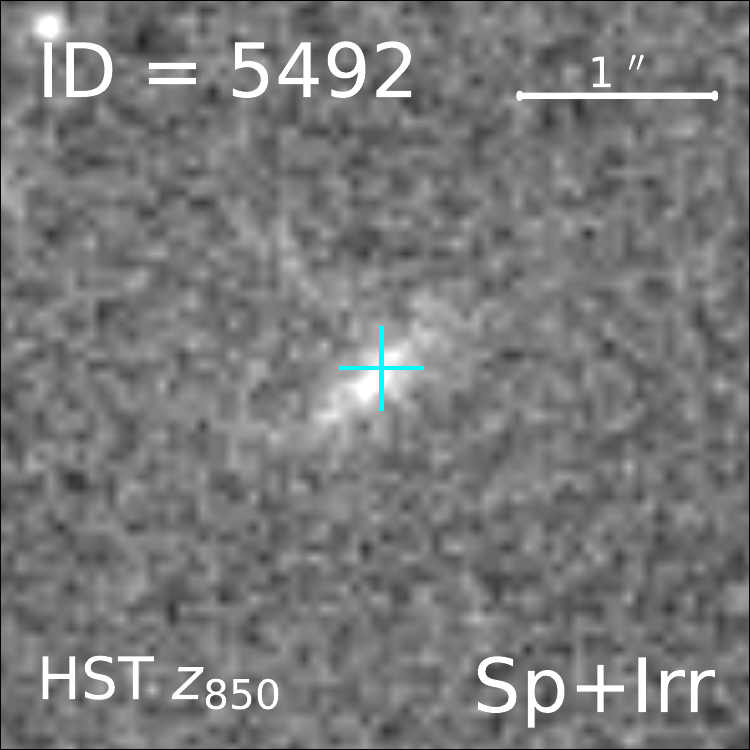} \\
	\includegraphics[width=2.5cm,height=2.5cm]{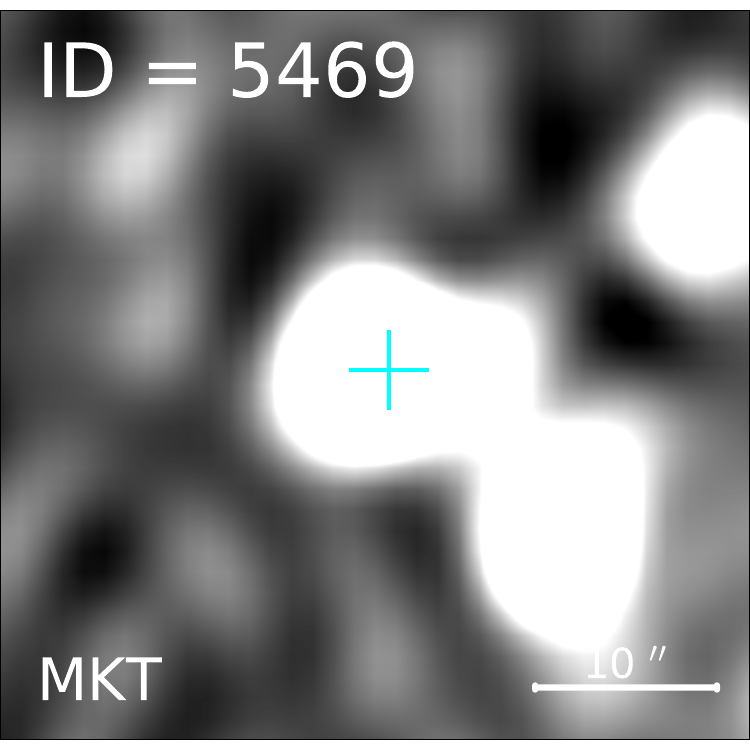}\hspace{-2.00mm}
	\includegraphics[width=2.5cm,height=2.5cm]{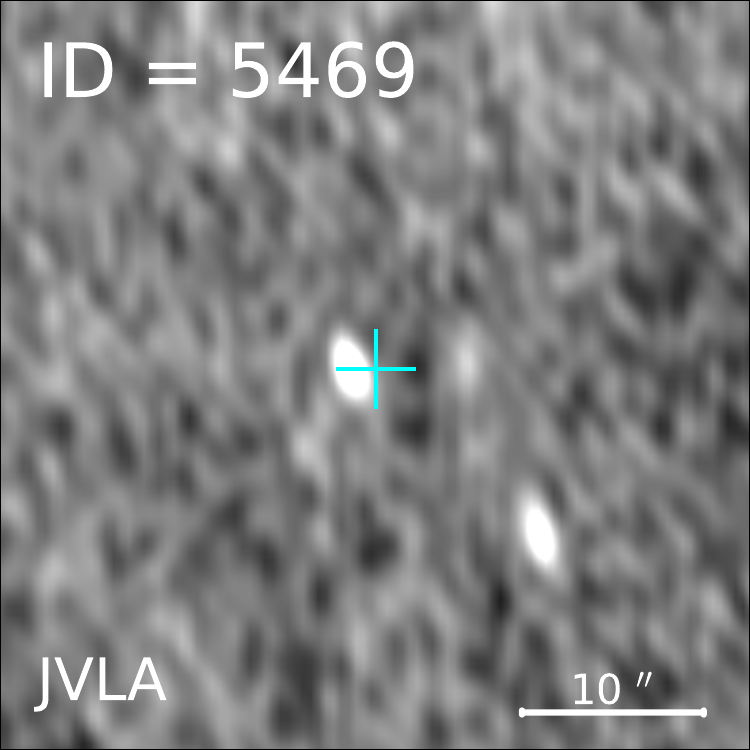}\hspace{-2.00mm}
	\includegraphics[width=2.5cm,height=2.5cm]{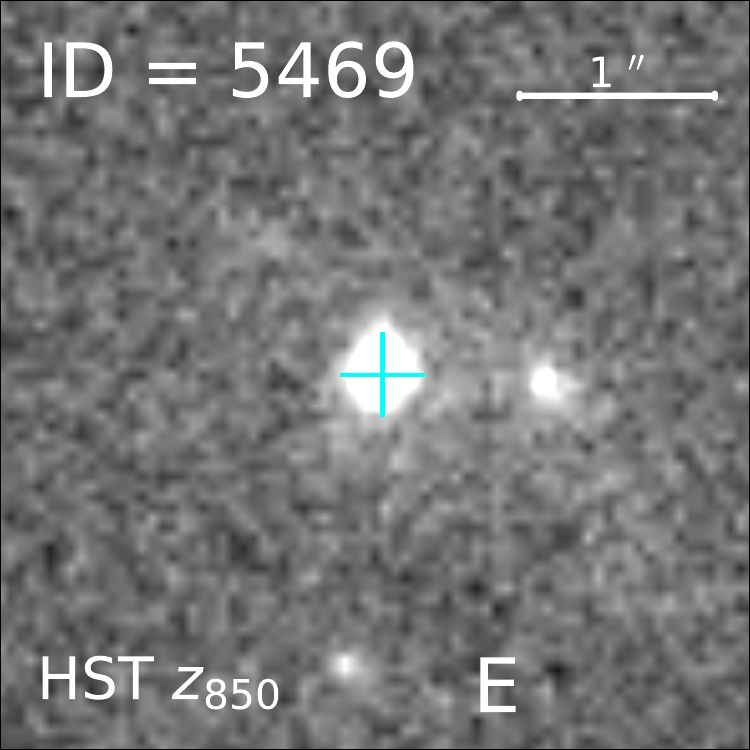}    
	\includegraphics[width=2.5cm,height=2.5cm]{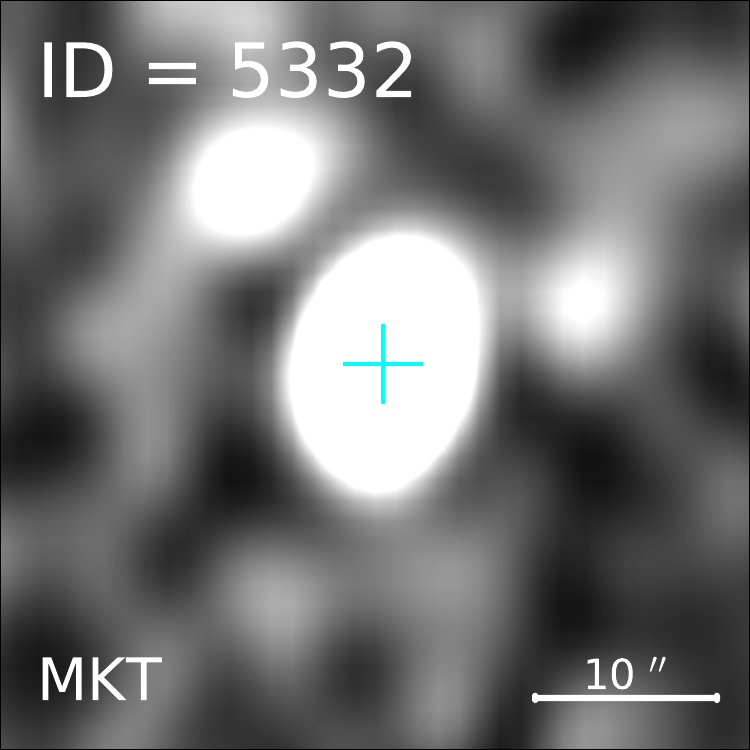}\hspace{-2.00mm}
	\includegraphics[width=2.5cm,height=2.5cm]{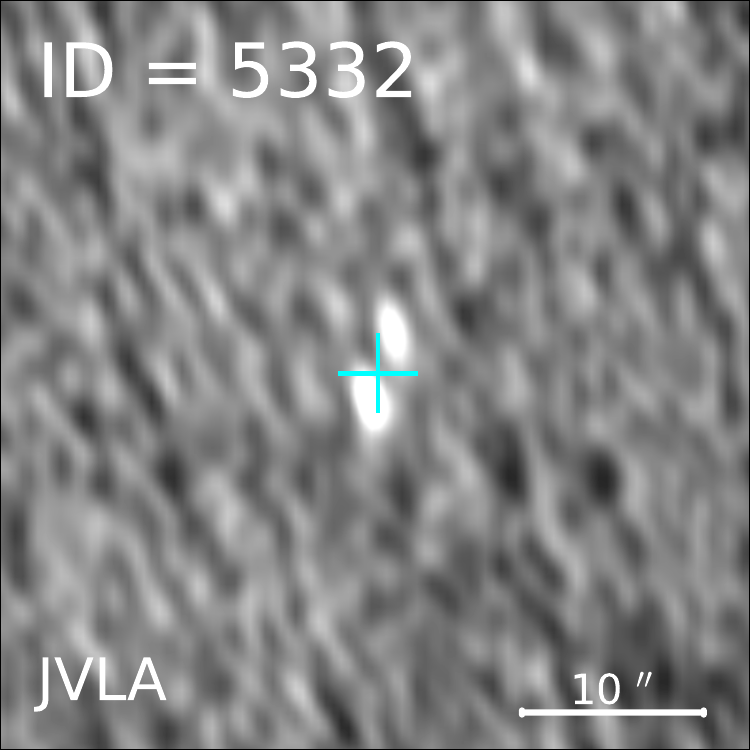}\hspace{-2.00mm}
	\includegraphics[width=2.5cm,height=2.5cm]{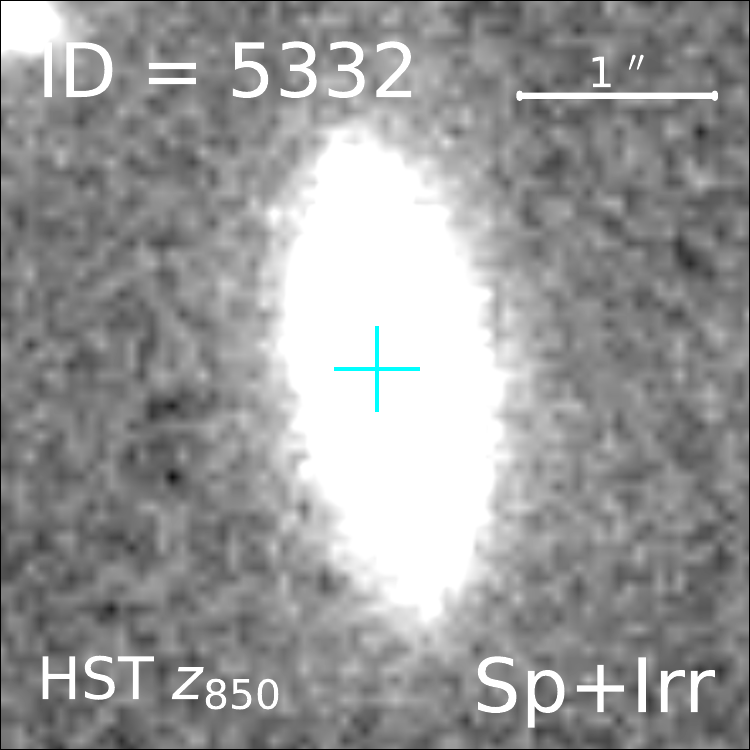} \\
	\includegraphics[width=2.5cm,height=2.5cm]{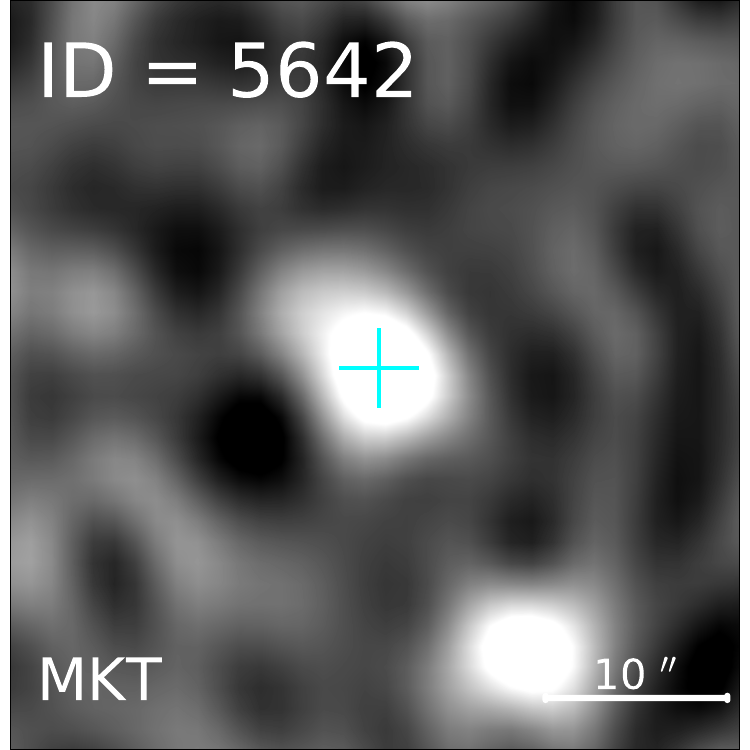}\hspace{-2.00mm}
	\includegraphics[width=2.5cm,height=2.5cm]{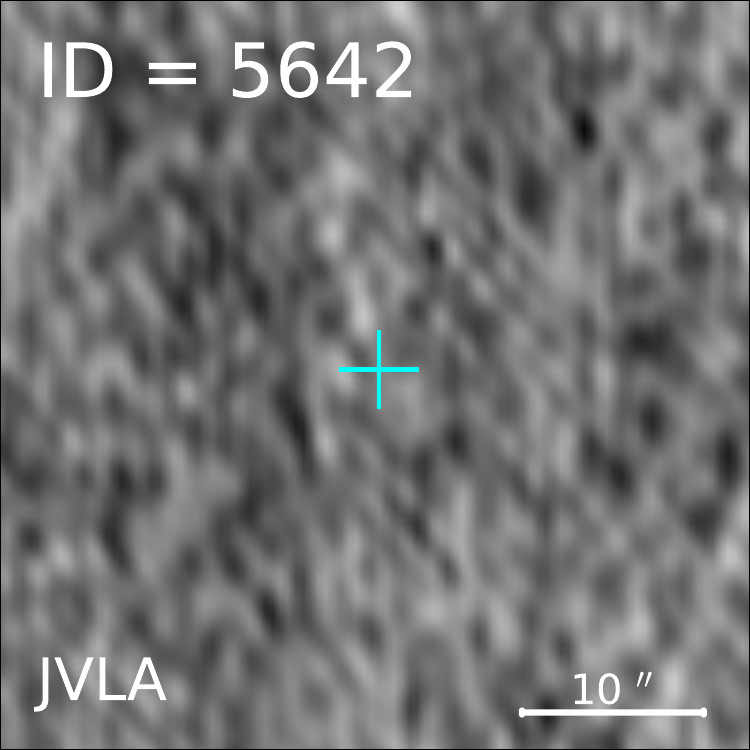}\hspace{-2.00mm}
	\includegraphics[width=2.5cm,height=2.5cm]{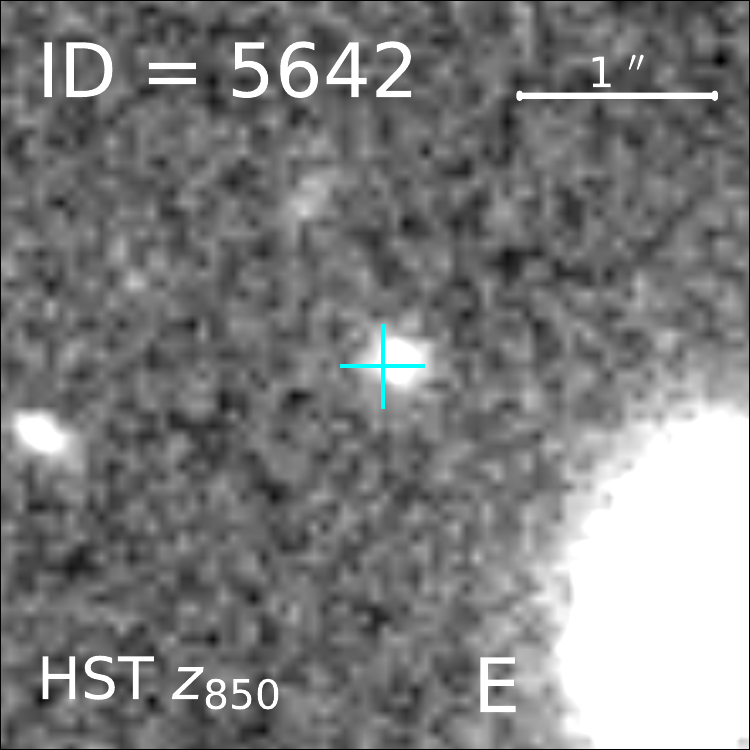}   
	\includegraphics[width=2.5cm,height=2.5cm]{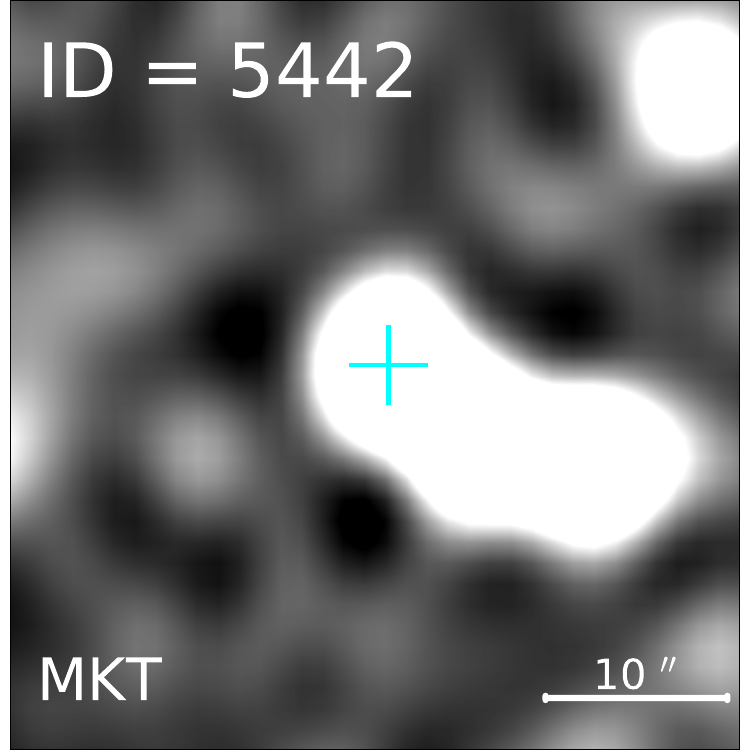}\hspace{-2.00mm}
	\includegraphics[width=2.5cm,height=2.5cm]{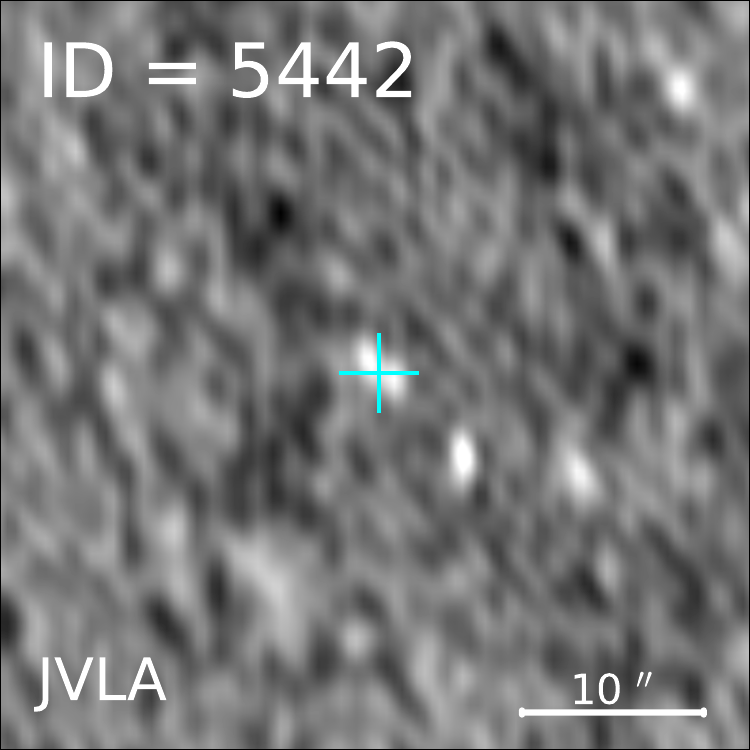}\hspace{-2.00mm}
	\includegraphics[width=2.5cm,height=2.5cm]{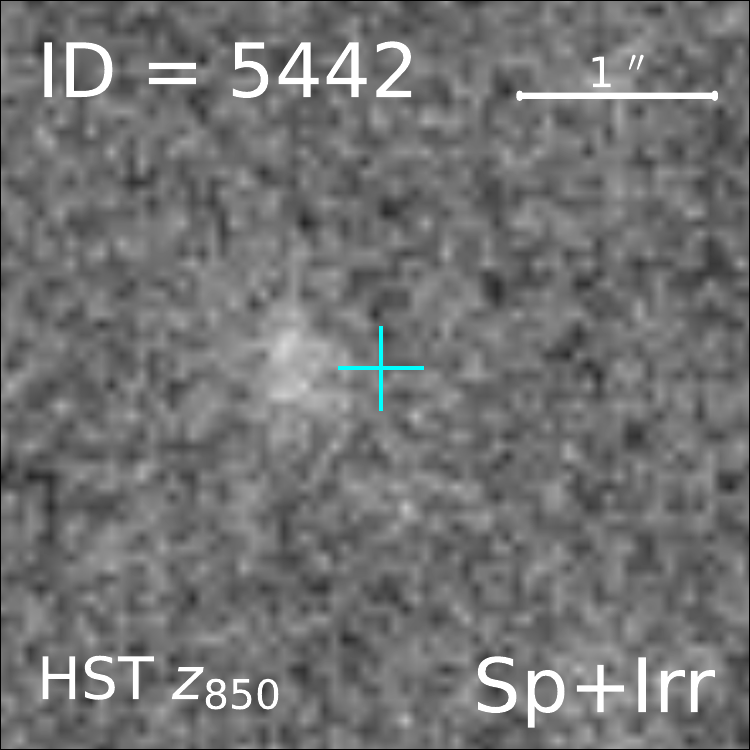} \\
	\includegraphics[width=2.5cm,height=2.5cm]{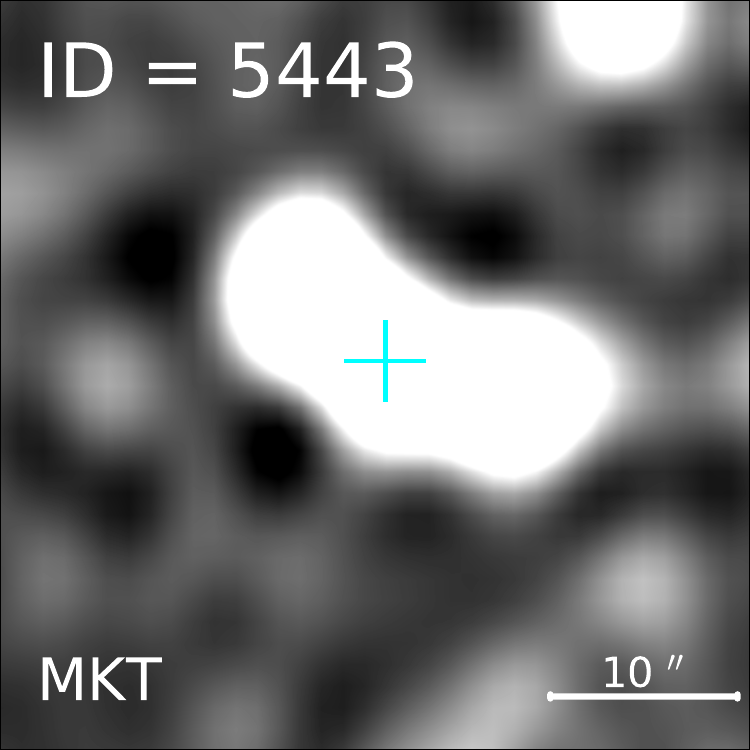}\hspace{-2.00mm}
	\includegraphics[width=2.5cm,height=2.5cm]{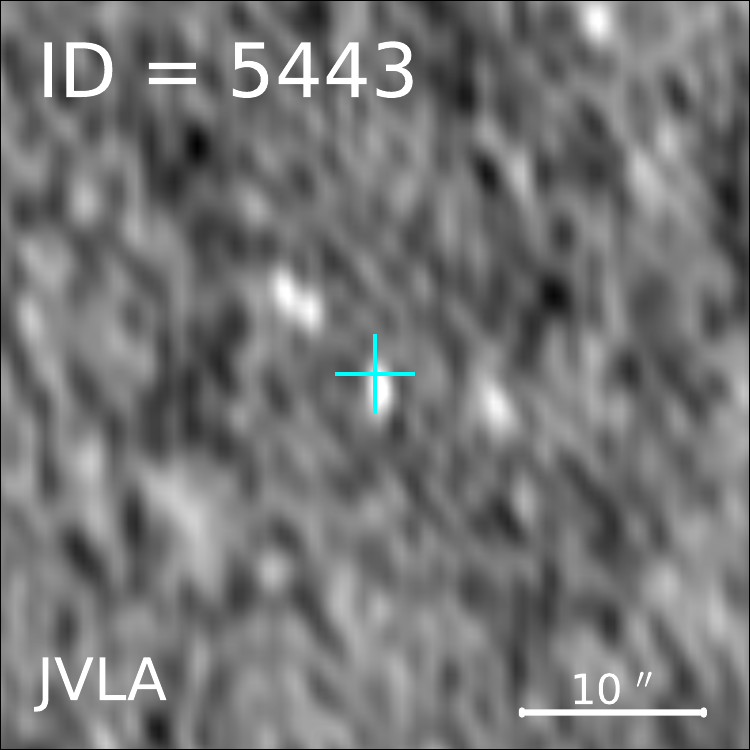}\hspace{-2.00mm}
	\includegraphics[width=2.5cm,height=2.5cm]{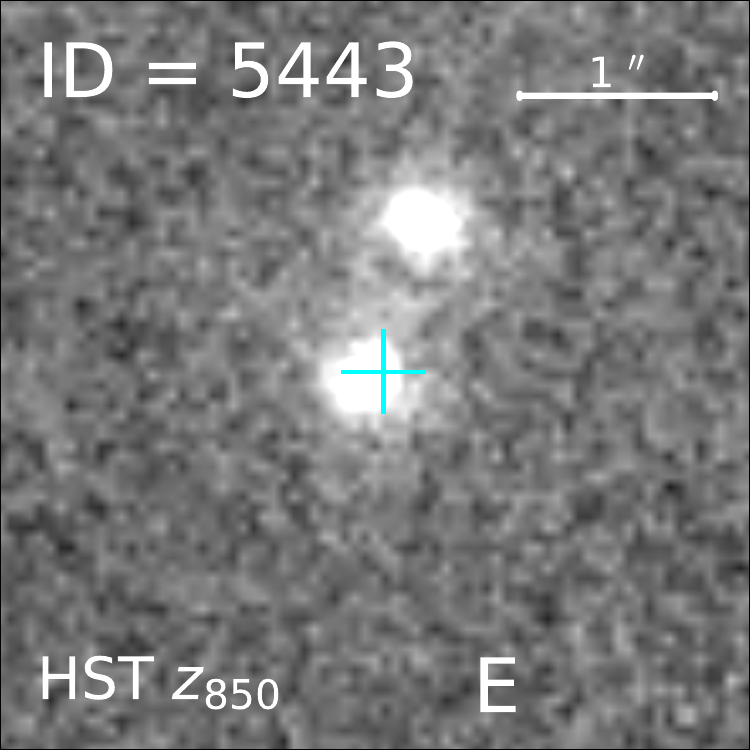}   
	\includegraphics[width=2.5cm,height=2.5cm]{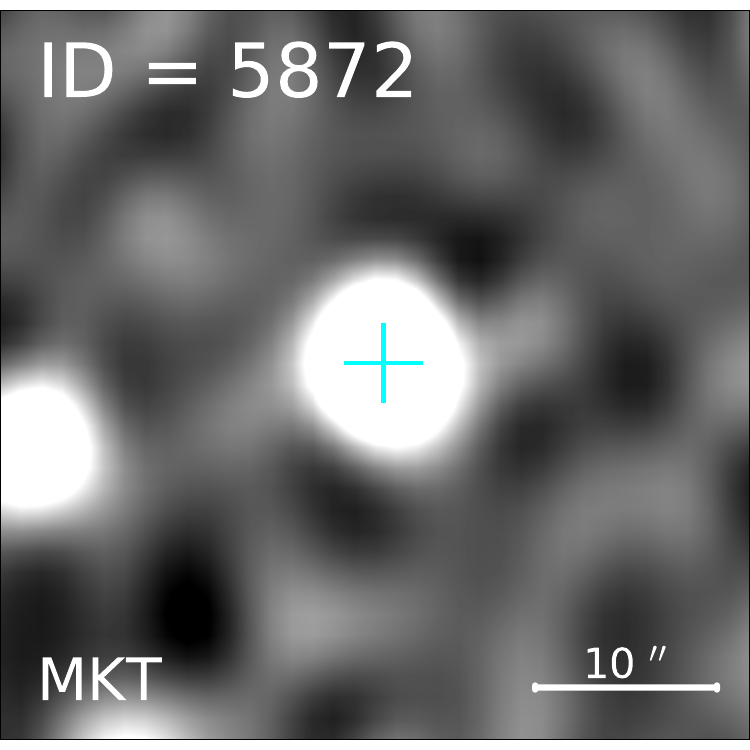}\hspace{-2.00mm}
	\includegraphics[width=2.5cm,height=2.5cm]{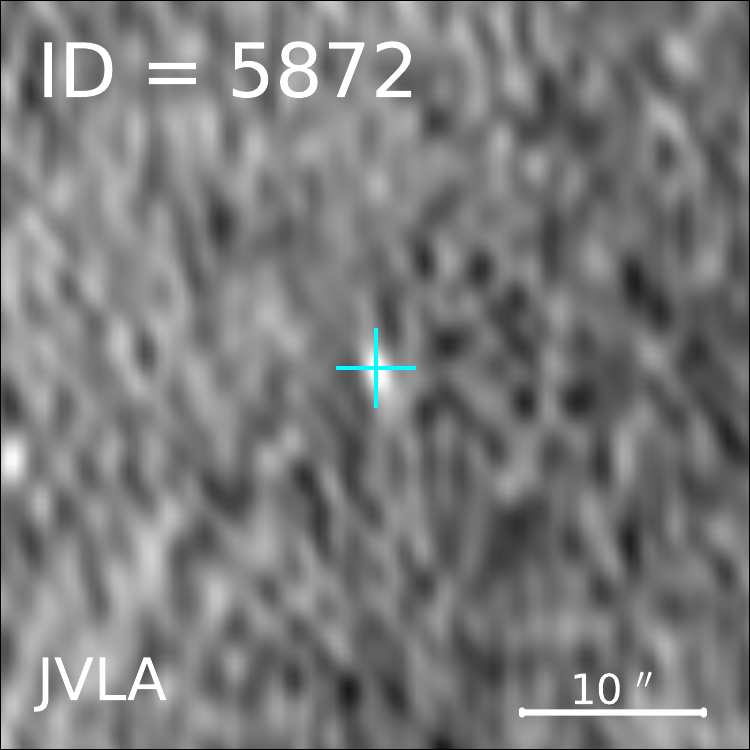}\hspace{-2.00mm}
	\includegraphics[width=2.5cm,height=2.5cm]{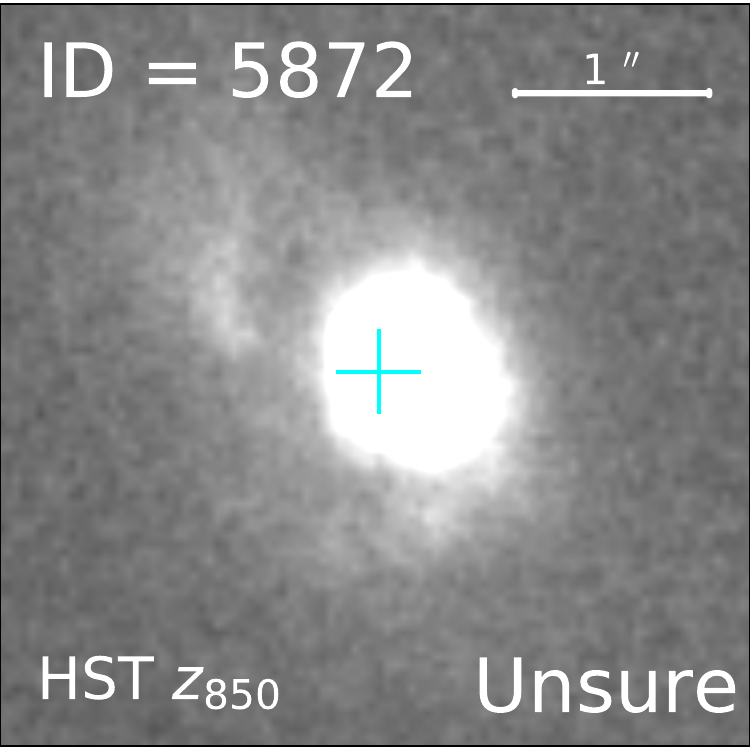} \\
	\includegraphics[width=2.5cm,height=2.5cm]{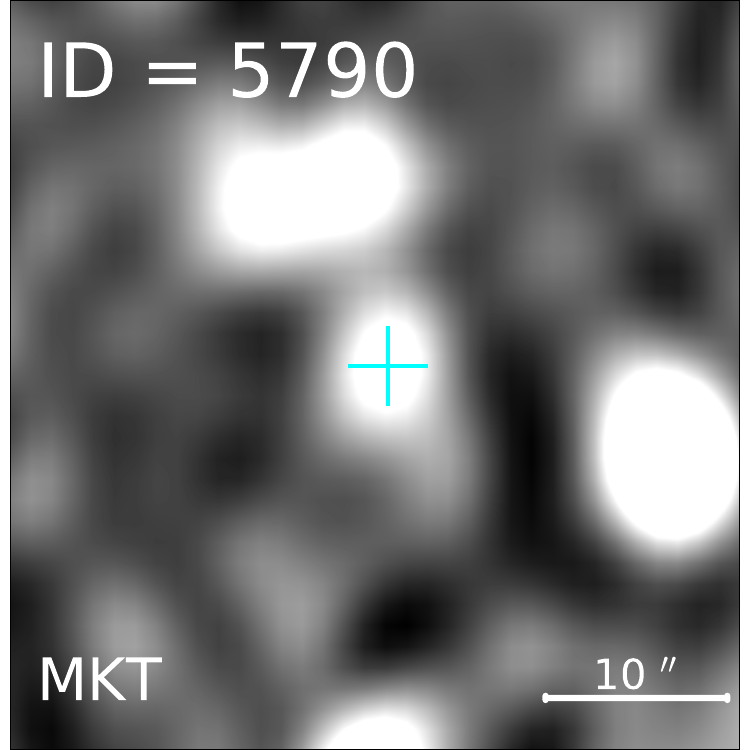}\hspace{-2.00mm}
	\includegraphics[width=2.5cm,height=2.5cm]{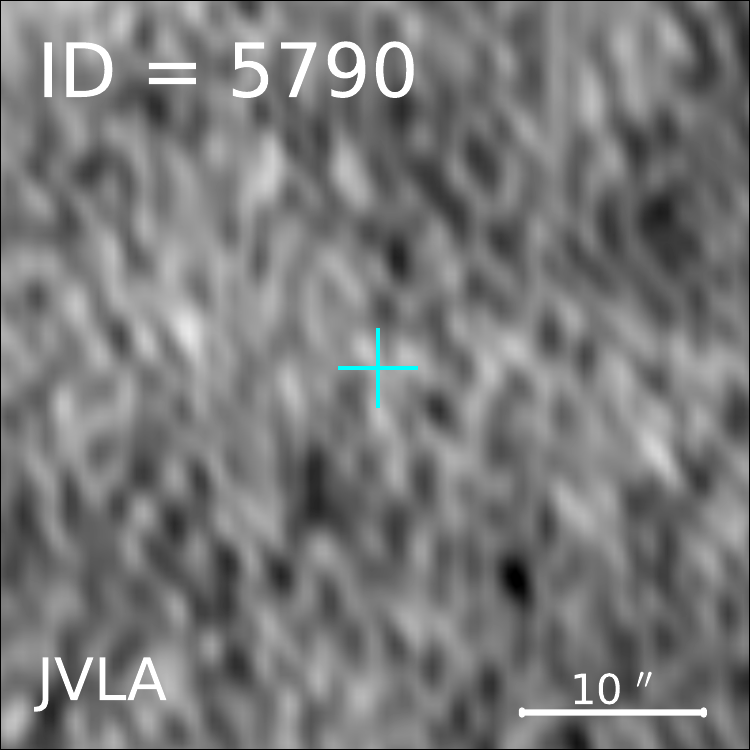}\hspace{-2.00mm}
	\includegraphics[width=2.5cm,height=2.5cm]{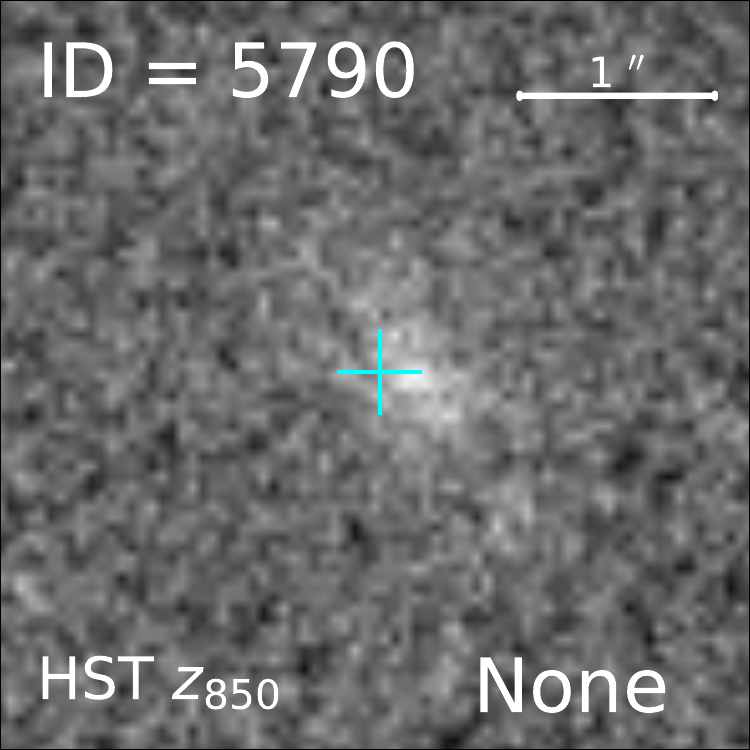}    
	\includegraphics[width=2.5cm,height=2.5cm]{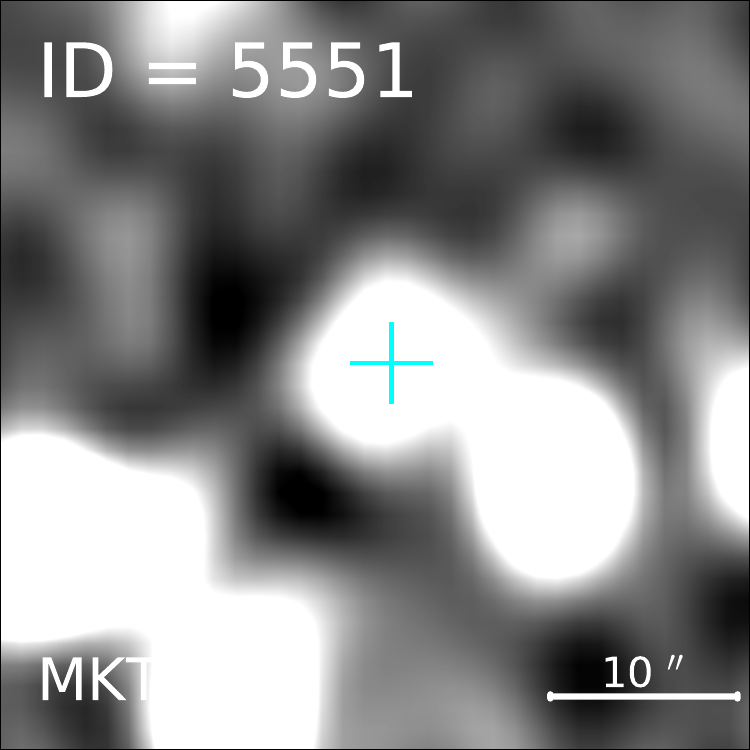}\hspace{-2.00mm}
	\includegraphics[width=2.5cm,height=2.5cm]{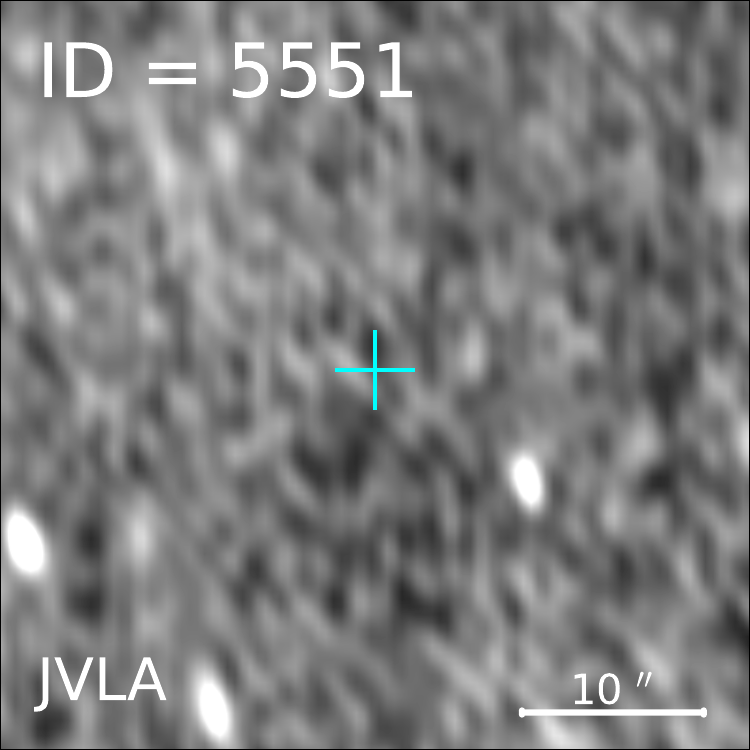}\hspace{-2.00mm}
	\includegraphics[width=2.5cm,height=2.5cm]{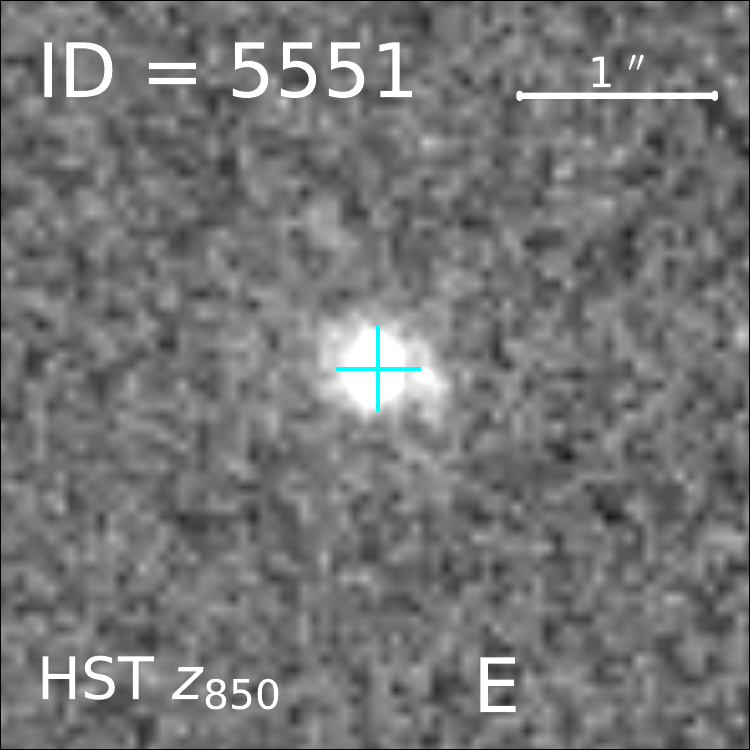} \\
	\includegraphics[width=2.5cm,height=2.5cm]{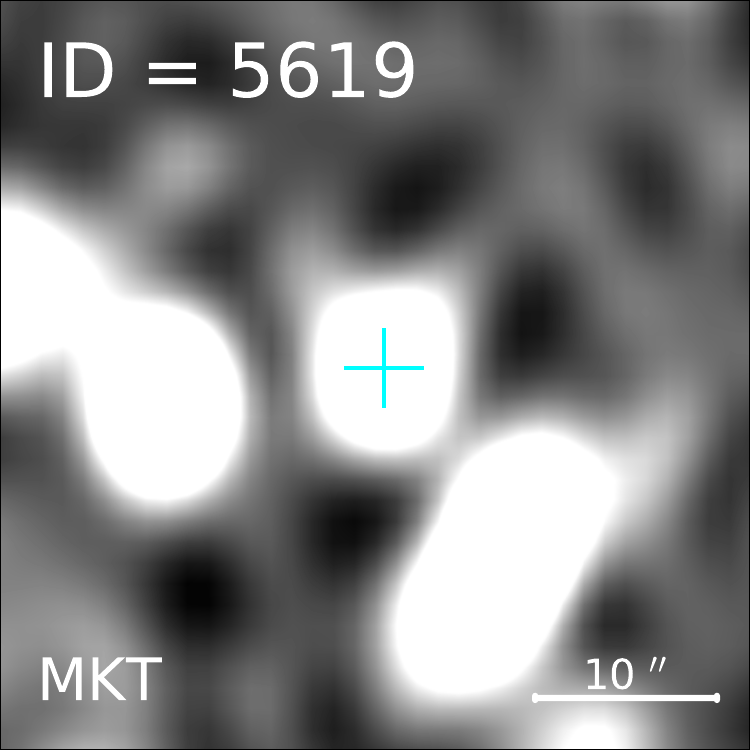}\hspace{-2.00mm}
	\includegraphics[width=2.5cm,height=2.5cm]{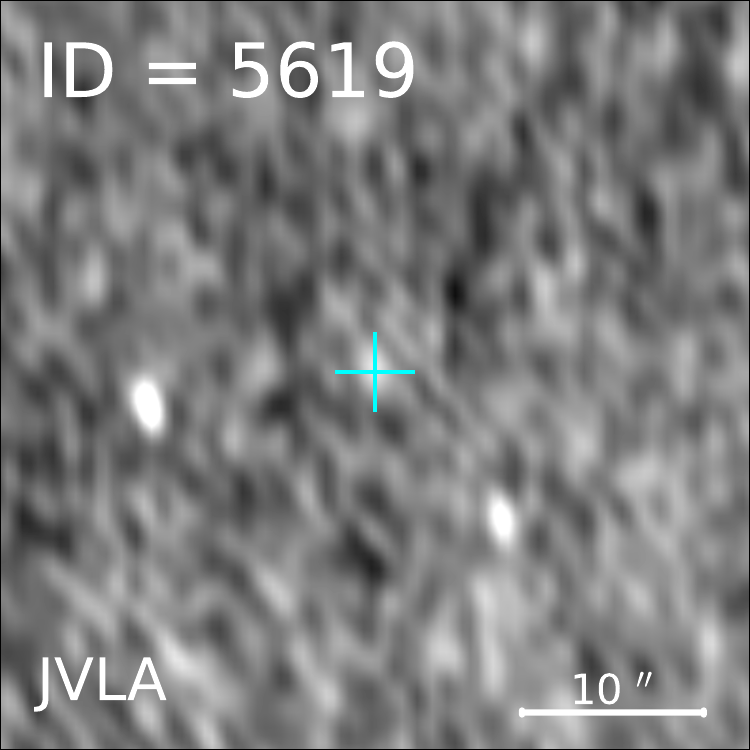}\hspace{-2.00mm}
	\includegraphics[width=2.5cm,height=2.5cm]{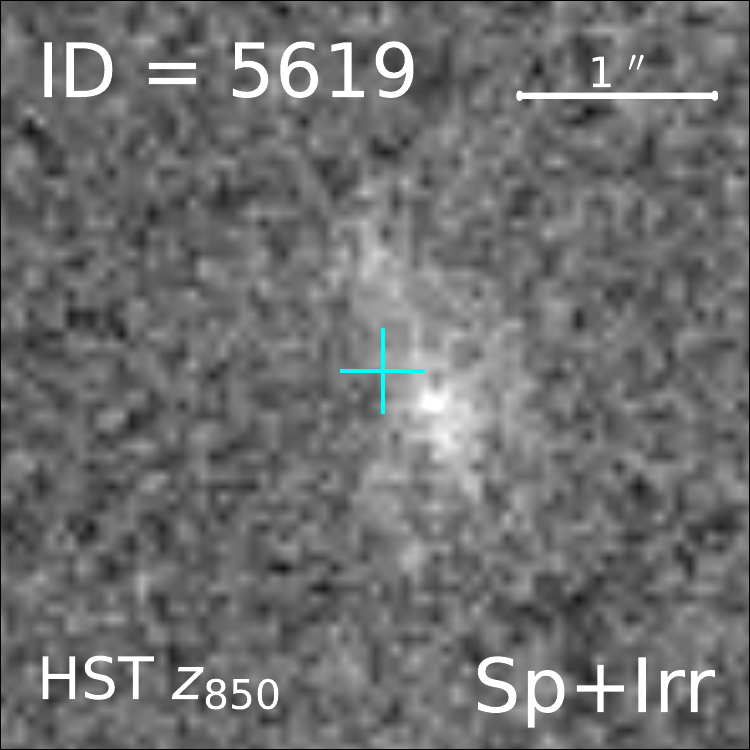}    	
	\includegraphics[width=2.5cm,height=2.5cm]{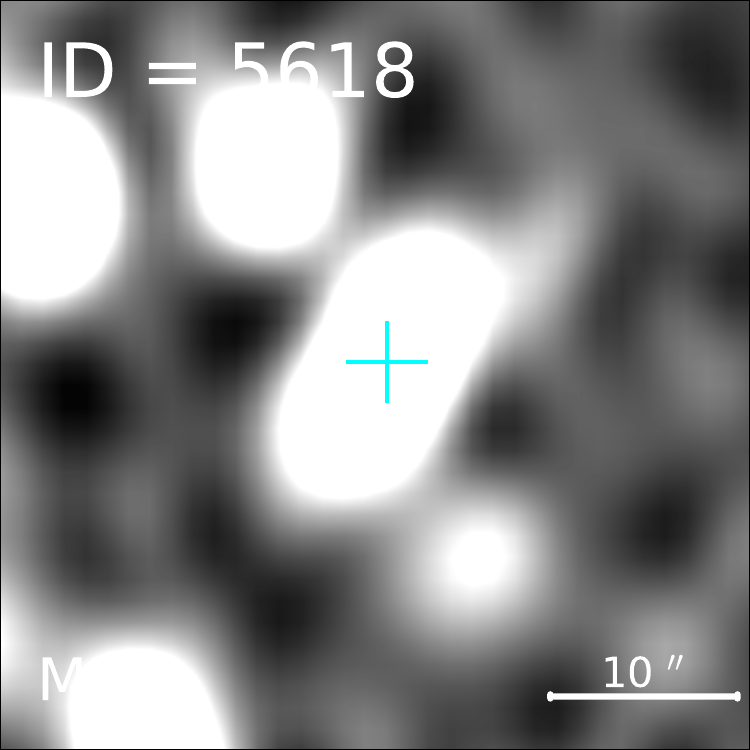}\hspace{-2.00mm}
	\includegraphics[width=2.5cm,height=2.5cm]{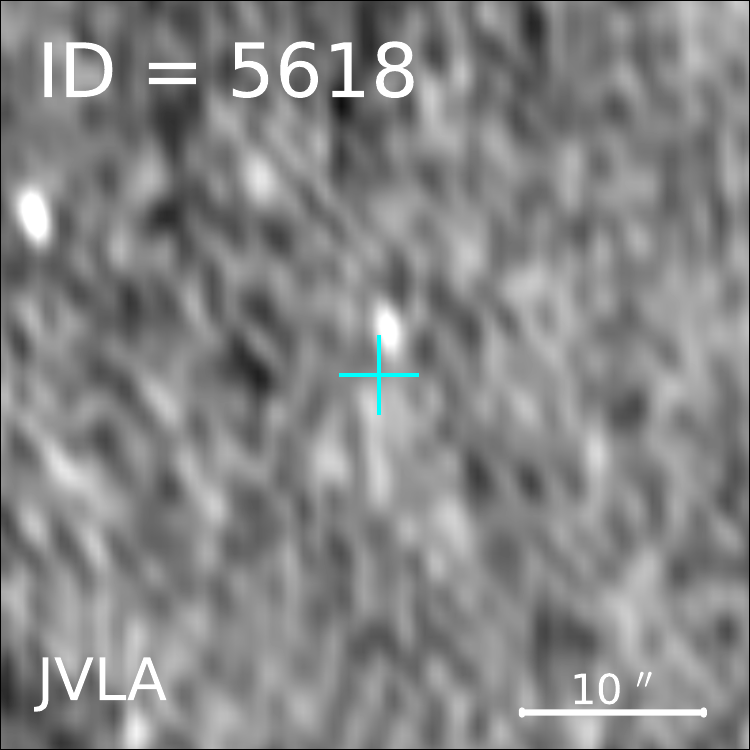}\hspace{-2.00mm}
	\includegraphics[width=2.5cm,height=2.5cm]{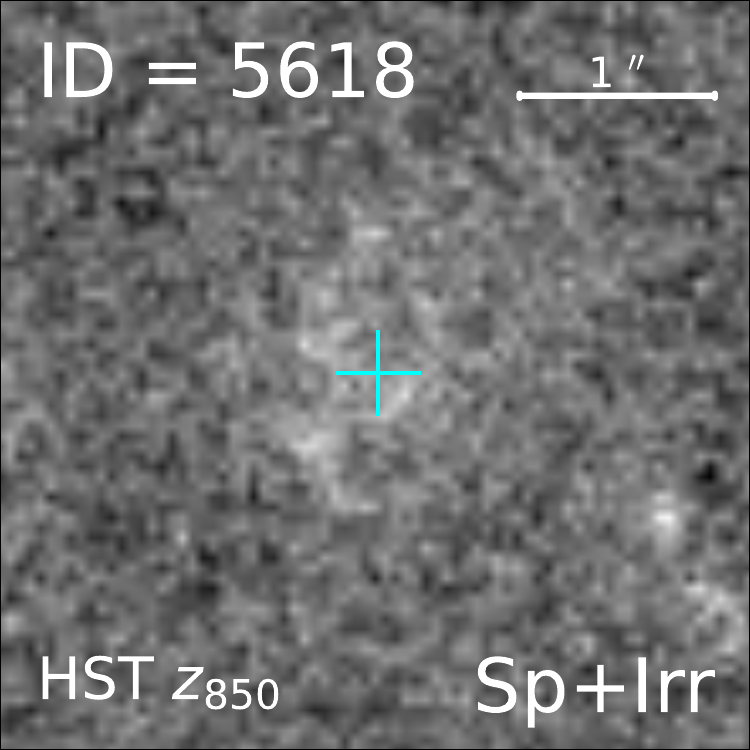}
	\caption{The postage stamp images of the selected members of the J2215 cluster within a radius of 0.8 Mpc of the original  X-ray cluster position.~The source ID number presented in Table  \ref{table:table1SFR}  is  located in the upper left corner of each extracted postage stamp image. The cyan marker represents the  MeerKAT source position. Each set of the postage stamp images was taken from  the MeerKAT $L$-band image (left panel),  the JVLA $L$-band image in A configuration \citep{2015ApJMa} (middle panel) and the  \textit{Hubble Space Telescope} ACS $z_{850}$-band image \citet[right panel image from which the initial classification of the cluster galaxy morphology was performed in][]{2009ApHilton}.~The MeerKAT  and JVLA postage stamp images are 40$^{\prime \prime}$, whilst the HST image is 3.75$^{\prime \prime}$ on a side with east at the left.} 
	\label{fig:Morphology}
\end{figure*}

We used the catalogue of morphologically classified J2215 cluster members in \citet{2009ApHilton} for this work.~We cross-matched the catalogue in \citet{2009ApHilton} with that of the MeerKAT-detected cluster members within a radius of 6$^{\prime \prime}$.~In \citet{2009ApHilton} morphological classification of the galaxies was performed on magnitudes brighter than $z_{850}$ $\leq$ 24  via visual inspection by five human classifiers under four different morphological bins i.e., the elliptical (E),~lenticular (S0),~spiral (S) and~irregular (Irr).~They determined the morphologies using an existing training set of galaxies compiled by \citet{2005Postman} from the morphological studies of high redshift cluster galaxies i.e., 0.8 $< z  <$ 1.3.~An in-depth description of the morphological classification of the J2215 cluster galaxies can be found in \citet{2009ApHilton}.~The postage stamp images of the selected galaxies from the MeerKAT $L$-band image  listed in Table \ref{table:table1SFR} can be seen in Figure \ref{fig:Morphology}.~Each set of the postage stamp images was taken from the MeerKAT $L$-band image (left panel), the JVLA $L$-band image in A configuration \citep{2015ApJMa} (middle panel) and  the $z_{850}$-band counterpart images (right panel) from which the initial morphologies were determined \citet[see Section 3.2  of ][]{2009ApHilton}.

Only 17/24  MeerKAT-detected cluster members had  their morphological  counterparts in the \citet{2009ApHilton} catalogue, 7/24 did not have any morphological classification.~Out of the 17 MeerKAT-detected cluster members having morphological classification  12 had Spiral/Irregular morphology (6/12 - SFGs , 2/6 - ISFGs, 4/6 - AGNs) whilst 5 had elliptical/lenticular galaxy morphology (3/12 - SFGs, 1/6 - ISFGs and 1/6 AGNs). Therefore, the majority of the MeerKAT sources correspond to host galaxies with late type morphologies.

\begin{figure*}
	\centering
	\includegraphics[width=2.5cm,height=2.5cm]{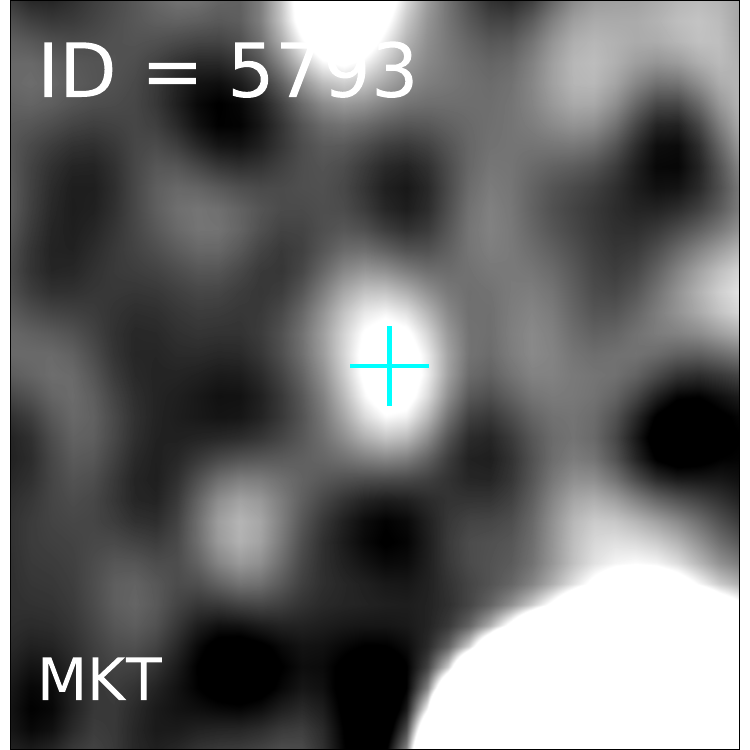}\hspace{-2.00mm}
	\includegraphics[width=2.5cm,height=2.5cm]{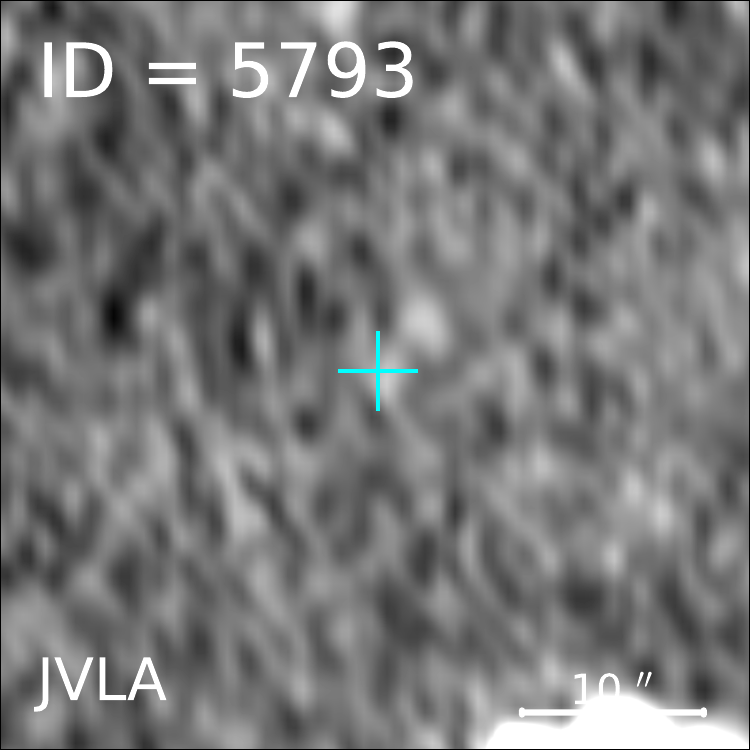}\hspace{-2.00mm}
	\includegraphics[width=2.5cm,height=2.5cm]{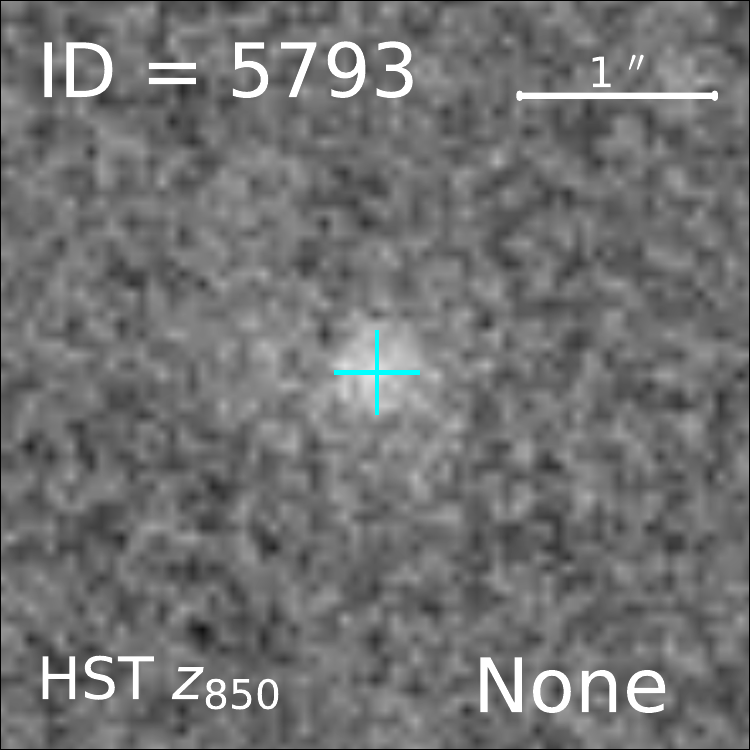}    
	\includegraphics[width=2.5cm,height=2.5cm]{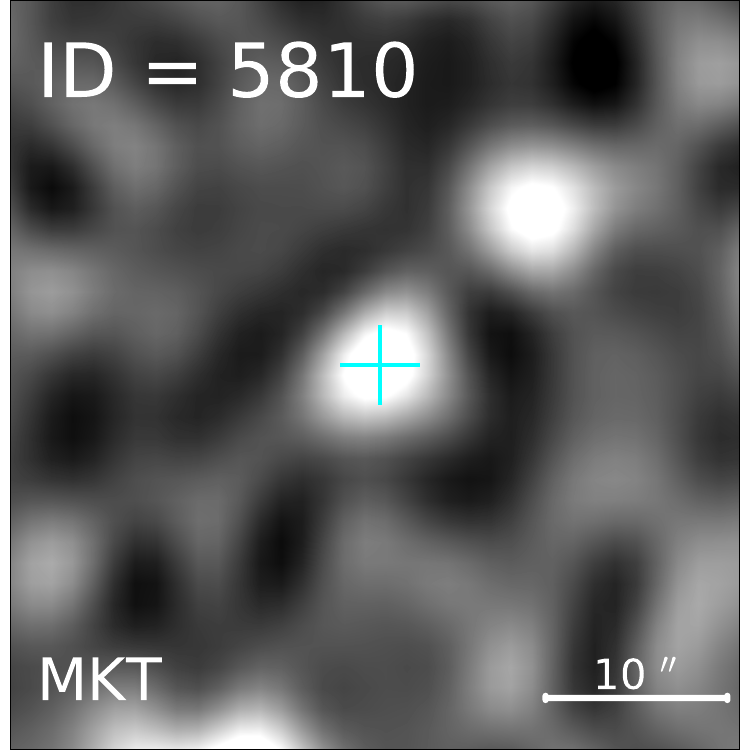}\hspace{-2.00mm}
	\includegraphics[width=2.5cm,height=2.5cm]{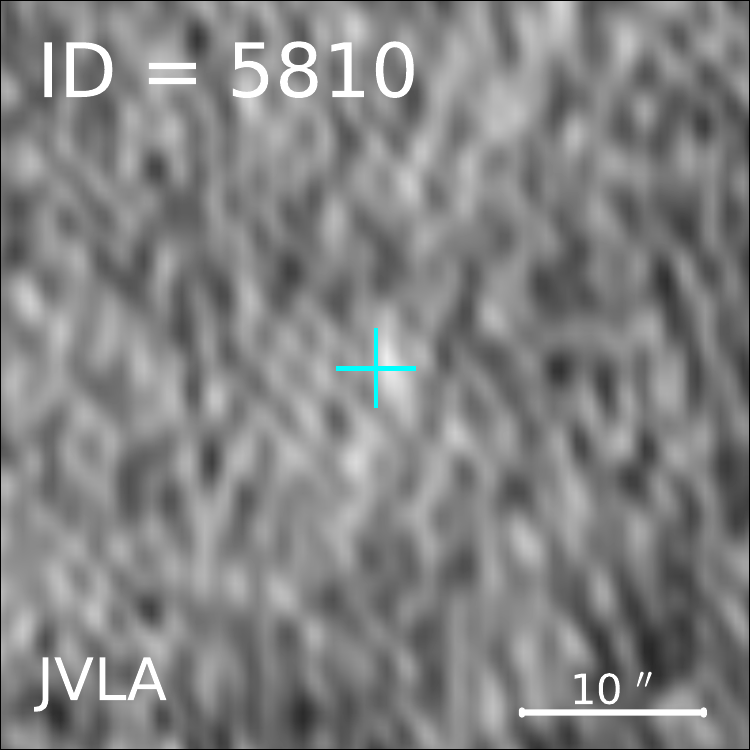}\hspace{-2.00mm}
	\includegraphics[width=2.5cm,height=2.5cm]{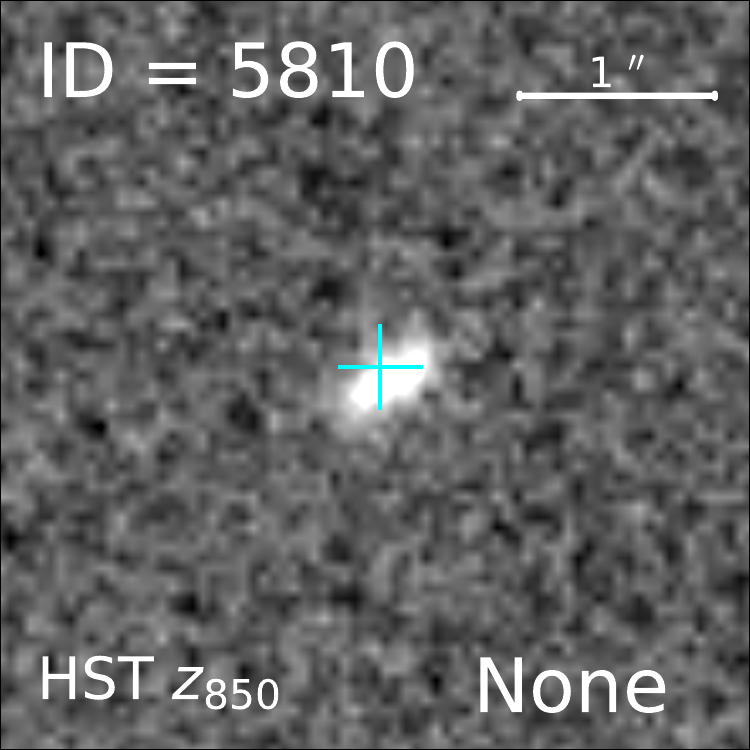}  \\
	\includegraphics[width=2.5cm,height=2.5cm]{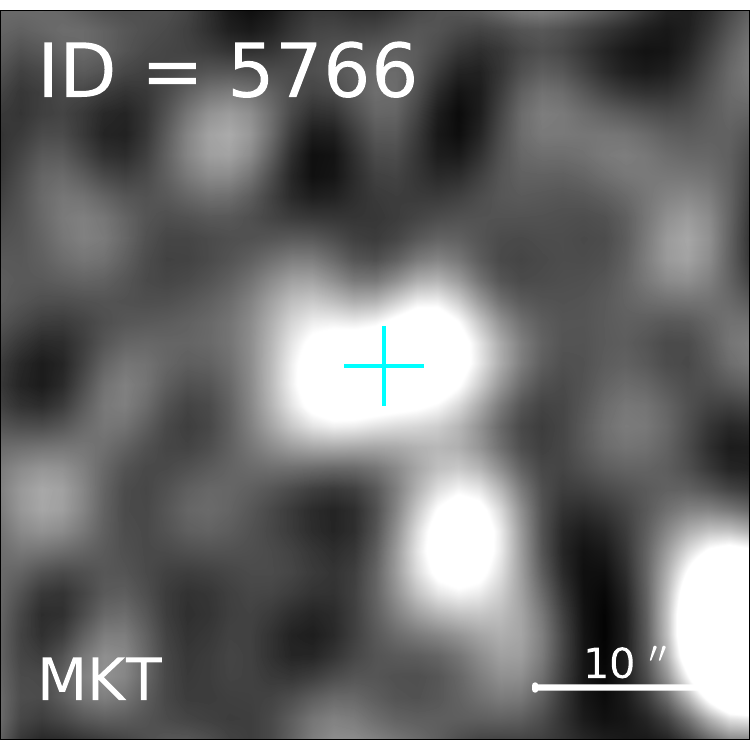}\hspace{-2.00mm}
	\includegraphics[width=2.5cm,height=2.5cm]{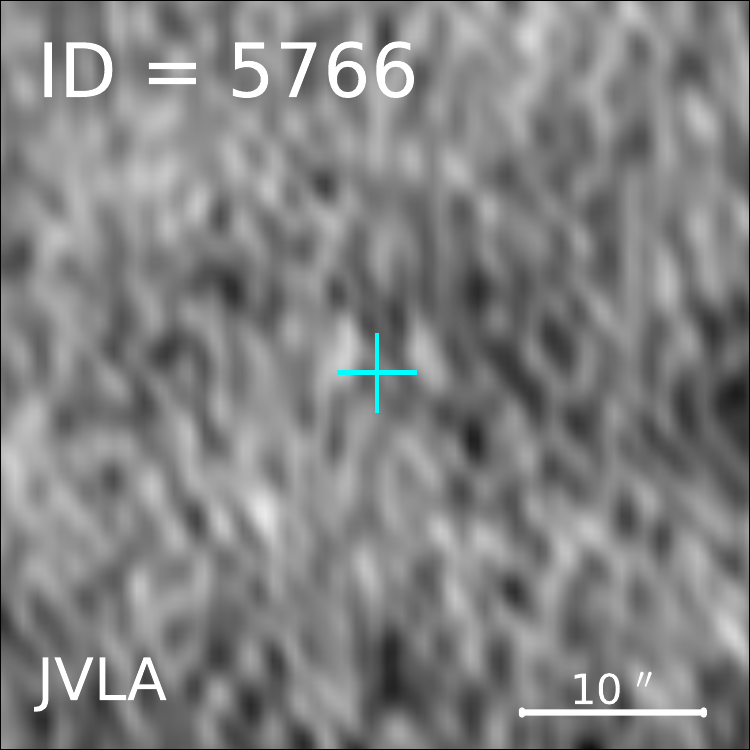}\hspace{-2.00mm}
	\includegraphics[width=2.5cm,height=2.5cm]{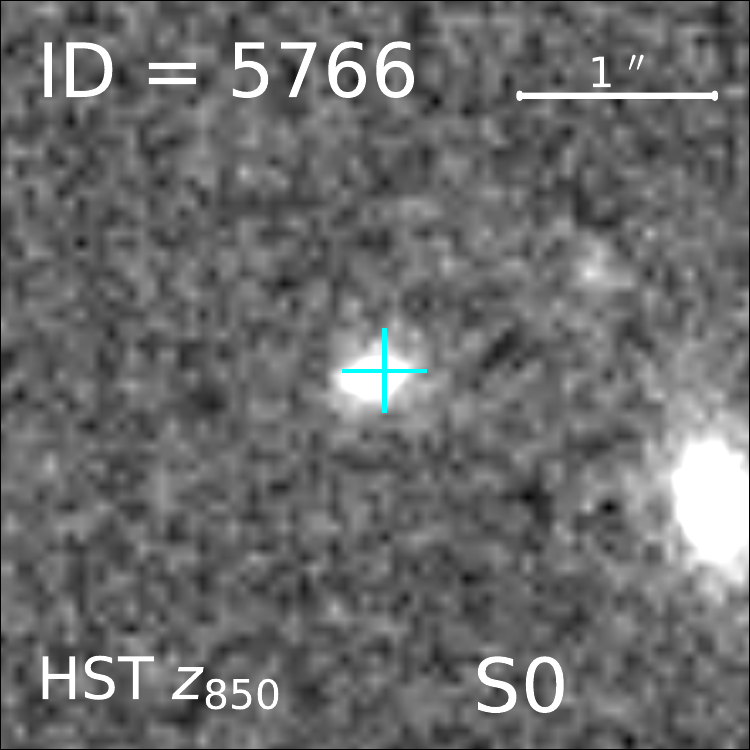}
	\includegraphics[width=2.5cm,height=2.5cm]{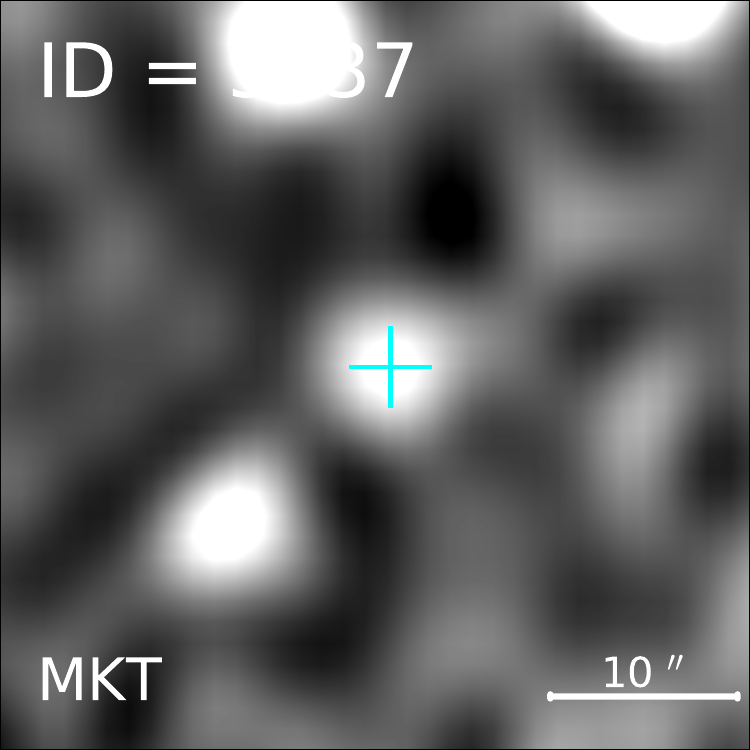}\hspace{-2.00mm}
	\includegraphics[width=2.5cm,height=2.5cm]{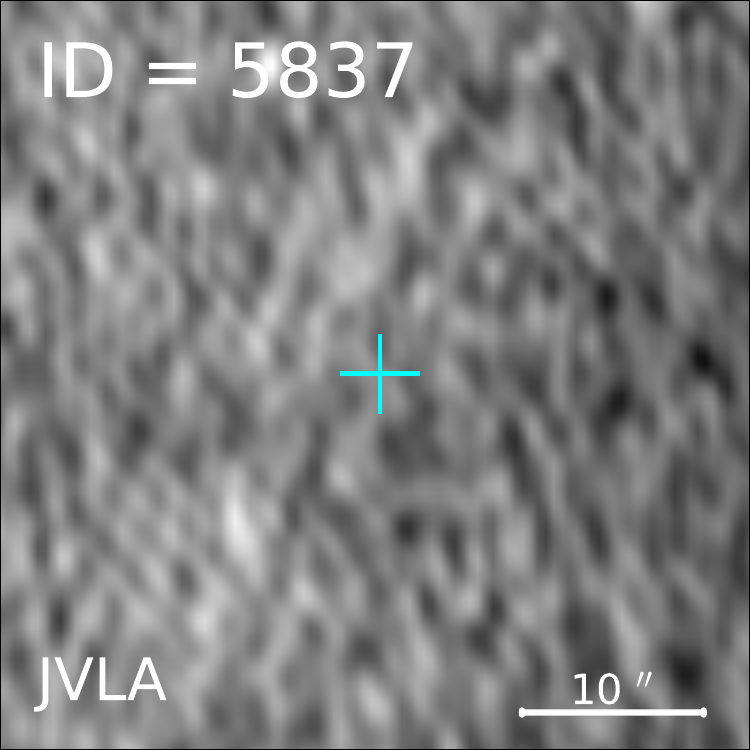}\hspace{-2.00mm}
	\includegraphics[width=2.5cm,height=2.5cm]{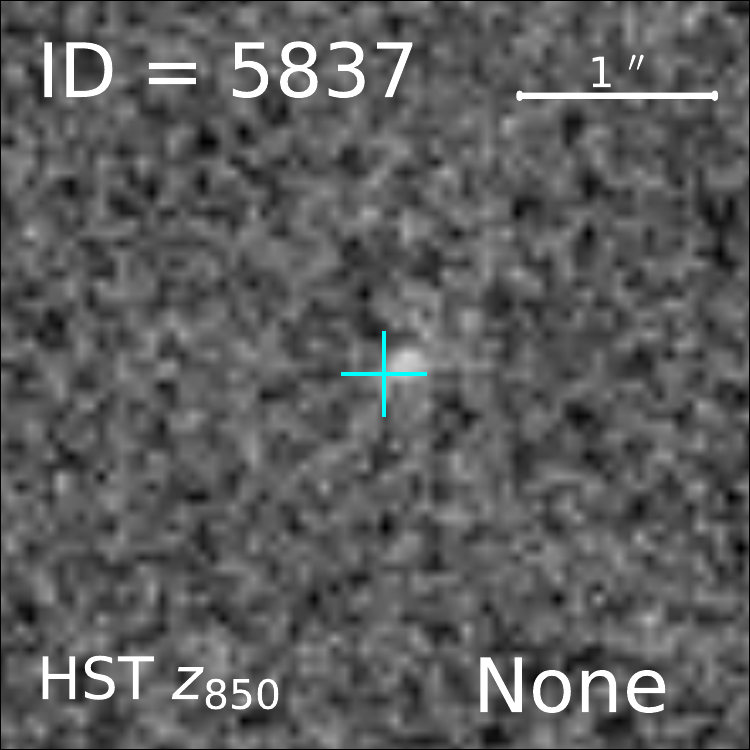}    \\
	\includegraphics[width=2.5cm,height=2.5cm]{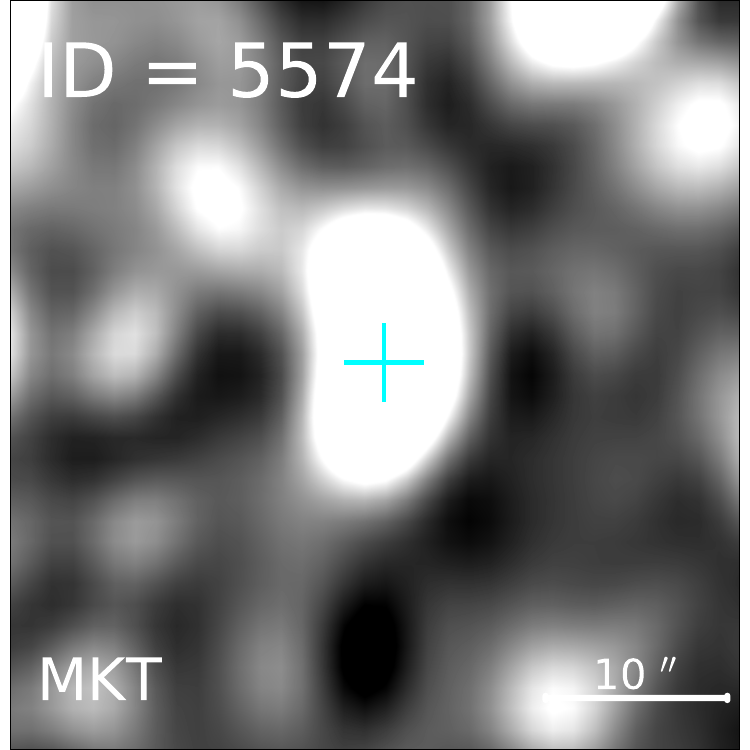}\hspace{-2.00mm}
	\includegraphics[width=2.5cm,height=2.5cm]{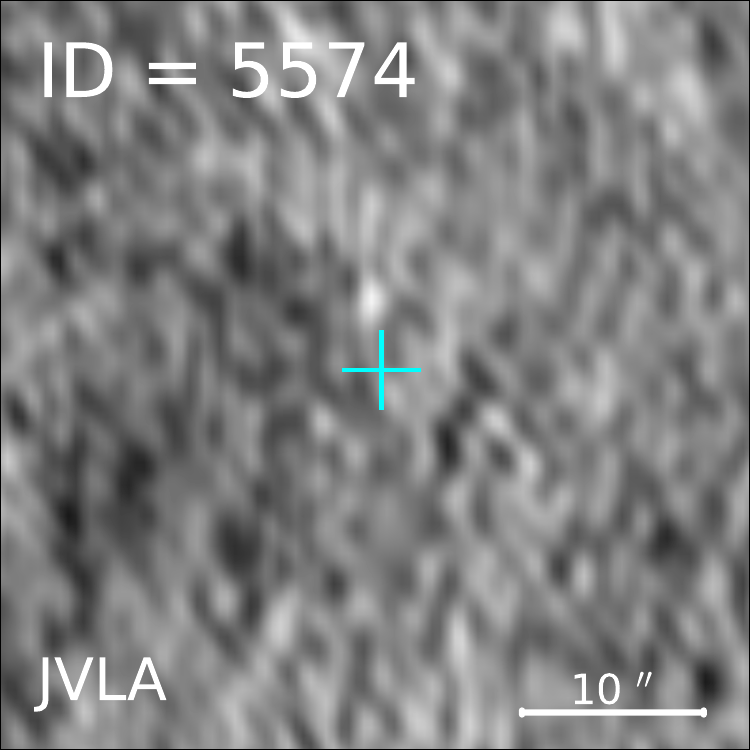}\hspace{-2.00mm}
	\includegraphics[width=2.5cm,height=2.5cm]{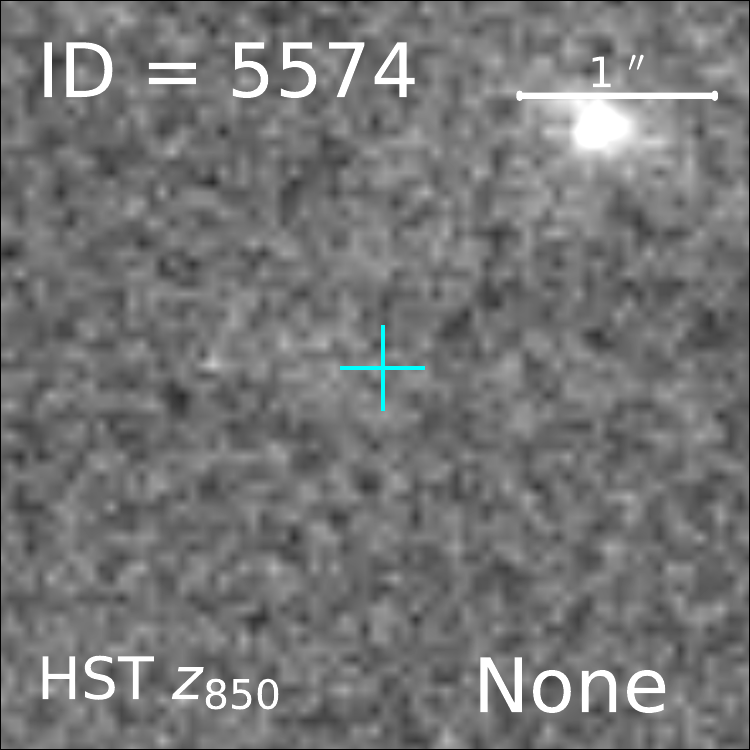}    
	\includegraphics[width=2.5cm,height=2.5cm]{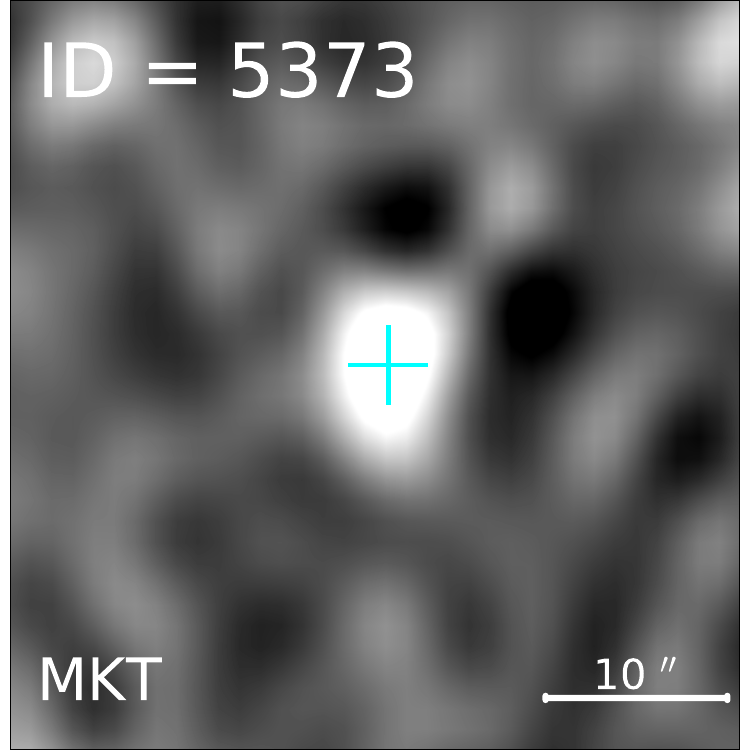}\hspace{-2.00mm}
	\includegraphics[width=2.5cm,height=2.5cm]{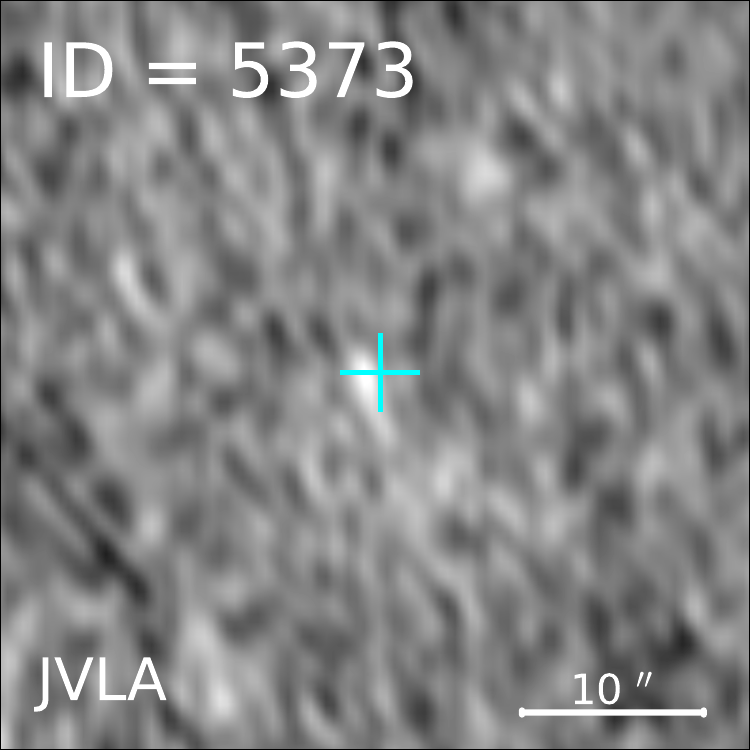}\hspace{-2.00mm}
	\includegraphics[width=2.5cm,height=2.5cm]{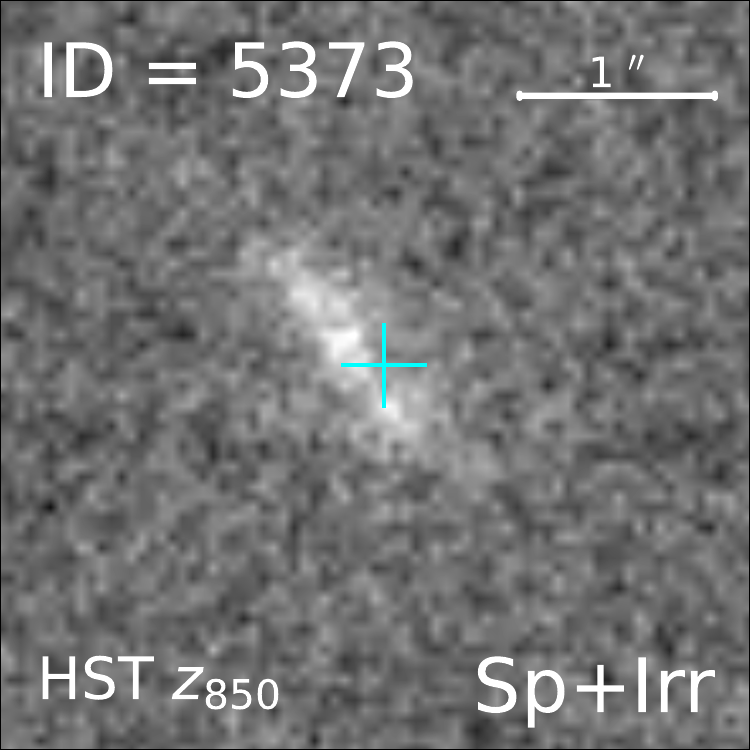}    \\
	\includegraphics[width=2.5cm,height=2.5cm]{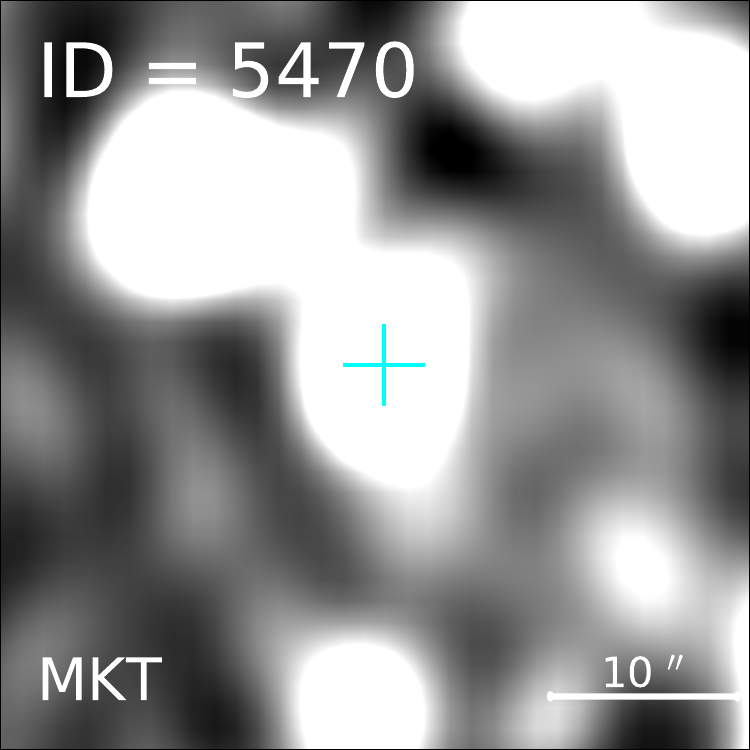}\hspace{-2.00mm}
	\includegraphics[width=2.5cm,height=2.5cm]{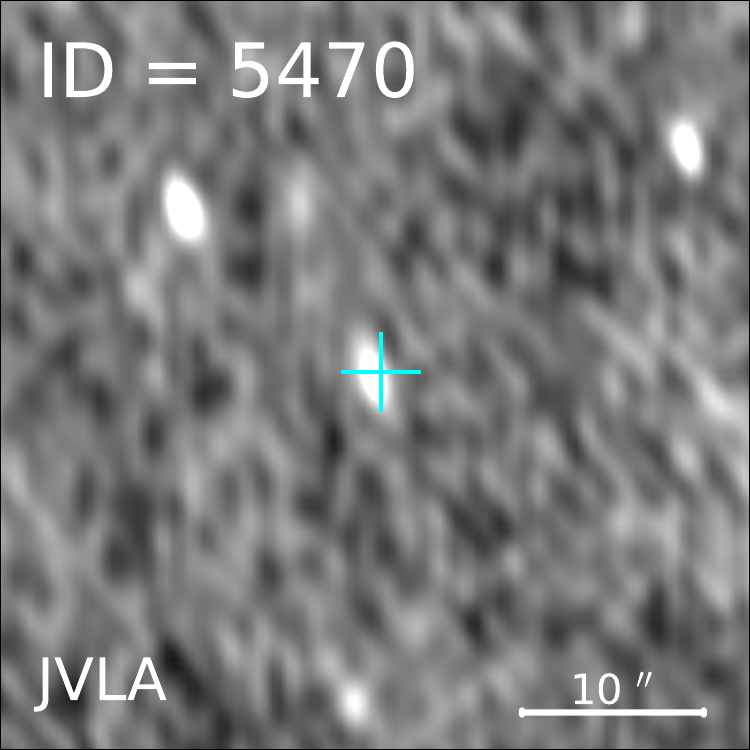}\hspace{-2.00mm}
	\includegraphics[width=2.5cm,height=2.5cm]{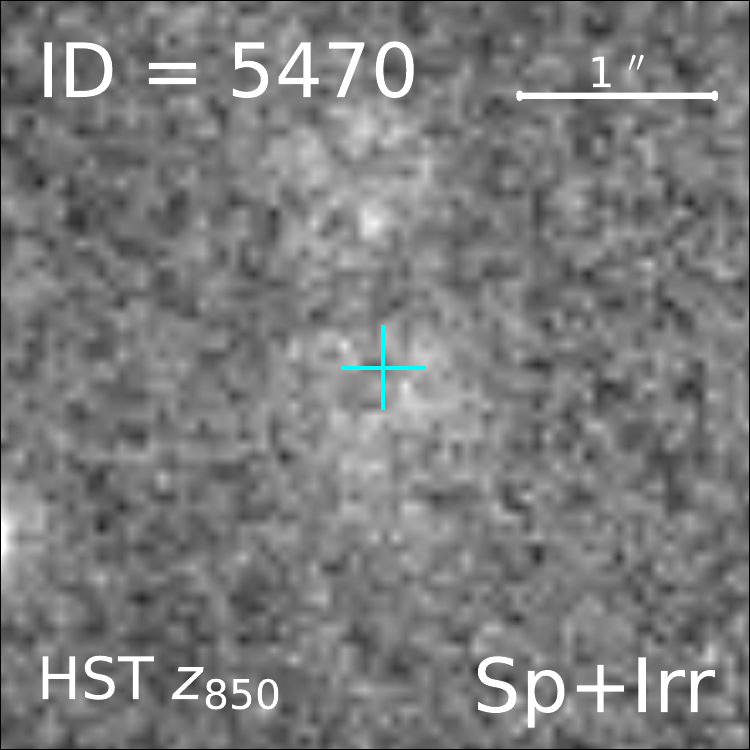}
	\includegraphics[width=2.5cm,height=2.5cm]{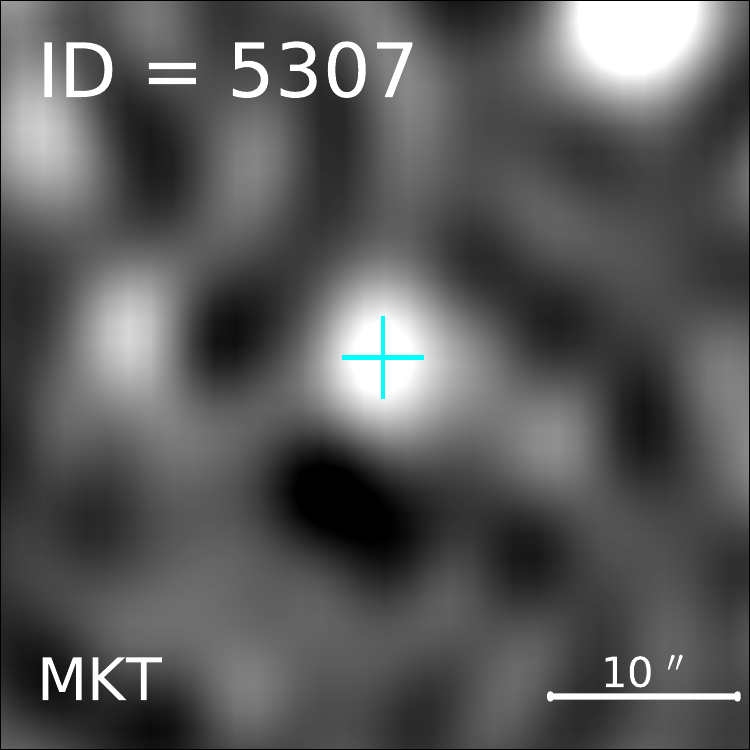}\hspace{-2.00mm}
	\includegraphics[width=2.5cm,height=2.5cm]{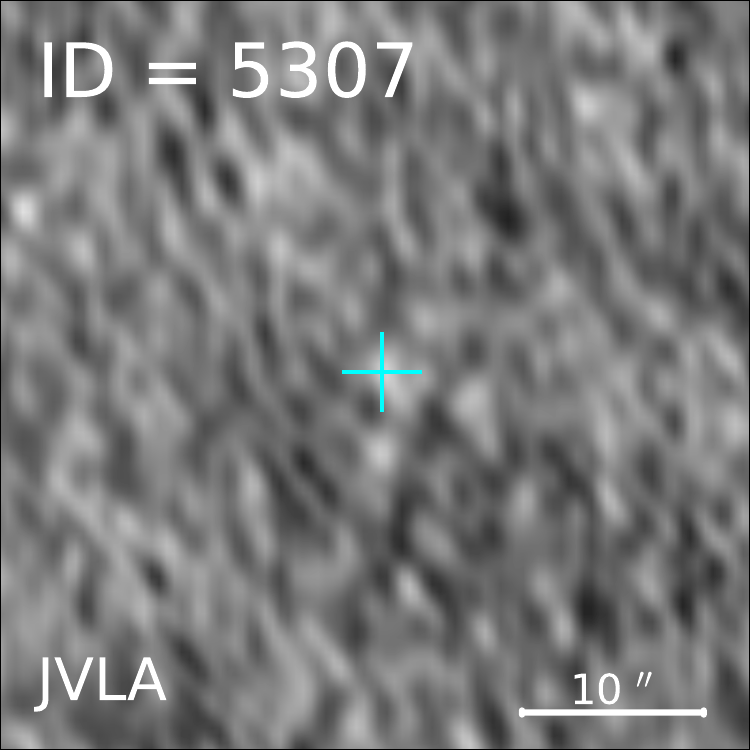}\hspace{-2.00mm}
	\includegraphics[width=2.5cm,height=2.5cm]{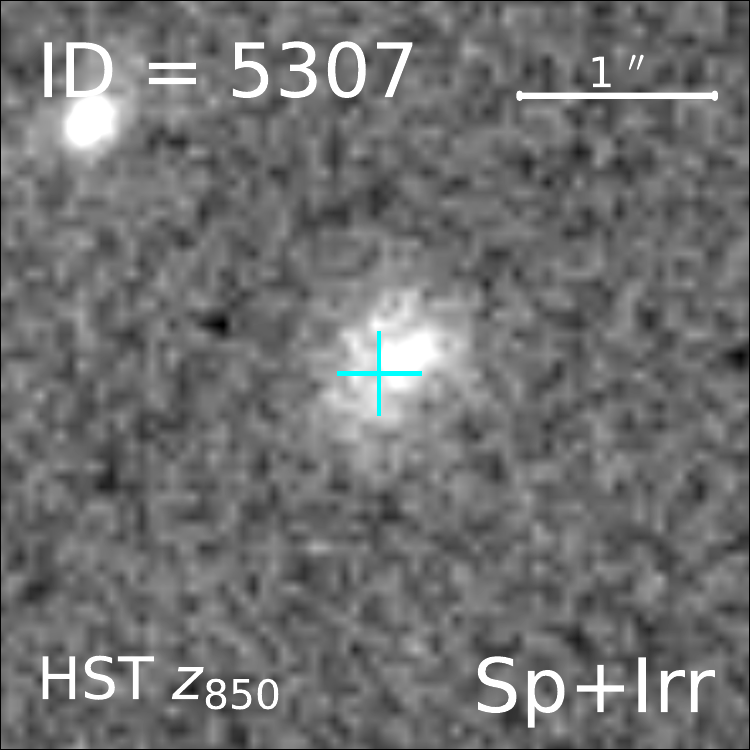}    \\
	\includegraphics[width=2.5cm,height=2.5cm]{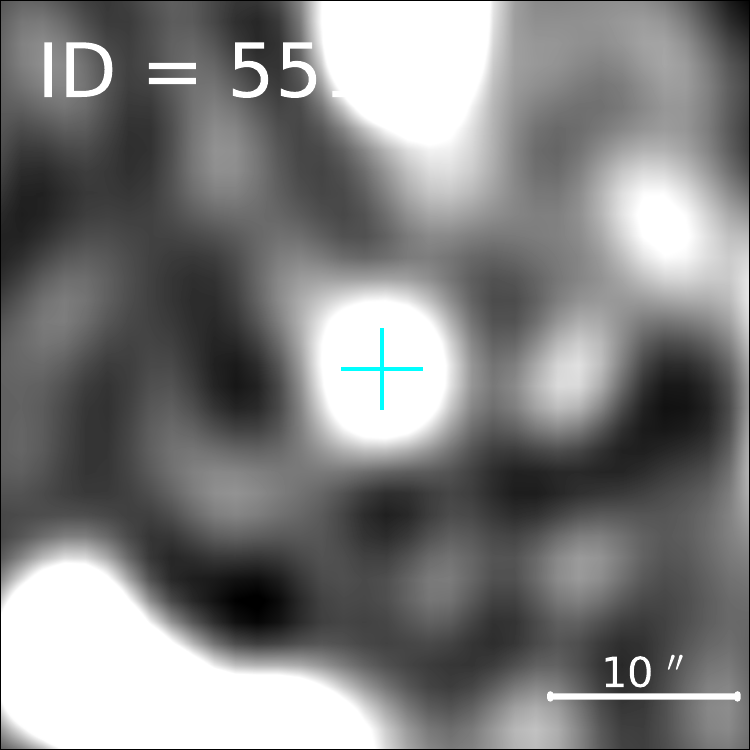}\hspace{-2.00mm}
	\includegraphics[width=2.5cm,height=2.5cm]{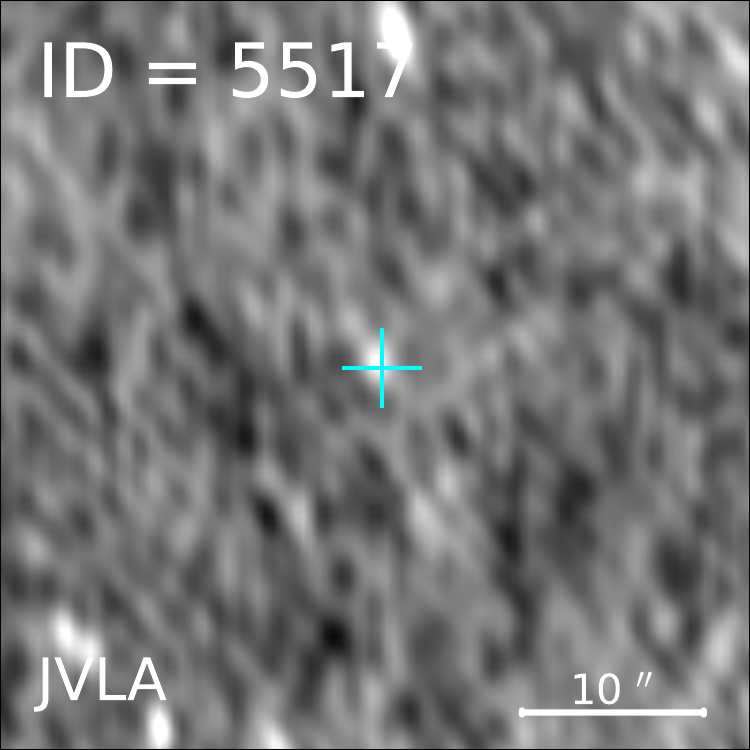}\hspace{-2.00mm}
	\includegraphics[width=2.5cm,height=2.5cm]{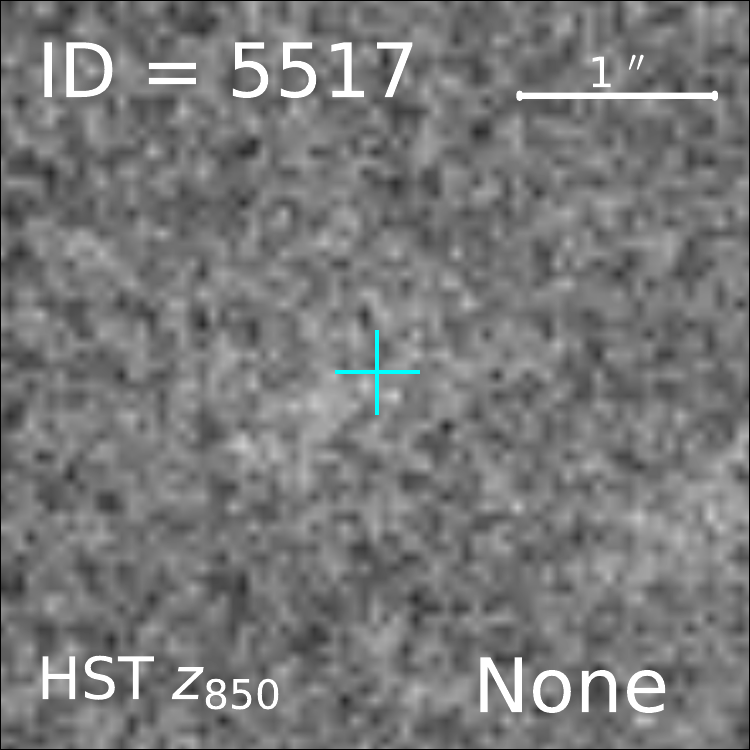}
	\includegraphics[width=2.5cm,height=2.5cm]{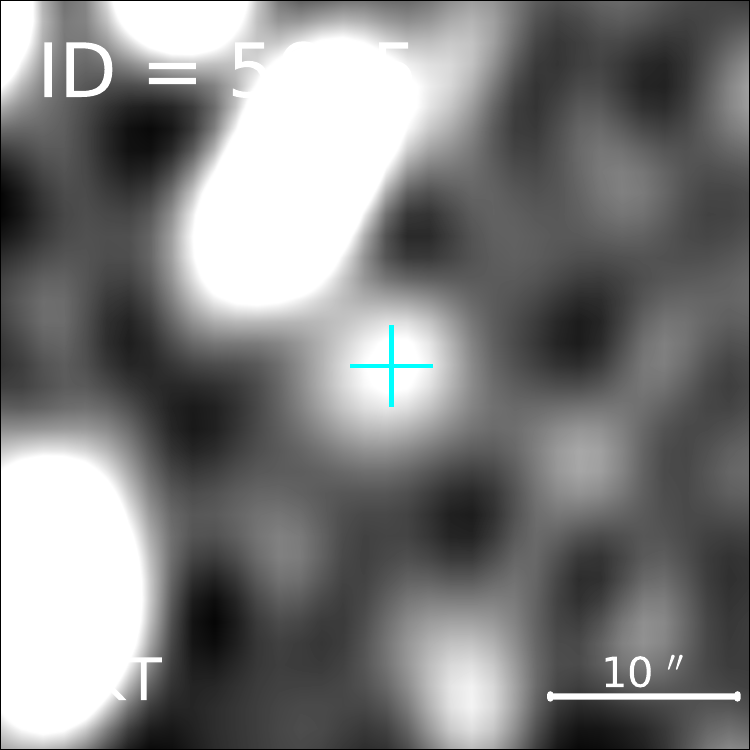}\hspace{-2.00mm}
	\includegraphics[width=2.5cm,height=2.5cm]{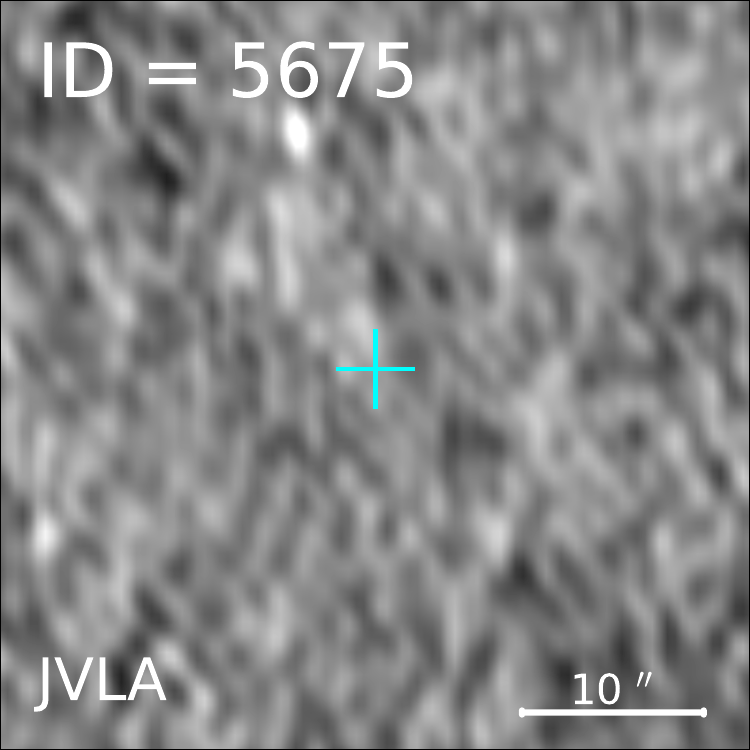}\hspace{-2.00mm}
	\includegraphics[width=2.5cm,height=2.5cm]{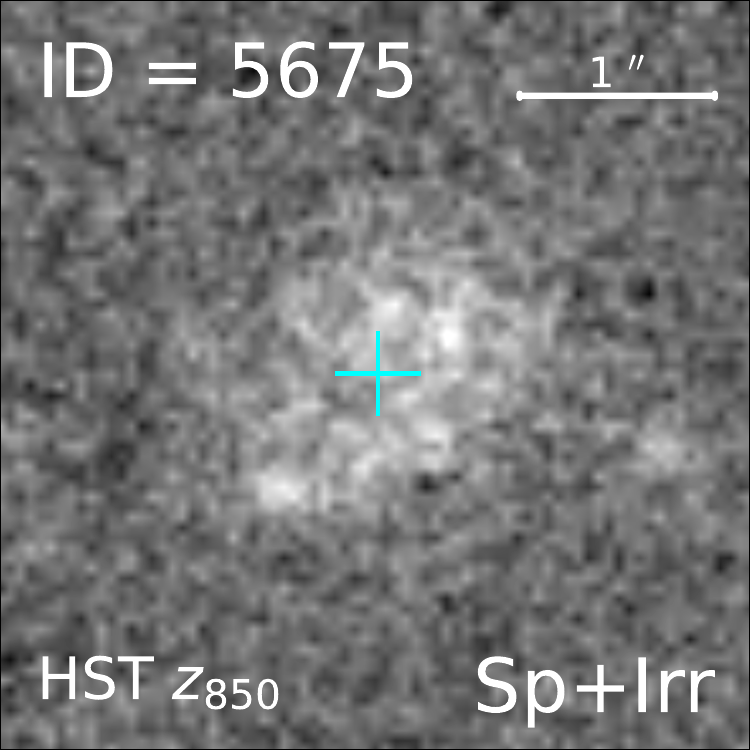}   
	\vspace{20pt}	
	\caption{Caption as per  Figure	\ref{fig:Morphology} }
\end{figure*}

\vspace{10pt}

\subsection{Colour Magnitude Relation} \label{subsec:ColourMagnitude}
\begin{figure}
	\centering
	\vspace{-0.05cm}
	\includegraphics[width=8.5cm]{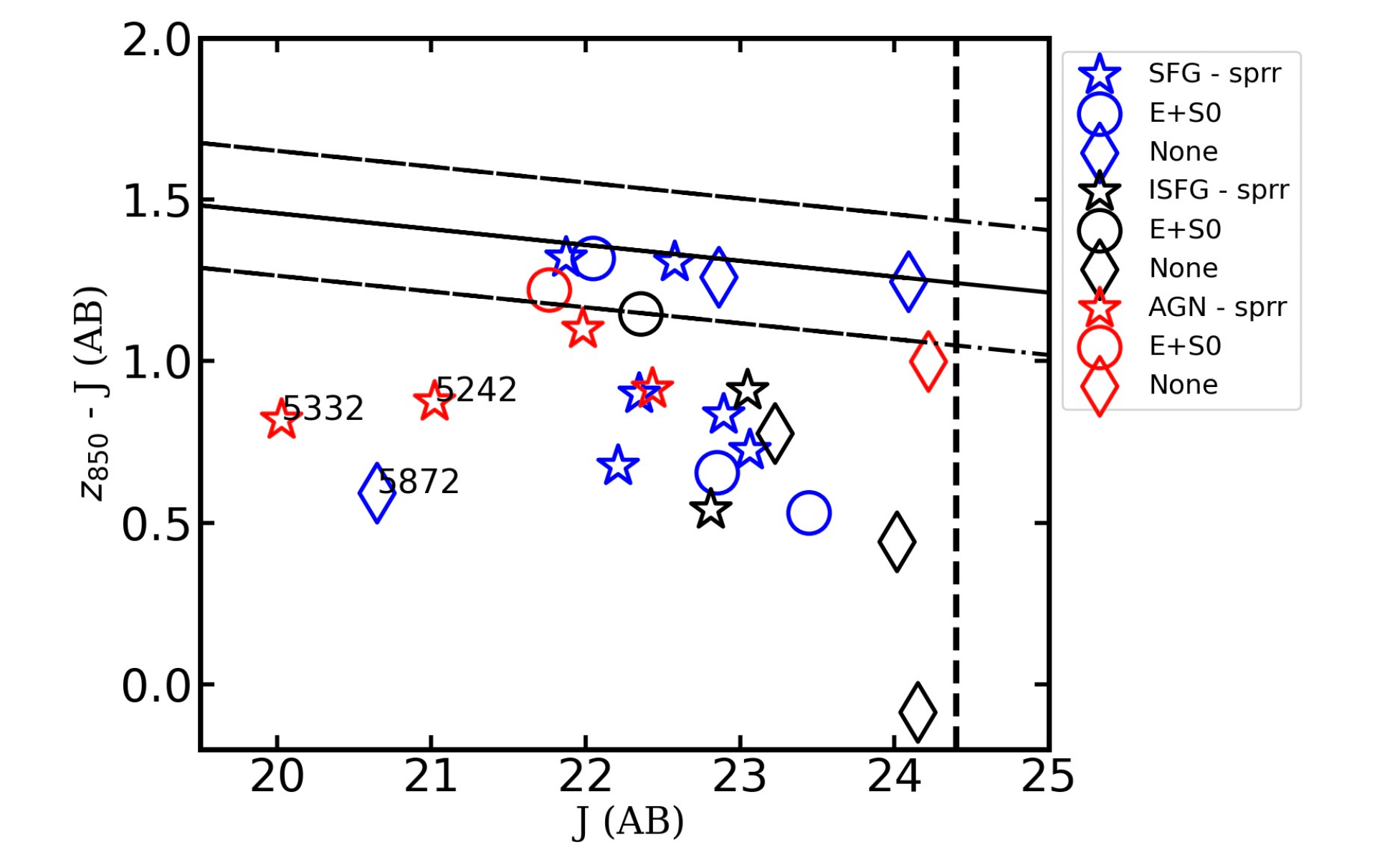}
		\caption{The $z_{850}-J$ colour–magnitude diagram of all the  selected cluster members detected in the MeerKAT $L$-band image located within 0.8~Mpc of the cluster centre.~Elliptical/Lenticular (E+S0) galaxies are shown as circles; spiral/irregular galaxies (Sp+Irr) with stars; galaxies without any morphological classification are marked with a diamond shape (None). The red markers signify the MeerKAT sources classified as AGNs whilst the blue  and black markers represent star-forming galaxies (SFGs) and intermediate SFGs respectively (ISFGs).~The solid black line denotes the red sequence path derived from a fit to the colour-magnitude relation of the elliptical/lenticular galaxies (early-type) detected in \citet{2009ApHilton}.~The black dotted dashed lines show the 3-$\sigma$ deviations above and below the red sequence.~It can be seen that the majority of our cluster members are located below the red sequence line and 12 out of the 17  MeerKAT-detected cluster members with morphological classification have spiral/irregular morphology ($\approx$ 71$\%$).~This may imply that the majority of all the MeerKAT-detected cluster galaxies may be actively forming stars irrespective of whether they have been classified as AGNs, ISFGs or SFGs.} 
	\label{fig:colourmagnitude}
\end{figure}

Figure \ref{fig:colourmagnitude} shows the ($z_{850}-J$) colour versus magnitude ($J$) diagram of all cluster members selected in the MeerKAT $L$-band radio image that had counterparts in the optical and infrared catalogue compiled by \citet{2009ApHilton}.~The solid black line shows a fit to the colour-magnitude relation of the early-type galaxies detected in \citet{2009ApHilton} i.e., the red sequence. We show the  early-type galaxies (i.e., Elliptical/lenticular - E+S0) and late-type galaxies (Spiral/Irregular - Sp+Irr) with circles and stars respectively whilst sources without morphological classification and uncertain morphologies are shown as diamond markers.~The red, blue  and black markers signify MeerKAT-detected cluster members classified as AGNs, SFGs and intermediate SFGs (ISFGs) respectively (see Section \ref{sec:Results} ).~A significant number of our sources fall below the red sequence line i.e., in the blue cloud. 

The majority of these sources are blue and faint, however, three outliers with source IDs \#5332 ,  \#5242 and  \#5872 are blue but very bright compared to the other cluster members.~IDs \#5332  and  \#5242  have dual morphologies i.e., spiral/irregular suggesting that they are blended in the infra-red image ($J$) \citep{2009ApHilton}  and the JVLA image (see Figure \ref{fig:Morphology} ) thus their photometry and  estimated SFR may not be reliable. We also classified them as radio AGNs based on the three selection schemes mentioned in Section \ref{sec:Results}.  

Also, we detected both early-type and late-type galaxies within 3-$\sigma$ of the red sequence in this work (Figure \ref{fig:colourmagnitude}).~The elliptical galaxies within the red sequence of J2215 are very bright  ($J <$ 22.5)  compared to other ellipticals  outside the red sequence. Is it likely that J2215 is  experiencing the final episode of star-formation  within its early-type galaxies ? 

\vspace{10pt}

\vspace{15pt}

\section{Classification of radio sources}\label{sec:Results}
\subsection{Mid-Infrared Colour--Colour Criteria} \label{subsection: Mid  Infra-red Colour-Colour Criteria}
The mid-infrared photometry provides a robust approach to the identification of AGNs.~AGN SEDs are redder than star-forming galaxies in the mid-infrared bands where stellar populations dominate the red, falling portion of the stellar spectra.

~\citet{2005Stern} found that  simple mid-infrared colour criteria could robustly separate AGNs from normal star-forming galaxies and stars (this method could identify $> $ 90$\%$ of spectroscopically selected quasars and Seyfert 1 galaxies). 

\begin{figure}
	\includegraphics[scale = 0.35]{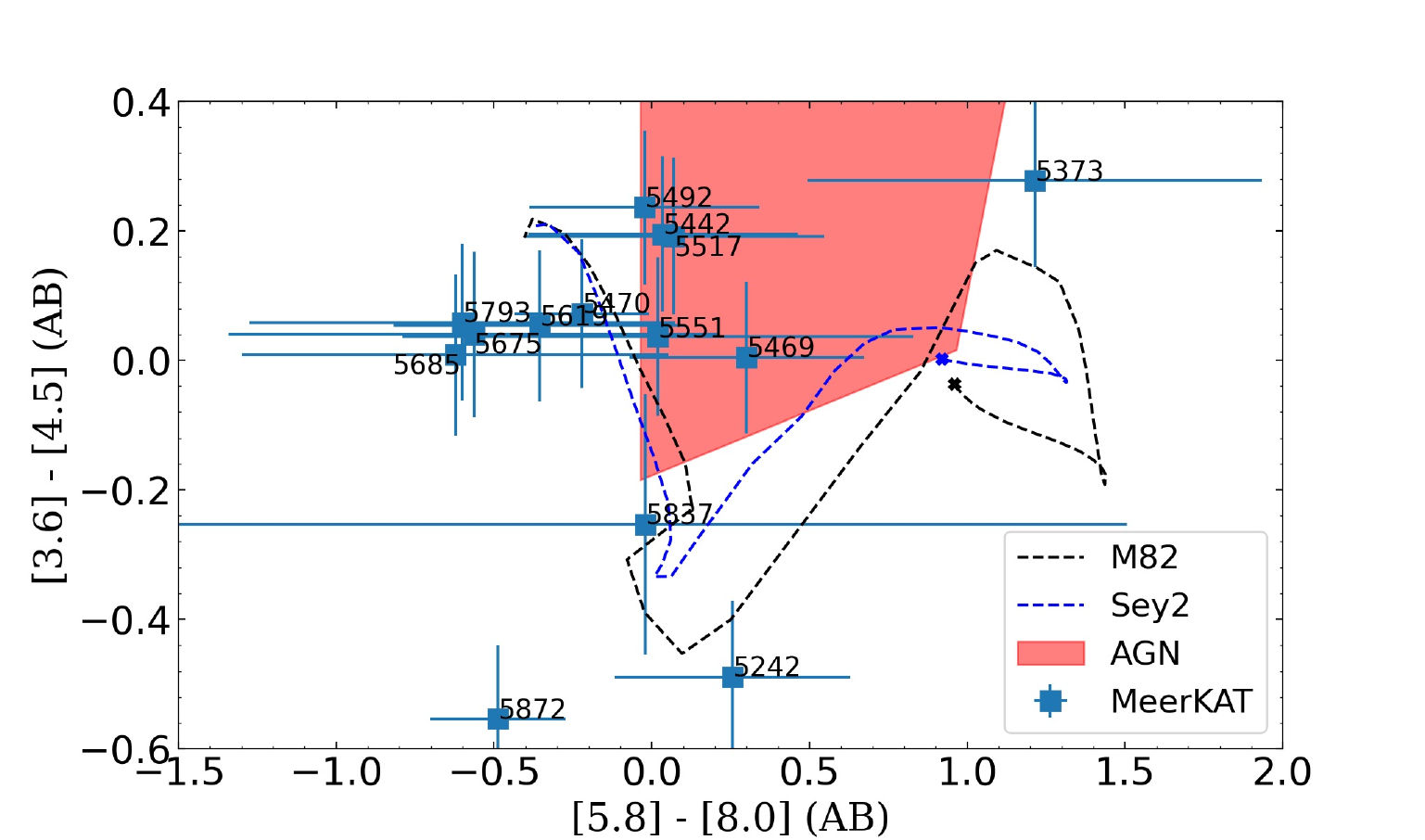}
	\caption{The IRAC colour -- colour plot of the selected cluster members detected with the MeerKAT telescope, with some sources falling within the \enquote{AGN zone} (red-shaded region). Overlaid are two evolutionary tracks of spectral templates from the library of  \citet{2007Polletta}; a starburst galaxy (M82) and an active galaxy (Seyfert 2) as they are redshifted from $z =$ 0 (the crosses in each template track) to $z =$ 2 (the open end of each track). The majority of the cluster members are all located outside the AGN wedge which implies that using only the IRAC colour -- colour plot as diagnostic tool, the majority of the radio emission is mostly powered by star-formation activity rather than AGNs.} 
	\label{fig:Color-color_plot}
\end{figure}

We adopted the same approach from \citet{2005Stern}.
~Figure \ref{fig:Color-color_plot} shows the mid-infrared colour -- colour plot of the MeerKAT-detected cluster members obtained from the IRAC catalogue.~We show an overlay of two non-evolving tracks of spectral templates obtained from the library of \citet{2007Polletta} covering a wavelength range of 8–1000-$\mu$m.~The Figure shows these tracks as they evolve from $z$ = 0 to $z$ = 2.~The shaded region depicts the area within the colour --colour space dominated by broad-lined AGNs \citep{2005Stern}. Any source located outside the \enquote{AGN wedge} is likely to be a normal star-forming galaxy or another type of AGN apart from quasars and Seyfert 1 \citet[since the mid-infrared colour criteria can identify only quasars and Seyfert 1 efficiently,][]{2005Stern}.  
From Figure \ref{fig:Color-color_plot}, it can be observed that at least five cluster members fall within this region, 
IDs:~\#5492,~\#5442,~\#5517,~\#5551 and~\#5469.~Sources with no IRAC data were not included in the colour -- colour AGN diagnostic plot. i.e., 8/24 cluster members did not have IRAC photometry for the two longest wavelength IRAC channels (i.e., 5.8$\mu$m and 8.0$\mu$m ) because the IRAC images at those bands are shallow. 

Due to the large error bars in both the [3.4] $-$ [4.5] and [5.8] $-$ [8.0] colours we can not draw a firm conclusion based on only this criterion (i.e., the mid-infrared colour -- colour AGN diagnostic plot).~The Far Infrared Radio Luminosity Ratio, $q_{\rm{IR}}$ and the Far Infrared Radio Correlation (FIRRC) are alternatives to distinguish normal star-forming galaxies from radio AGN via the so-called \enquote{radio excess} approach \citep{2013DelMoro}.

\begin{figure}
	\includegraphics[scale = 0.35]{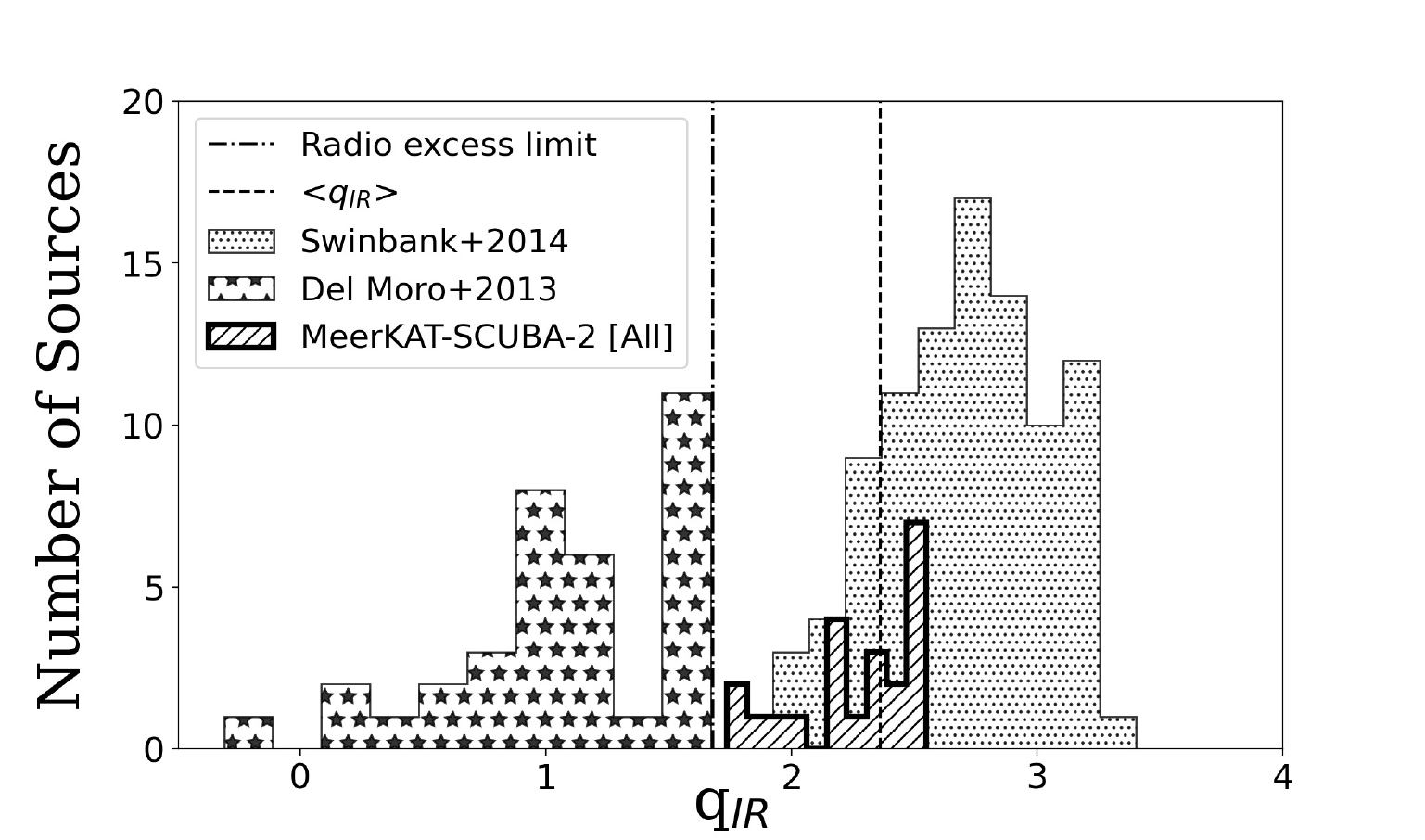} 
	\caption{The distribution of $q_{\rm{IR}} $ (Far Infrared Radio Luminosity Ratio) for all sources in the MeerKAT/FIR sample (hashed histogram. The black vertical dashed line denotes the mean $q_{\rm{IR}} $ value, $<q_{\rm{IR}}>$ greater than 2 for the entire MeerKAT/FIR sample  (i.e.,  $<q_{\rm{IR}}>  =  2.36 \pm 0.04$). 
	The  black vertical dot-dashed line represents the threshold below which sources are classified as radio excess AGN i.e., $q_{\rm{IR}}  = $ 1.68 adopted from \citet{2013DelMoro}. We compared our work with other surveys of high redshift galaxies from the ALESS survey of the  sub-millimetre galaxies in the Extended Chandra Deep Field South observation with ALMA 870--$\mu$m and JVLA  \citet[dotted histogram,][]{2014Swinbank} and the GOODS-Herschel (North) field with the VLA, 24~$\mu$m Spitzer/MIPS, Chandra X-ray and Herschel infrared data \citet[star-filled histogram,][]{2013DelMoro}.~All the MeerKAT/FIR sources fall within the radio normal region using  only the $q_{\rm{IR}} $ criterion.} 
	\label{fig:radioexcess}
\end{figure}

\begin{figure}
	\includegraphics[scale = 0.36]{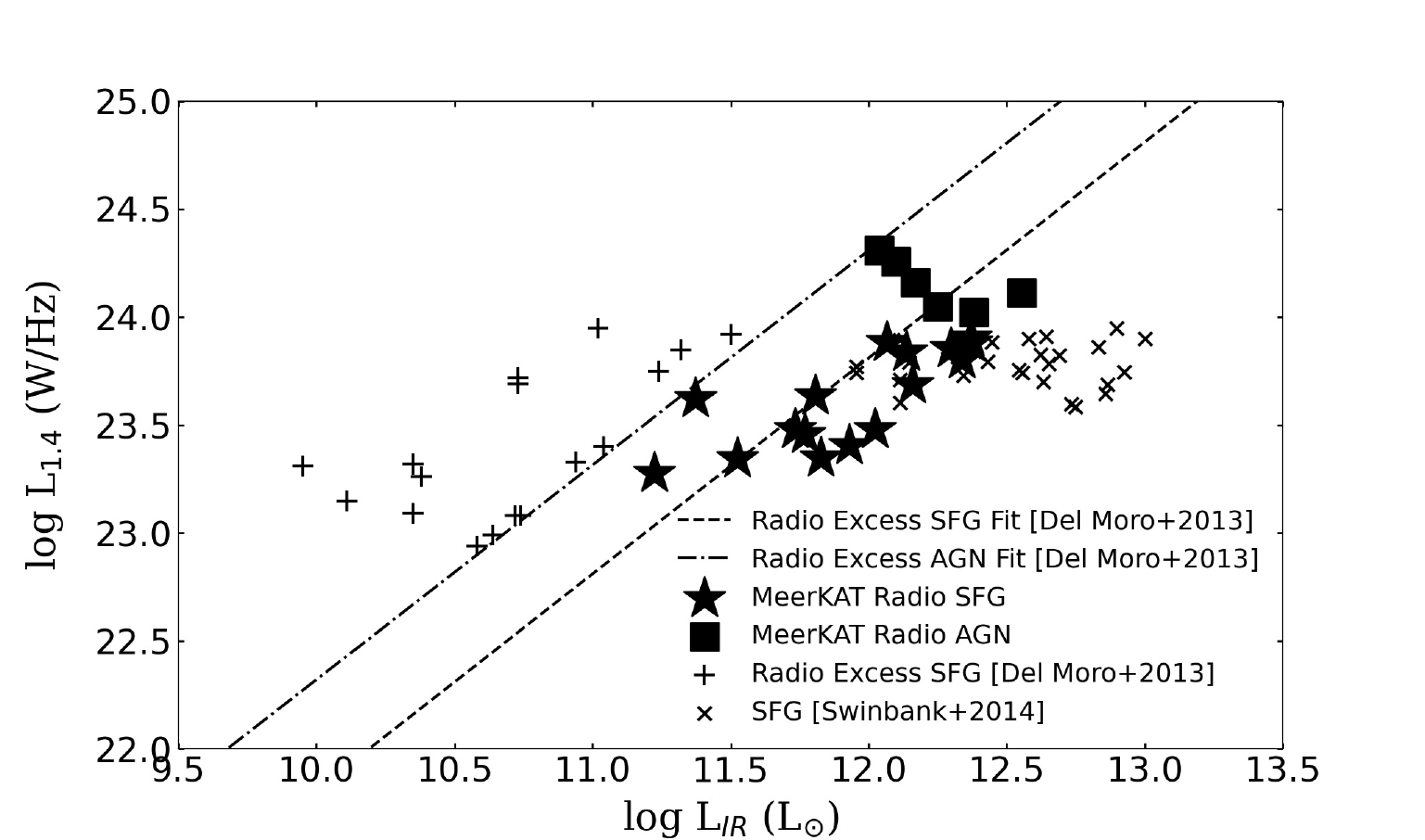}
	\caption{The Far Infrared Radio Correlation (FIRRC) for all the J2215 MeerKAT/FIR galaxies located within 0.8 Mpc of the cluster centre.
	~It can be seen that all the MeerKAT radio normal star-forming galaxies 18/24 (solid star markers)  and  the  radio loud  AGNs 6/24 (solid square markers) are all located within the radio normal region i.e., below the radio excess AGN fit.~We compare our work with other high redshift galaxies from the GOODS- North field, \citet[Del Moro+2013 radio excess star-forming galaxies -- plus symbol;][]{2013DelMoro} and the Extended Chandra Deep Field South (ECDFS) star-forming galaxies \citet[Swinbank+2014 SFG -- cross marker;][]{2014Swinbank}.~The  dashed  and dotted dashed lines represent  the best-fit models to the radio excess star-forming galaxies  and AGNs  from \citet{2013DelMoro}  respectively.}
	\label{fig:RIFC}
\end{figure}

\subsection{Far Infrared Radio  Luminosity Ratio}\label{subsection: The infra-red radio ratio}
The Far Infrared Radio Luminosity Ratio, $q_{\rm{IR}} $ value is used to distinguish normal star-forming galaxies from AGNs \citep{2013DelMoro,2021Delvecchio,2021Radcliffe}.~This $q_{\rm{IR}} $ is defined similarly to \citet[eg.,][]{1992Condon,2003ApJBell,2015ApJMagnelli,calistro2017lofar,2020ApJAlgera} as 
\begin{equation}
	q_{\rm{IR}}    = \log_{10} \left(\frac{L_{IR}}{3.75 \times 10 ^{12}}\right) - \log_{10} (L_{1.4}),
\end{equation}\label{eq:Qvalue1} where $L_{1.4}$ is the rest frame radio luminosity.~$L_{\rm{IR}}$ is the total infrared luminosity obtained from  the best-fit model to the observed  SED using galaxy dust emission templates from \citet{2014ApJDraine}, see Section \ref{subsection: The Far infra-red radio Correlation} for details.

Figure \ref{fig:radioexcess} shows the $q_{\rm{IR}}$ distribution of all  the MeerKAT/FIR selected cluster members (i.e.,  hashed histogram) and other high redshift field galaxies from the VLA/ALMA 870-$\mu$m observations \citep[dotted histogram,][]{2014Swinbank} and VLA/24-$\mu$m \citep[star-filled histogram,][]{2013DelMoro}. 
The cut-off/separation boundary adopted from \citet{2013DelMoro} was defined at $q_{\rm{IR}} = $ 1.68 (The black vertical dot-dashed line).  
Sources with  $q_{\rm{IR}} > $  1.68 were defined as \enquote{radio normal} whilst those below the cut-off were defined to be \enquote{radio excess} sources. The \enquote{radio normal} sources according to  \citet{2013DelMoro} are mostly dominated by star-forming galaxies and low luminous radio AGNs, whilst the \enquote{radio excess} sources are most likely to be dominated by AGNs.

~Comparing our work with other high redshift field galaxy surveys in Figure \ref{fig:radioexcess}, it can be seen that, all the MeerKAT/FIR  detected  cluster galaxies  fall within the radio  normal  region. 

\vspace{15pt}

\subsection{Far Infrared Radio Correlation}\label{subsection: The Far infra-red radio Correlation}
Although there is a vast difference between the emission mechanism of radio (synchrotron) and infrared galaxies (dust re-emission), there is a tight correlation between the radio and far-infrared (FIR) luminosities of star-forming galaxies \citep{1985ApJelou,1991Condon,2001Yun,2003ApJBell,2004appletonfar,2010Ivison,2015Basu,2020ApJAlgera}.
The FIRRC has been found to be well correlated for several galaxy samples irrespective of their stellar activities or morphology excluding radio-loud active galaxies or some other interacting galaxies \citep{2010lisenfeldshock,2015lisenfeldfar}.~Previous works have established the validity of the FIRRC out to $z \approx 2 $ \citep{2002Garrett,2004appletonfar,2010sargentvla,sargent2010no, 2020ApJAlgera}, with only a few uncertain deviations at higher redshifts $\left(3 \leq z \leq  6 \right)$ \citep{2009murphyfar,2009seymourinvestigating,2022ApJShen}.  
The FIRRC for all the J2215 members from the MeerKAT $L$-band survey (this work) and its far-infrared (FIR) counterpart can be seen in Figure \ref{fig:RIFC}. 

We compare our work with other high redshift radio excess star-forming galaxies from the GOODS-Herschel North field  \citet[plus marker,][]{2013DelMoro} and the ALMA LESS (ALESS) survey of sub-millimetre Extended Chandra Deep Field South (ECDFS) observation \citet[cross marker,][]{2014Swinbank} in Figure \ref{fig:RIFC}.~All of the MeerKAT/FIR sources are located within the radio  normal star-forming region.

\vspace{15pt}

\subsection{Source Classification} \label{subsec:Source Classification}
We employed three selection criteria to distinguish star-forming galaxies (SFGs) from AGNs i.e., sources with  $L_{1.4}$ $<$ 10$^{24}$ WHz$^{-1}$,  $q_{\rm{IR}}$   values $>$ 1.68 and finally sources detected outside the IRAC AGN zone were all considered as potential SFG candidates.~However, sources that passed at least two out of the 3 selection criteria mentioned earlier are considered  intermediate SFGs (ISFGs) otherwise, they are selected  as AGNs.~In total, we classified 12/24 sources as SFGs, 6/24 as ISFGs and 6/24 as AGNs.

\section{AGN activity within J2215} \label{subsec:AGN Activity in higher and low redshift surveys}
Based on the three selection criteria described in  Section \ref{sec:Results} we classify all the other non-SFGs as potential AGNs ie., 6/24 this implies that $25\%$ of the MeerKAT sources that are cluster members are AGN hosts and there is a possibility of a higher AGN activity ongoing in the core of this high redshift cluster.

\citet{2016ApJAlberts}  found that in a sample of $\approx$ 250 galaxy clusters obtained from the  IRAC Shallow Cluster Survey (ISCS) and IRAC Distant Cluster Survey (IDCS) at 0.5 $< z <$ 2, AGN-composite (moderately dominated AGN fraction) cluster galaxies increased with respect to redshifts above the field galaxies at 1.0 $< z <$ 1.5. They attributed the rise in the AGN-composite cluster galaxies at these redshifts to a conducive cluster environment that allows black hole growth or activities. 

Also, a multi-wavelength study of AGNs using three different selection criteria i.e.,  mid-IR colour, radio luminosity, and X-ray luminosity of galaxy clusters in the $8.5^{\circ} \times 8.5 ^{\circ}$ Boots field by \citet{2009ApJGalametz} showed that a higher  fraction of AGN is observed in cluster centres than in the field at  $ z \gtrsim$ 1.0. 

Further, \citet{2013ApJMartini} studied the AGN evolution of 13 galaxy clusters from the Spitzer/IRAC Shallow Cluster Survey and their surrounding field galaxies at 1 $< z <$ 1.5. They observed that there was an order of magnitude increase in the evolution of AGN fraction in clusters from  0 $< z <$ 1.25 than in the field of these clusters whilst a reverse is seen at lower redshift i.e., AGN fraction is a factor of $\approx$ 6 times higher in the field than in the clusters at lower redshift. 

High redshift studies of two proto-clusters and field galaxies at $z >$ 2 showed an increase in AGN fractions in proto-cluster than in field galaxies \citep{2009ApJLehmer,2010MNRASDigbyNorth}. 

Given the above examples in relation to the high redshift clusters and their AGN fraction, the J2215 galaxy cluster is no different from other higher redshift clusters investigated by previous surveys.

\vspace{30pt}

\section{STAR FORMATION ACTIVITY WITHIN J2215}\label{sec:STAR_FORMATION_WITHIN_J2215}
\subsection{Star Formation Rate vs  Stellar Mass} \label{subsec:Star Formation Rate vs Stellar Mass}
Observations suggest a  rough correlation between the star formation rates in galaxies and their stellar mass (M$_{\ast}$), the so-called \enquote{main sequence} \citep{2004MNRASBrinchmann,2007ApJNoeske,2007Elbaz}. This correlation claimed to hold from $z = $ 0 to $z = $ 4 \citep{2004MNRASBrinchmann,2007ApJNoeske,2007Elbaz,2007ApJDaddiA,2009ApJPannella,2015ASchreiber}. This has resulted in the \enquote{dual mode} of star formation evolution i.e., the normal star-forming mode and the star-burst mode depending on the position of the galaxy in the SFR--M$_{\ast}$ plot although there are increasing doubts about the reality of this distinction  \citep{2018Elbaz, 2021MNRASPuglisi}. 

Galaxies found within the \enquote{main sequence} region are termed  the \enquote{normal star-forming} galaxies, whilst those above the main sequence regions are termed  the \enquote{star-burst} galaxies \citep{2011Elbaz}. The \enquote{normal star-forming} mode is associated with the gas accretion process \citep{2009ApJDekel,2010MNRASDav} whilst the \enquote{star-burst} mode is often associated with the major merger interactions \citep{2011ApJRodighiero,2010ApJDaddi,2010MNRASGenzel}. 

\begin{figure}
	\includegraphics[trim = 0cm 0cm 0cm 0cm, scale = 0.35]{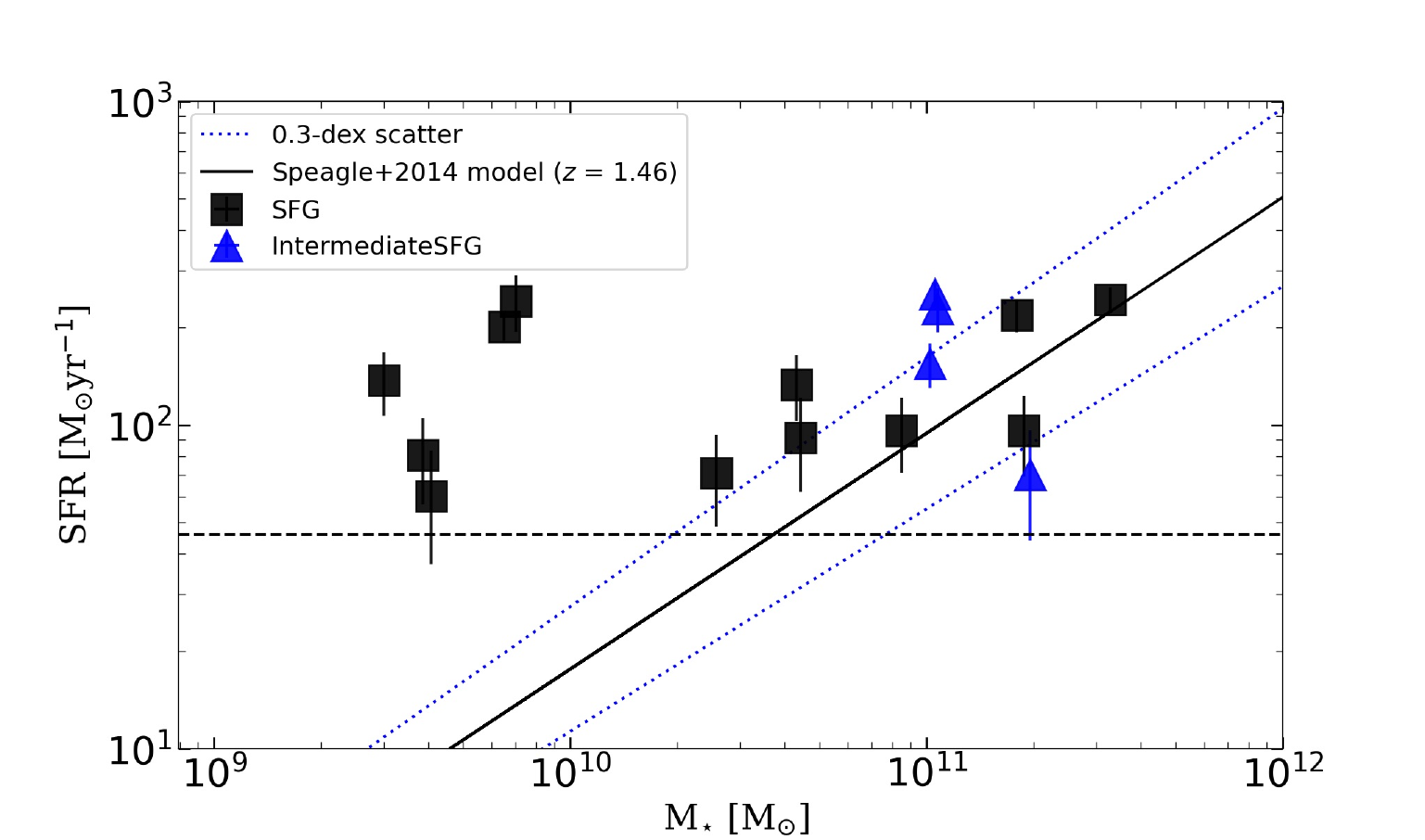} 
	\caption{The SFR vs M$_{\ast}$ of the  SFGs (black solid square) and the  intermediate SFGs (blue solid triangle) detected in the J2215 MeerKAT image. The black line represents the best-fit to the main sequence (MS) relations derived from observations in literature  \citep[i.e., equation 28 of ][]{2014ApJSpeagle} at the cluster redshift ($z =$ 1.46). The $\pm$ 0.3~dex scatter about the adopted MS fit to our data is represented as the blue dotted line. The 4-$\sigma$ radio flux detection limit equivalent to a SFR value of 46 M$_{\odot}$yr$^{-1}$ is represented as the black horizontal dashed line.}
	\label{fig:SFRMassRelation}
\end{figure}

In Figure \ref{fig:SFRMassRelation} we show how the SFR of the cluster galaxy varies with the galaxy stellar mass. The stellar masses used in the work were obtained by fitting SED models described in Section  \ref{subsection:SEDCIGALE} to the observed data.~The majority of the SFGs and  ISFGs  are located on or above the best-fit to the main sequence (MS) relations derived from observations in literature  \citet[black line,][]{2014ApJSpeagle}.
It can also be seen in Figure \ref{fig:SFRMassRelation} that, our SFR detection limit of 46 M$_{\odot}$yr$^{-1}$ lies above the MS for all but the most massive galaxies in our sample.

\citet{2018MNRASCoogan} showed that only 2/8 of the Cl J1449+0856 cluster galaxies at $z =$ 2 were found to reside above the main sequence relation (see Figure 8 of that paper) although the majority of the cluster galaxies exhibited star-burst-like characteristics.  

A similar trend was also observed by \citet{2019MNRASmith} during the study of  CLJ1449 at $ z = $ 2. They found that most of the cluster members lay within the MS region (see Figure 10 of that paper) with only a few cluster galaxies located about a factor of 2 or 3 above the predicted MS region at the cluster redshift. They suggested that the  star-burst activities within those cluster galaxies could be caused by a merger event 
and this goes to support the claim made by \citet{2018MNRASCoogan} that the star-formation activity of the same cluster is mostly merger-driven.

\enquote{Star-burst} galaxies have also been observed by \citet{2013ApJHung} via visual morphological analysis of the 2084 COSMOS field galaxies (Herschel-PACS and SPIRE observation) at z = 0 - 1.5  they observed that about 50$\%$ of the galaxies were undergoing a merging process and these systems were seen to be deviating from the main sequence relation supporting the claim that galaxies that are located above the MS relation are mostly merging systems.They also observed that some main sequence galaxies ($\approx$ 18 $\%$) were also undergoing merging events.
Is it possible that the majority of J2215 cluster galaxies are undergoing a merger episode at this high redshift?  This is because in Figure  \ref{fig:SFRMassRelation}  it can be observed  that  the majority of  the  cluster galaxies fall on or above the main sequence relation. 

Also, in a CO (2-1) emission line survey of the J2215 cluster galaxies conducted by \citet{2017ApJHayashi} they showed that, two of the cluster galaxies located at $R\approx$ 0.5 Mpc were undergoing a merger activity \citet[i.e., \# ALMA.B3.15 and \# ALMA.B3.16 - this is visible as two bulges in the intensity maps of the individual cluster galaxies in Figure 3 of][the \# ALMA.B3.16 counterpart is \# 5442 in this work]{2017ApJHayashi}. 
Again, \citet{2022ApJIkeda} classified 6 out of 17 CO emitters located within $\approx$ 0.5 Mpc of the J2215 cluster centre as early-stage mergers  i.e.,  \#ALMA.B3.06, \#ALMA.B3.09 and \#ALMA.B3.16
which correspond to IDs  \#5470, \#5619 and \#5442 respectively in this work. Therefore J2215 may be undergoing a high rate of  galaxy-to-galaxy merger episode at this high redshift.

\vspace{15pt}

\subsection{Integrated Star Formation Rate}\label{subsec:Integrated Star Formation Rate}
~We estimated a cluster-wide integrated star formation rate for 12/24 galaxies selected to be star-forming based on the three criteria mentioned in Section  \ref{subsec:Source Classification}.  

This resulted in $\sum$~SFR = 1700  $\pm$ 330 M$_{\odot}$yr$^{-1}$ at $\lessapprox$ 0.8~Mpc and  1100 $\pm$ 210 at $\lessapprox$ 0.5~Mpc for 7/24 SFGs. Our measurement of the integrated SFR is approximately two times higher than the value reported by \citet{2015ApJMa} within 0.8~Mpc of the cluster centre i.e., $\sum$~SFR value of $800 ^{+ 360} _{-250}$  M$_{\odot}$ yr$^{-1}$  but it is in agreement with the value reported  by  \citet{2017ApJStach}  (i.e., 840~M$_{\odot}$yr$^{-1}$ within the central $\approx$ 0.5~Mpc of the J2215 cluster). 

The integrated SFR that we obtain for J2215 is of similar magnitude to that seen in other clusters at similar redshift.~\citet{2014MNRASantos} obtained an $\sum$~SFR value of $780 \pm$ 90 M$_{\odot}$ yr$^{-1}$ for a high redshift cluster at 1.62 (i.e., CLG0218.3-0510) within 0.5~Mpc radius.~\citet{2010ApJTran} also estimated an $\sum$~SFR of $\approx$ 1370 M$_{\odot}$ yr$^{-1}$ within $<$ 0.5 Mpc of the same cluster at that redshift emphasizing the possibility of an underestimation due to the low sensitivity of the 24-$\mu$m image, which implies that there may be a possibility of higher star formation activity ongoing in this high redshift cluster that has not been captured. 

The 0.5 Mpc central region of another high redshift cluster CLJ1449 at $z =$ 2 was also seen to be actively forming stars with an $\sum$~SFR value of 470 $\pm$ 120 M$_{\odot}$ yr$^{-1}$ \citep{2019MNRASmith}. Similar values were also estimated within the core of this cluster by \citet{2018Strazzullo} and \citet{2018MNRASCoogan}.

\citet{2019Cooke} estimated a median $\sum$ SFR  = 750 $\pm$ 190 M$_{\odot}$ yr$^{-1}$ within 1 Mpc of the cluster centre of eight submillimetre galaxy clusters at $z\approx$  0.8--1.6. This value is also in agreement with the residual field contamination corrected integrated SFR of eight high redshift ($z=$ 1.6 -- 2.0) submillimeter galaxy clusters studied by \citet{2024Smail}  i.e., a median  $\sum$ SFR = 530 $\pm$ 80 M$_{\odot}$ yr$^{-1}$.  

All these high redshift galaxy cluster studies gives an indication that there is indeed an increase in the star formation activity within the core of some high redshift clusters at $z =  $ 1.5 and beyond.     

\vspace{15pt}

\subsection{Mass--Normalized Integrated SFR}
~We determined the mass-normalized integrated SFR for all the 12/24 cluster members classified as non-AGNs, adopting a cluster mass M$_{\rm{cl}}$ of $3~\times$ 10$^{14}$~M$_{\odot}$ from \citet{2015ApJMa} and an integrated star-formation rate of  $\sum$~SFR of 1700 $\pm$ 330~M$_{\odot}$yr$^{-1}$. We obtained a value of $\sum$ (SFR)/M$_{\rm{cl}}$ = (570 $\pm 110 ) \times $ 10$^{-14}$~yr$^{-1}$. 

Previous works have suggested that the critical epoch of star formation is at $ z \approx$ 1.4; a study conducted via the IRACS Shallow Cluster Survey revealed that above $z = 1.4$ active star formation activities are dominant within the cluster core and the reverse is observed below this redshift \citep[eg.,][]{2013ApJBrodwin,2014MNRASAlberts}.

\begin{figure} 
	\begin{center}
		\includegraphics[width=9.0cm]{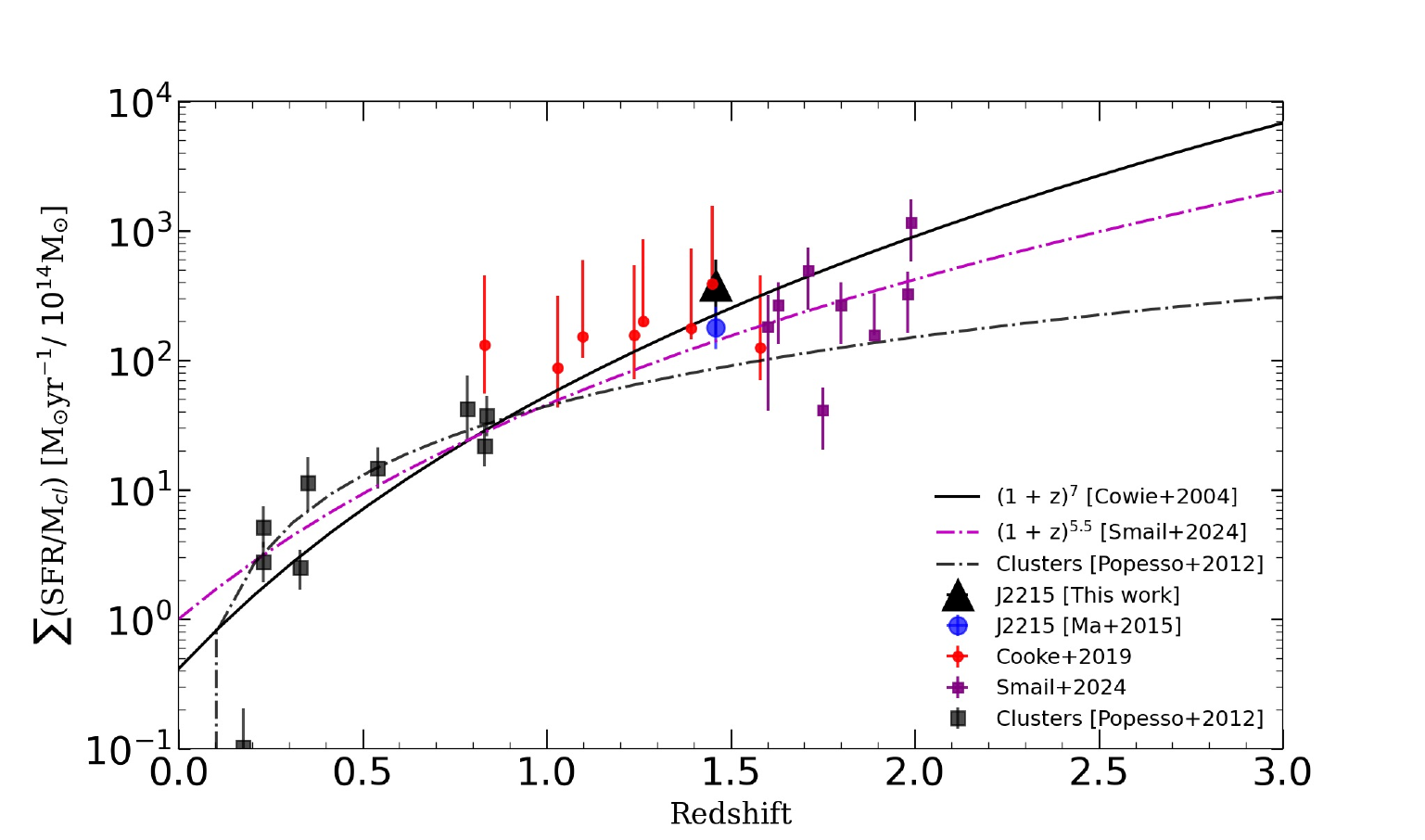}
		\caption{The $\sum$~(SFR)/M$_{\rm{cl}} - z$ relation for  low and high redshift clusters. The solid black line depicts the  evolutionary  model for  the number density of  star-forming ULIRGs radio sources out to $z \approx$ 1.5  that follows the $(1 + z)^{7} - z$ relation \citep{2004ApJCowie,2006ApJGeach}.~The purple dash-dotted line represents the best-fitting model  to the  $\sum$~(SFR)/M$_{\rm{cl}}$ of massive galaxy clusters studied in \citet{2024Smail} \citep[i.e., $(1 + z)^{5.5} - z$,][]{2013Webb}.~The black dash-dotted line denotes the best-fitting  model to the observed $\sum$~(SFR)/M$-z$ (the integrated star formation rate per unit total halo mass) relation for the 9 cluster sample detected by \citet{2012Popesso} including a blind extrapolation of the best-fitting model beyond the observed cluster redshifts (i.e.,  $0.15 < z < 0.85$). All the power law models have been normalized at $z =$ 0.8 and all measurements have been scaled to an equivalent FIR luminosity limit of 10$^{12}$~L$_{\odot}$ where necessary to ensure a uniform comparison with other high redshift clusters studied in \citet{2024Smail}. All  three power law models converge at the normalization redshift ($z =0.8$)  however beyond that redshift the $\sum$~(SFR)/M$_{\rm{cl}}$ increases sharply with high $z$ for the \citet{2004ApJCowie} and \citet{2024Smail} model with a relatively uniform pattern for the \citet{2012Popesso}  model which deviates by a factor of  $\approx$ 3 and 2  below the  \citet{2004ApJCowie} and \citet{2024Smail} model respectively.~The wider deviations from the two models at higher redshifts may be due to insufficient data at higher redshifts to better represent the best-fit model beyond the redshift range of the 9 cluster samples investigated by \citet{2012Popesso}. } 
		\label{fig:MassNormalized}
	\end{center}
\end{figure}

~This trend can be visualized in Figure \ref{fig:MassNormalized} where the $\sum$~(SFR)/M$_{\rm{cl}} - z$ relation for some high redshift clusters nearly follows the $(1 + z)^{7} - z$ relation.
i.e., the evolutionary model for the number density for radio star-forming galaxies whose luminosity values correspond to  local ULIRGs from $z \approx$  0 -- 1.5 proposed by \citet{2004ApJCowie}.

The black dash-dotted line denotes the best-fitting model to the observed $\sum$~(SFR)/M$_{\rm{cl}} - z$ relation for the nine cluster sample ($z =$ 0.15 -- 0.85) studied by \citet{2012Popesso}.

Comparing our work with previous surveys of high redshift clusters one can see that J2215 generally follows the \enquote{higher mass-normalized integrated SFR at higher redshift} trend.~When we correct our measurements to a corresponding FIR luminosity limit of  $\geq$ 10 $^{12}$~L$_{\odot}$  and normalize  all the power-law models in Figure  \ref{fig:MassNormalized}  at $z =$ 0.8 following  \citet{2024Smail} the $\sum$~(SFR)/M$_{\rm{cl}}$ value becomes a factor of $\approx$ 4.4, 1.7 and 3.5 higher than the value predicted by the  \citet{2012Popesso}, \citet{2004ApJCowie} and  \citet{2024Smail}  relation respectively. The J2215 cluster at $z =$ 1.46 is in relatively better agreement with the \cite {2004ApJCowie} power law and also consistent with the submillimetre  value of the same cluster studied by  \citet{2019Cooke}. Again, J2215 was among the two highest star forming clusters studied by \citet{2019Cooke}. They  also suggested that the mild evidence of a binomial velocity distribution of the cluster \citep{2010ApJHilton}  may be due to cluster to cluster merger event and that event may be responsible for the  extremely high star formation rate recorded in their work. Should this claim be true then J2215 may be undergoing a \enquote{twin} merger event  i.e.,  cluster galaxy to galaxy merger (see Section \ref{subsec:Star Formation Rate vs Stellar Mass}) and cluster to cluster merger event. This may imply that J2215 is at the peak of its star-formation phase at $z =$ 1.46.

\vspace{15pt}

\section{Summary} \label{section:summary}
We have studied star formation and AGN activity within the XMMXCS J2215.9-1738 cluster at $z = $ 1.46 using MeerKAT $L$-band radio data. We combined the radio data with other archival optical and infrared data to study the star formation and AGN activities within the cluster. We use for the first time in the study of the cluster three selection criteria; the radio luminosity, the far-infrared radio ratio $q_{\rm{IR}}$ and the mid-infrared colour, to classify sources as normal star-forming, intermediate star-forming and AGN galaxies. We classified 12/24 (50\%) as star-forming (SFGs) and 6/24 (25\%) as intermediate star-forming galaxies (ISFGs) and  6/24 (25\%) as radio AGNs. 

We further investigated the evolution of star-forming galaxies in clusters with redshifts.
We achieved this by comparing the mass-normalized integrated SFR of the MeerKAT-detected cluster members with other lower and higher redshift clusters.
 
We observed that XMMXCS J2215.9-1738 is consistent with the $(1 + z)^7 - z$  relation proposed by \citep{2004ApJCowie,2006ApJGeach} and these higher star formation activities occur within clusters in their youthful ages (high redshift) compared to their older ages (low redshift). 

Finally, we also showed that most of the cluster members that are radio sources detected by MeerKAT at this high redshift are late type-galaxies. This also gives an indication that this cluster at $z =$ 1.46 is actively star-forming. 

This work can be extended to improve upon the morphological classification and the visual identification of merger events with  higher resolution observations like that of the  \textit{James Webb Space Telescope} (JWST). 
We hope to also employ a more advanced technique to mitigate the blending effect in our MeerKAT radio images whilst we anticipate the arrival of the higher resolution and more sensitive near future radio telescopes such  as the  MeerKAT extension project (MK+) and the Square Kilometer Array (SKA). 

\vspace{14pt}

\section*{ACKNOWLEDGEMENTS}
We thank the referee for a number of useful suggestions that improved this work.
DYK acknowledges financial support from the South African Radio Astronomy Observatory (SARAO) via the Human Capital Development (HCD) programme. 
MH acknowledges financial support from SARAO and the National Research Foundation of South Africa.
IRS and AMS  acknowledge STFC(ST/X001075/1). 
We acknowledge the use of the University of Kwazulu-Natal UKZNs' high computing facility (\url{https://www.acru.ukzn.ac.za/~hippo/}) and South Africa's  National Integrated Cyberinfrastructure System (NICIS) Centre for High performance computing facility (CHPC) for our data processing. 
The primary data used for this work was from the MeerKAT radio telescope.
The MeerKAT telescope is operated by the South African Radio Astronomy Observatory, which is a facility of the National Research Foundation, an agency of the Department of Science and Innovation.

\section*{DATA AVAILABILITY} 
The MeerKAT continuum data used in this  work can be found in the MeerKAT data archive managed by the South African Radio Astronomy Observatory (\url{https://archive.sarao.ac.za/}). The data are associated with proposal ID SCI-20190418-MH-01.

\bibliographystyle{mnras}
\bibliography{references}

\appendix
\section{Spectral Energy distribution plots}\label{Appendix:sedplots}
We show the spectral energy distribution plots (SED) for some selected  MeerKAT sources that are J2215 cluster members. The best-fitting model for each of the cluster galaxies is shown as the black curve. 
The stellar attenuation and unattenuated models are shown as the yellow and the dashed blue curve respectively. The nebular, dust and AGN emission models are also shown with the green, red and orange curves respectively. The non-radio synchrotron emission for star-forming galaxies is shown as the brown curve. The red dots are the CIGALE-generated model fluxes, the purple open circles show the observed fluxes, whilst the green triangles represent the upper limit values for the radio sources that were not detected in either the optical or infra-red bands.~We assumed the following upper limits: A 4 sigma upper limit for the SCUBA-2 sources (i.e., 21.6~mJy -- 450~$\mu$m and 2.52~mJy -- 85~$\mu$m), 5 sigma upper limit for the 24~$\mu$m Spitzer MIPS (0.070~mJy) and for the IRAC channels we assumed a magnitude upper limit value of  $\approx$ 23 mag (Ch1 and Ch2) and  $\approx$ 20 mag (Ch3 and Ch4). The 5 sigma upper limit of the HST  $z_{850}$ band is $\approx$ 26 mag and the $i_{775}$ band $\approx$ 25.1 mag. 
$J$ band image is $\approx$ 24.4 whilst that of the $K_{s}$ -- band is $\approx$ 24.5.

\begin{figure*} \label{fig:MorphologySED1}
\includegraphics[width=6.0cm,height=6.0cm]{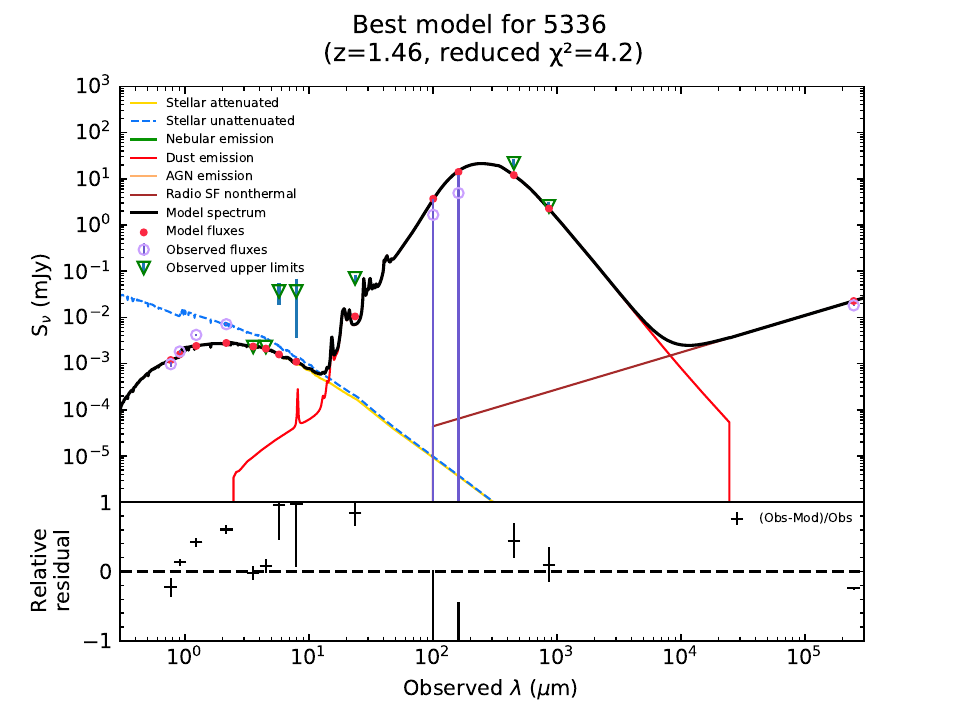} \hspace{-5.00mm}
\includegraphics[width=6.0cm,height=6.0cm]{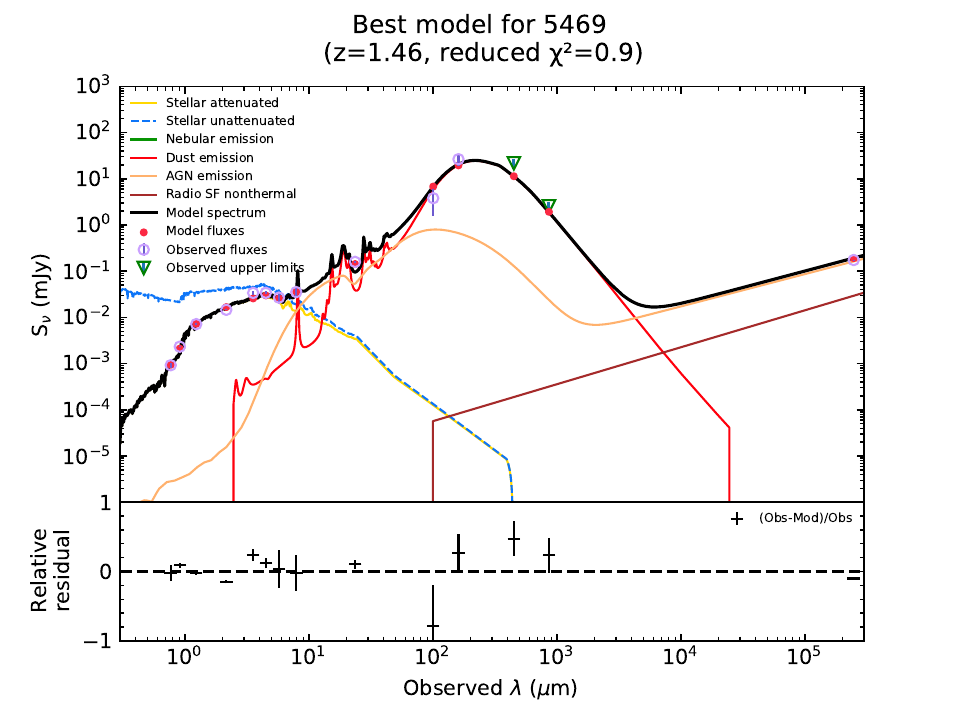} \hspace{-5.00mm}
\includegraphics[width=6.0cm,height=6.0cm]{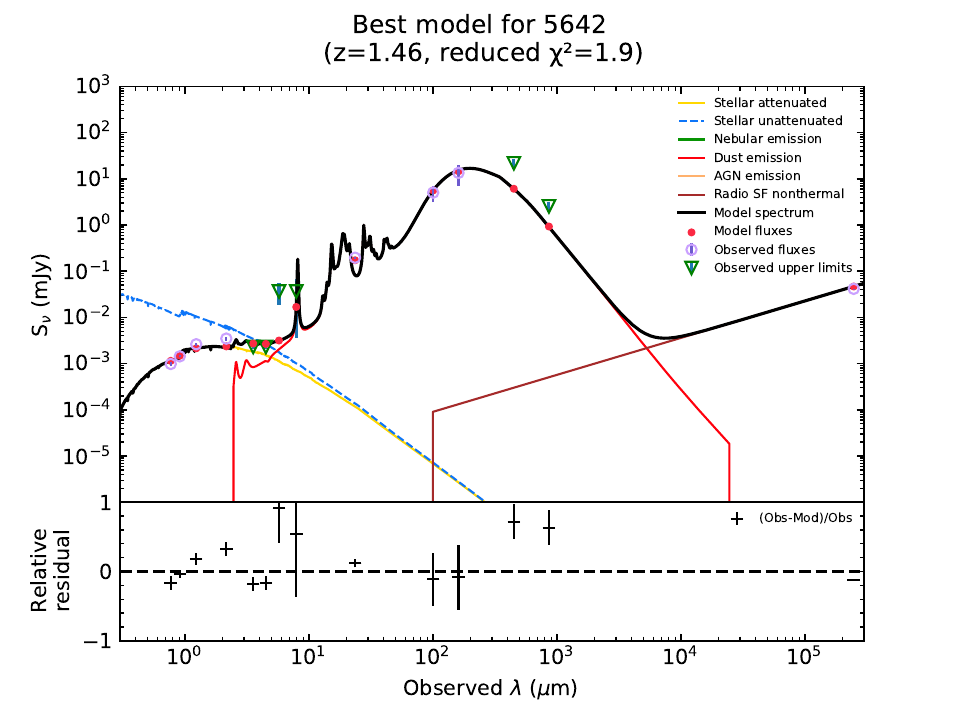} \hspace{-5.00mm}
\includegraphics[width=6.0cm,height=6.0cm]{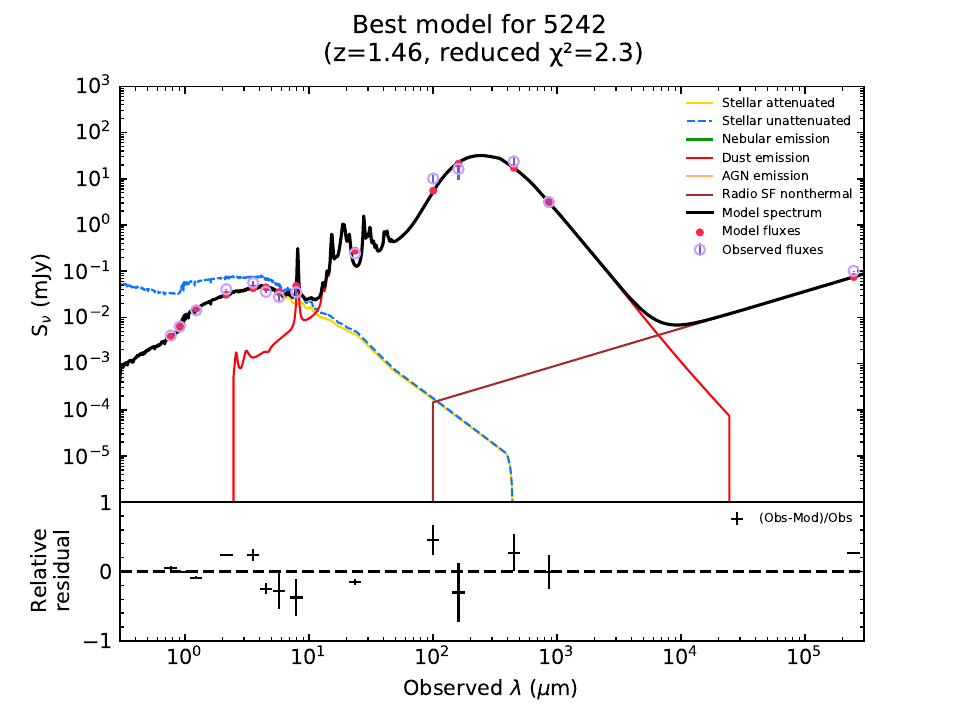} \hspace{-5.00mm}
\includegraphics[width=6.0cm,height=6.0cm]{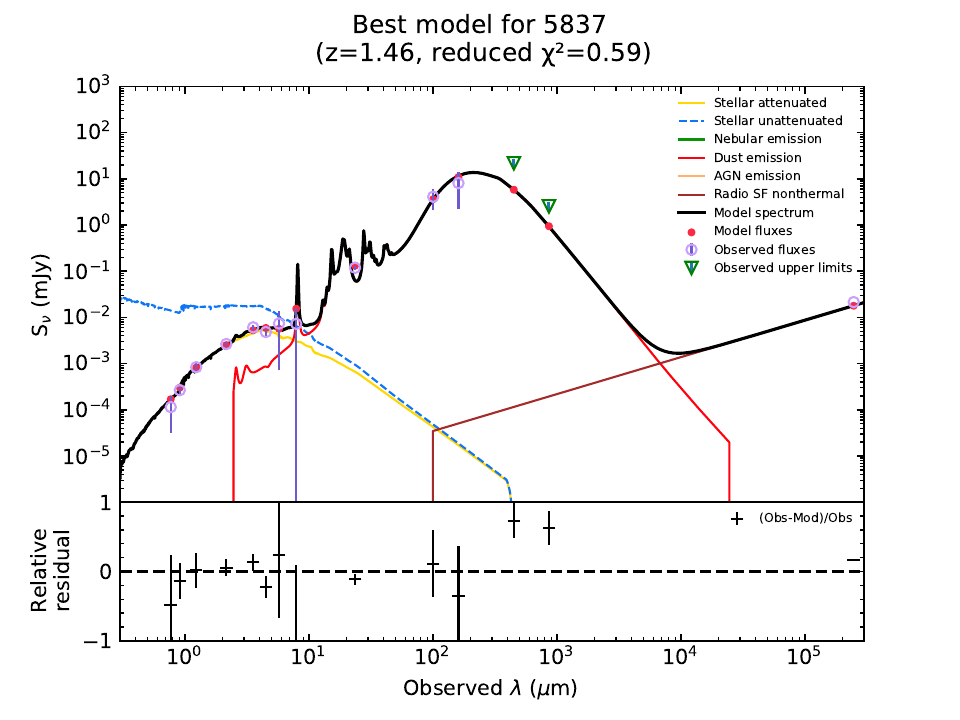} \hspace{-5.00mm}
\includegraphics[width=6.0cm,height=6.0cm]{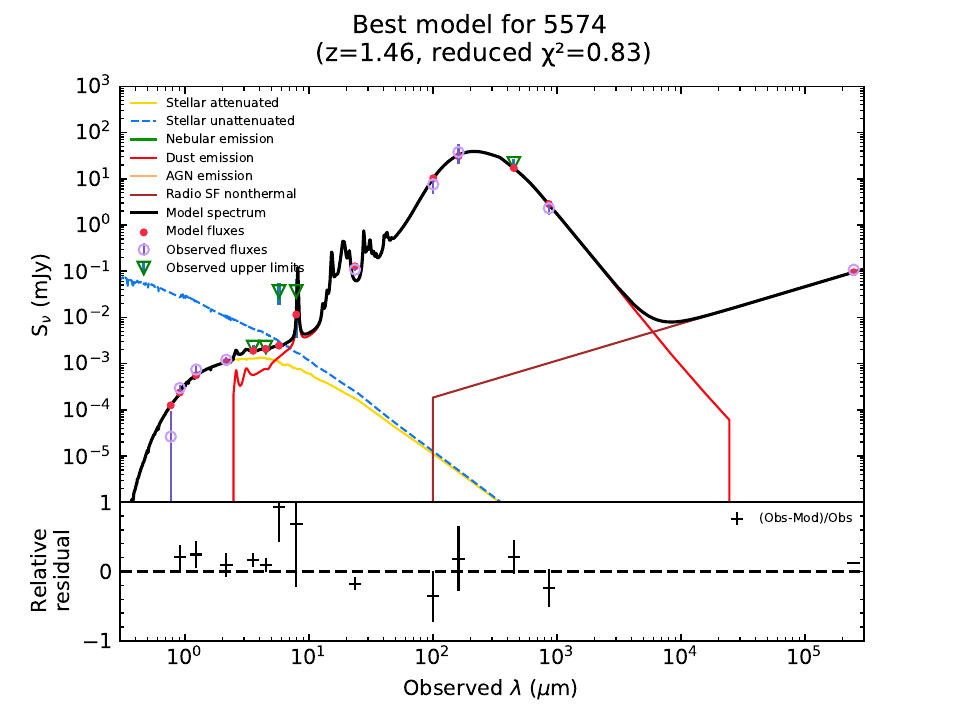} \hspace{-5.00mm}
\includegraphics[width=6.0cm,height=6.0cm]{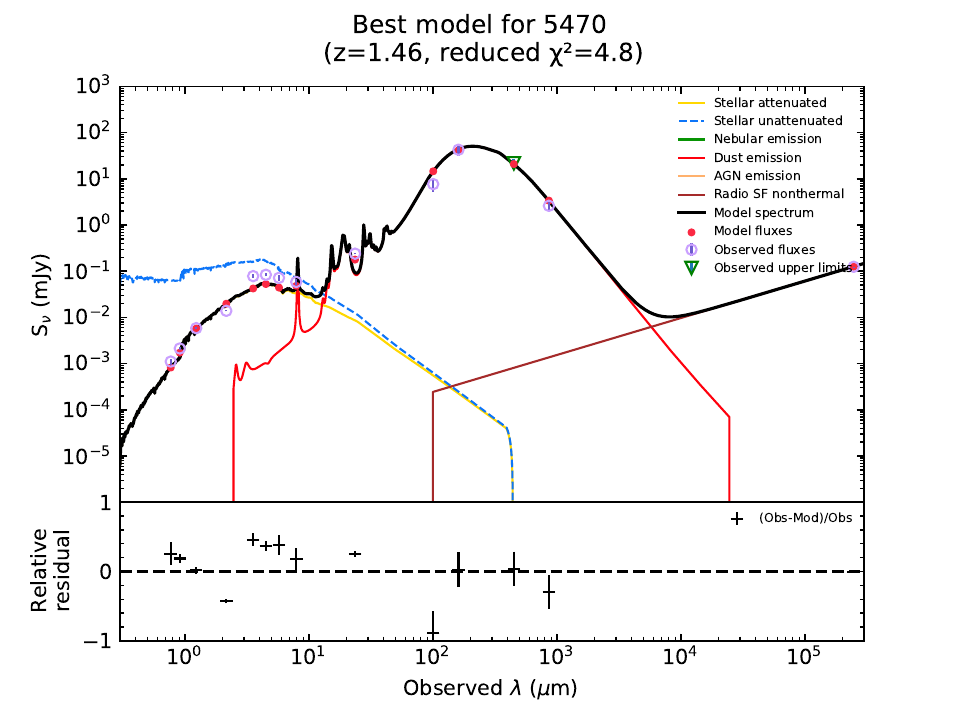} \hspace{-5.00mm}
\includegraphics[width=6.0cm,height=6.0cm]{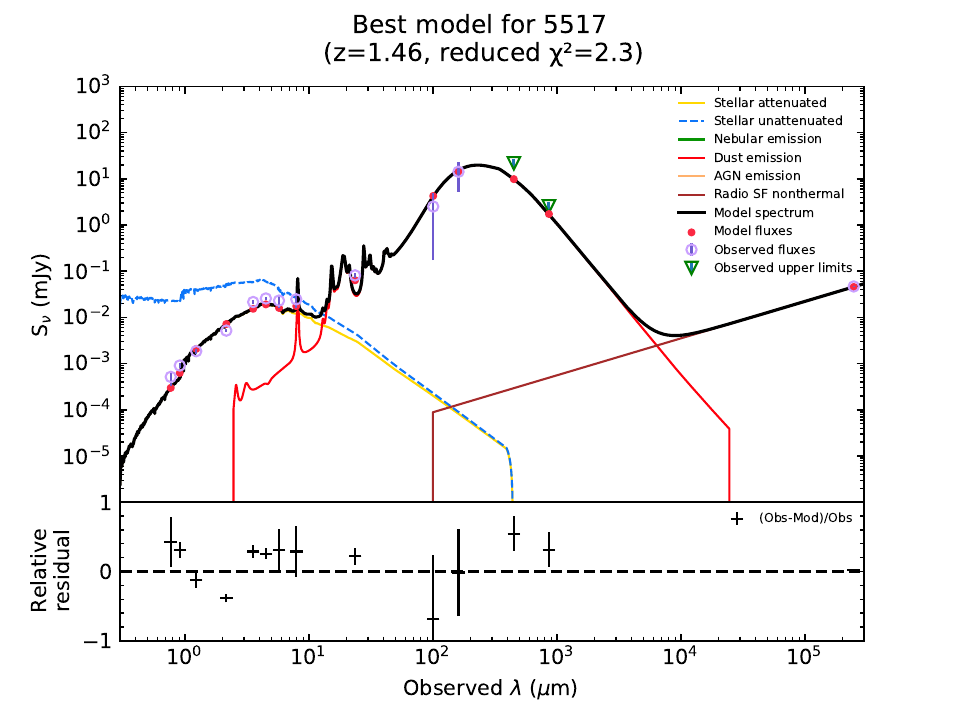} \hspace{-5.00mm} 
\includegraphics[width=6.0cm,height=6.0cm]{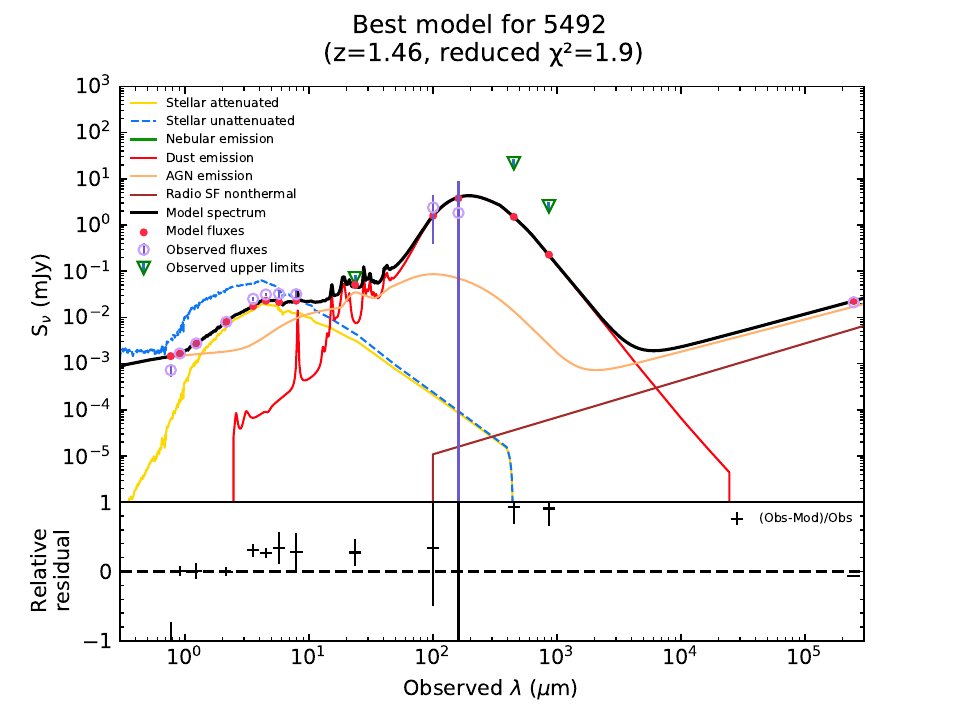} \hspace{-2.00mm}
\end{figure*}

\begin{figure*}  \label{fig:MorphologySED2}
	\centering 
\includegraphics[width=6.0cm,height=6.0cm]{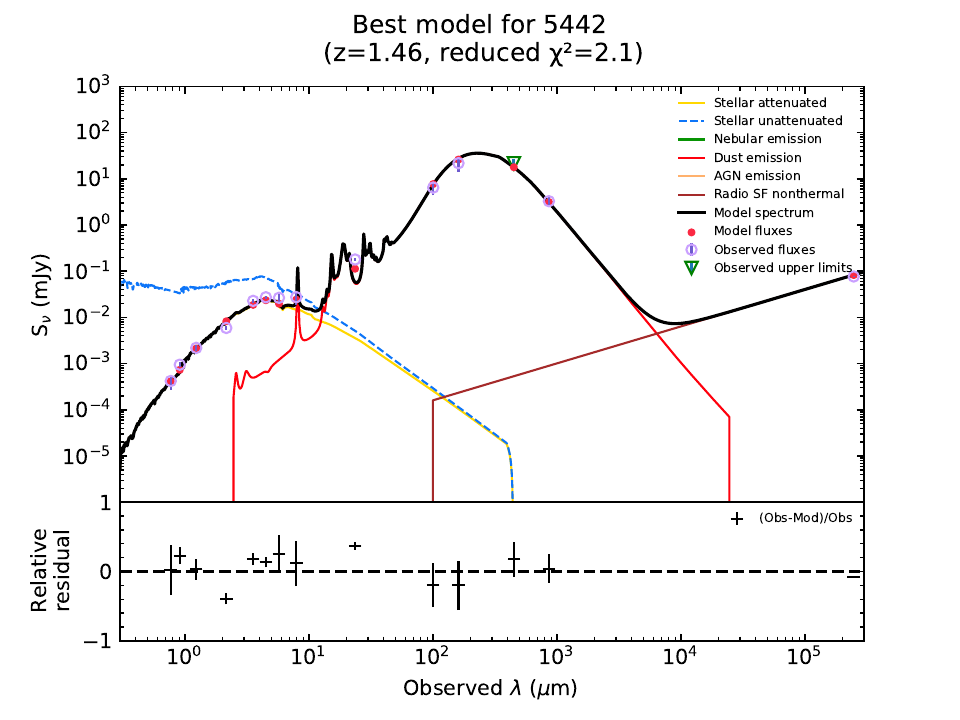} \hspace{-2.00mm}	\includegraphics[width=6.0cm,height=6.0cm]{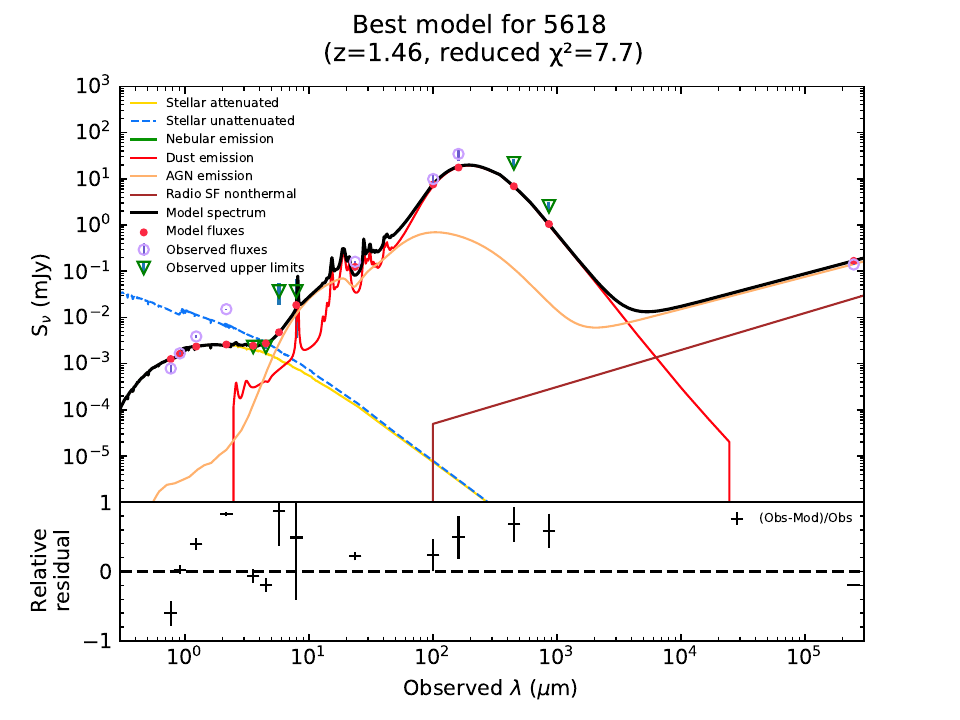} \hspace{-2.00mm}  
\includegraphics[width=6.0cm,height=6.0cm]{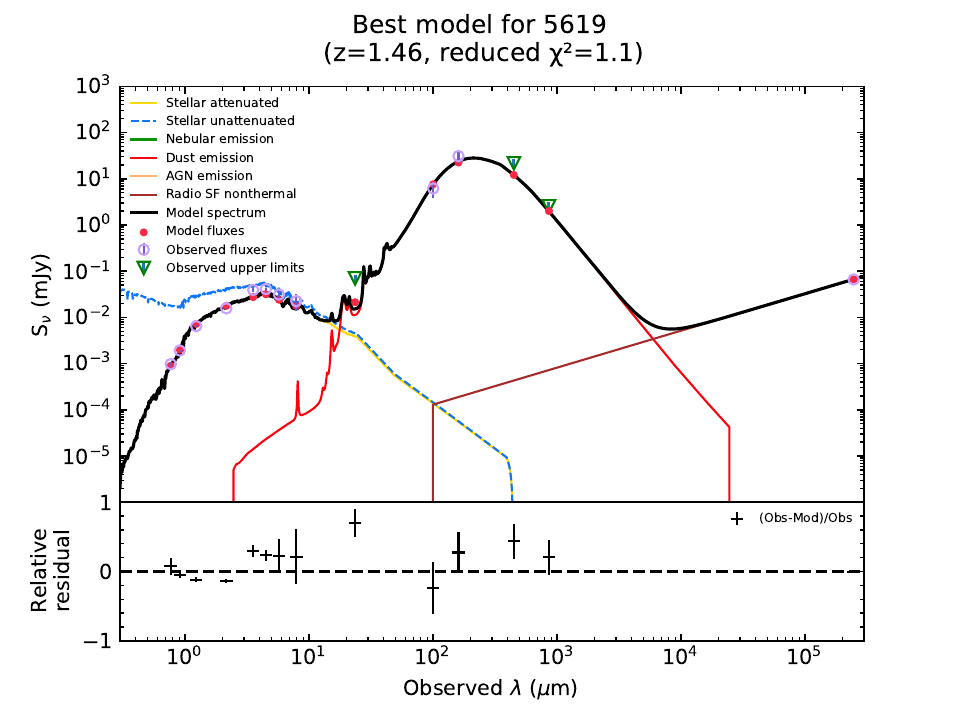} \hspace{-2.00mm}	
\includegraphics[width=6.0cm,height=6.0cm]{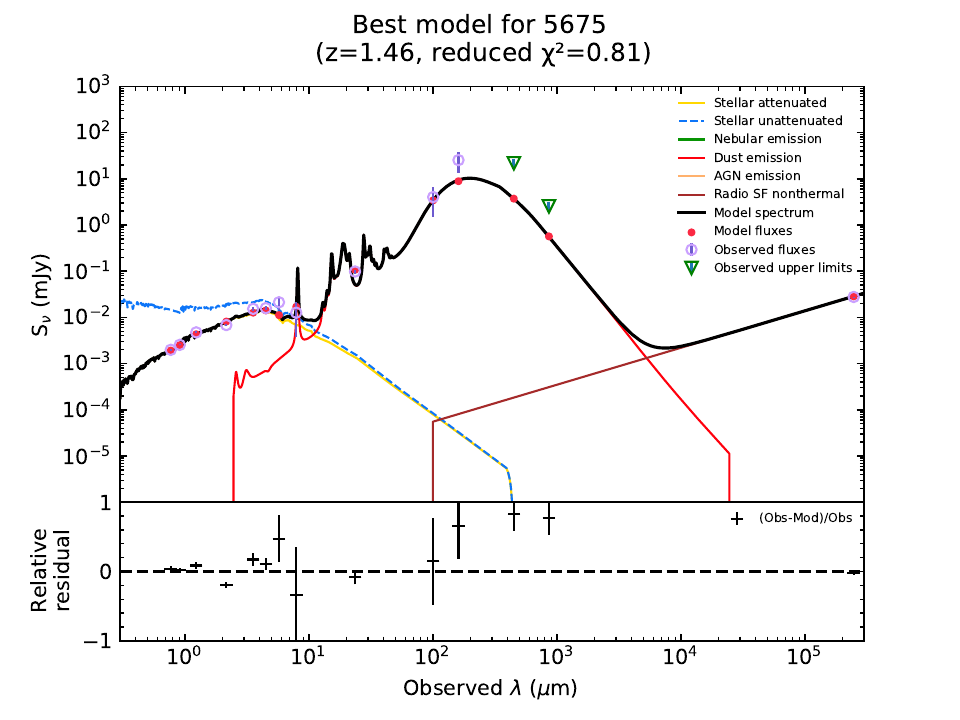} \hspace{-2.00mm}
\includegraphics[width=6.0cm,height=6.0cm]{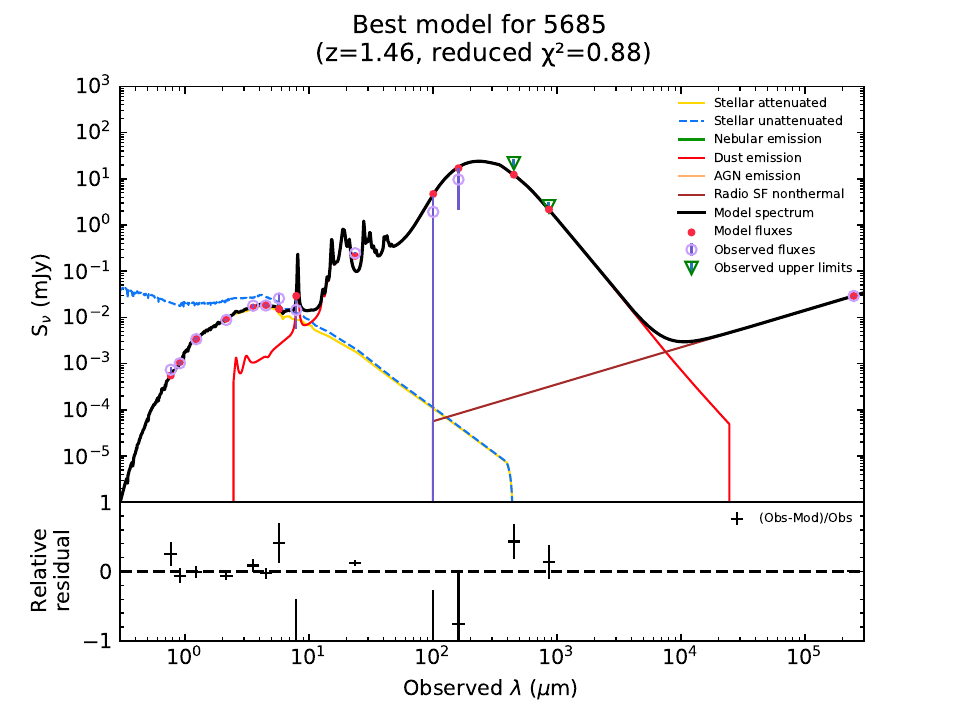} \hspace{-2.00mm}
\includegraphics[width=6.0cm,height=6.0cm]{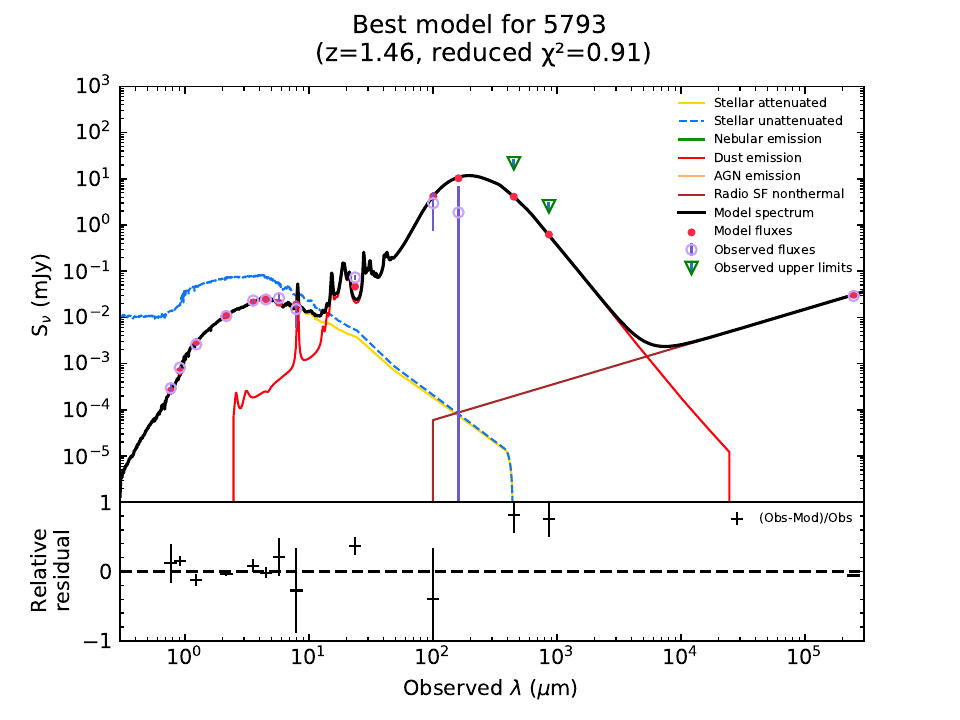} \hspace{-2.00mm}
\includegraphics[width=6.0cm,height=6.0cm]{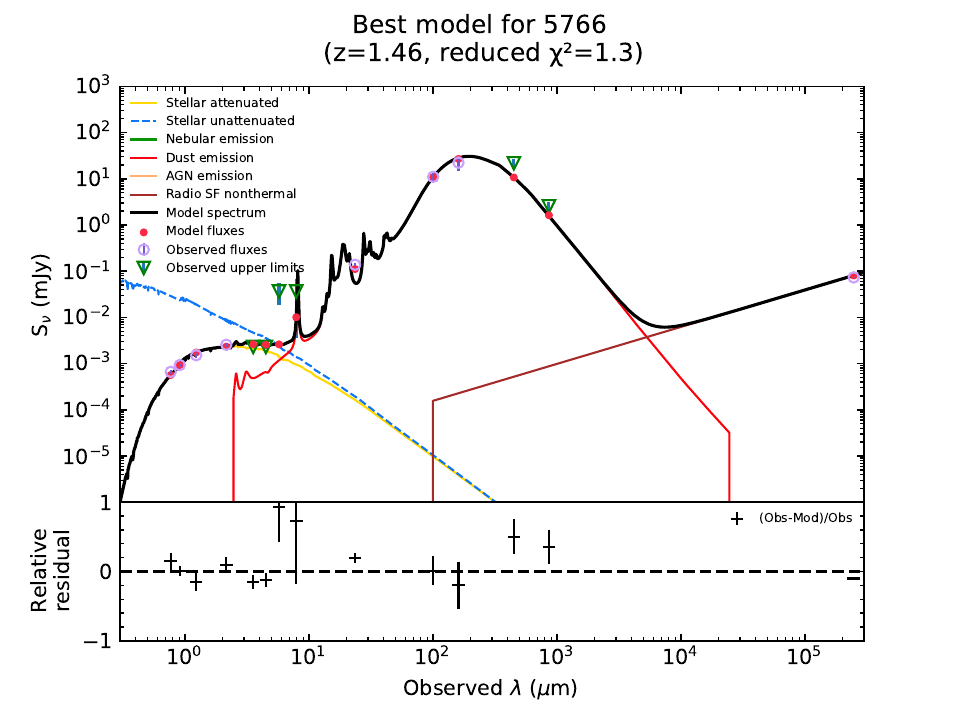} \hspace{-2.00mm}
\includegraphics[width=6.0cm,height=6.0cm]{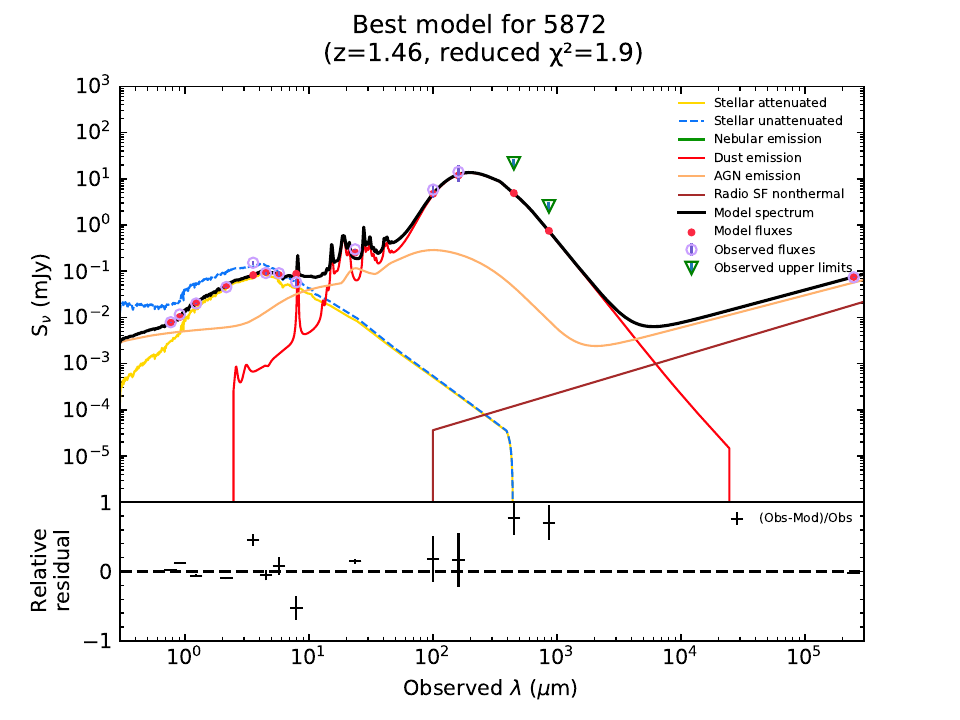} \hspace{-2.00mm}
\includegraphics[width=6.0cm,height=6.0cm]{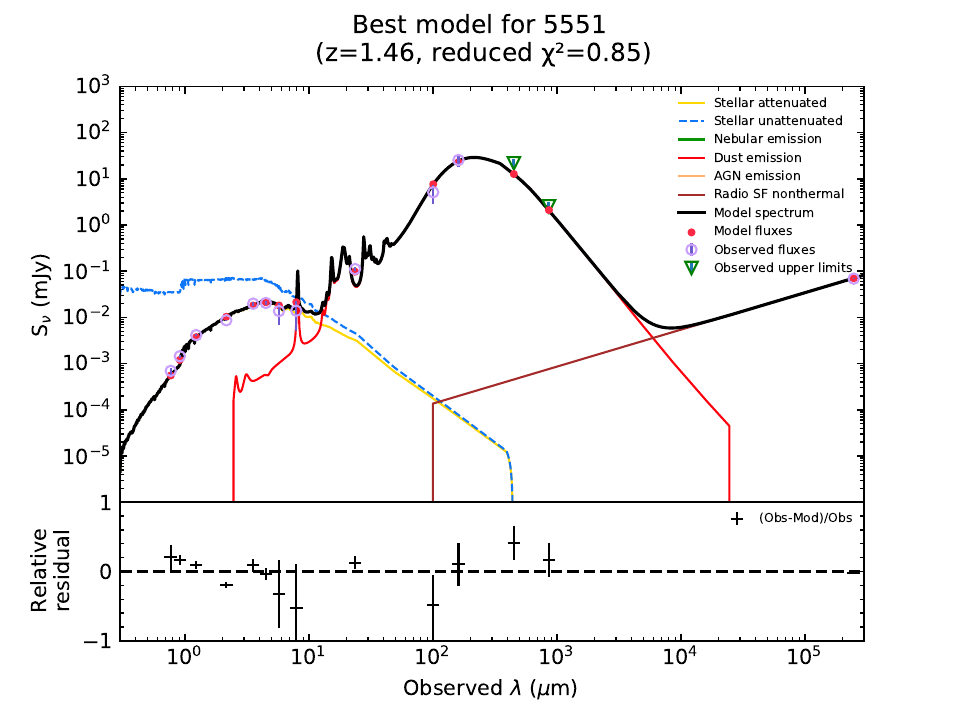} \hspace{-2.00mm}
\end{figure*}

\section{CIGALE Model Parameters}\label{subsect:CIGALE Model_Parameters}
\begin{table*}    
	\centering
	\addtolength{\tabcolsep}{5pt}
	\begin{tabular}{p{3cm}p{6cm}p{6cm}} 
		\hline  \hline 
		\multicolumn{1}{c}{Module} &
		\multicolumn{1}{c}{Parameter} & 
		\multicolumn{1}{c}{Values} \\
		\hline  		
	sfhdelayed & e-folding time of the burst of star formation  &  10, 16, 27, 46, 77, 129, 215, 359, 599, 1000 Myr   
		\\
		
		& Stellar age & 1000, 1291, 1668, 2154, 2782, 3593, 4641, 5994, 7742, 10000 Myr  \\

		&  Age of the last star-burst  &  5.0, 50.0, 500 Myr \\

		& Mass fraction of late star-burst population  &  0.0, 0.01, 0.05, 0.1, 0.25 \\  
		 \hline
		
		Simple stellar population (bc03) &  Initial mass function (imf)  &   1  (Chabrier) \\
		
		Bruzual \& Charlot (2003)  &  Metallicity   &  0.02 \\ \hline
		
		Nebular emission & Gas metallicity & 0.014 \\
		&Fraction of escaped photons  & 0.2  \\
		&Fraction of absorbed photons & 0.8 \\  \hline
		
		Dust attenuation  &  $V$-band attenuation in the interstellar medium & 0.15, 0.45, 0.60, 0.90, 1.20, 1.5, 1.8, 2.1, 2.4, 2.8, 0.50, 0.75   
		\\
		
		dustattmodifiedCF00 (modified Charlot \& Fall 2000 attenuation law)
		& attenuation ratio between older stars and younger stars, i.e., $\mu$ = $A_{v}^{\rm{ISM}}$ / A$_{v}^{ \rm{BC}}$ + $A_{v}^{\rm{ISM}}$ &   0.25, 0.50, 0.75, 0.44   \\	
		 \hline
				
		Dust emissions  &  Mass fraction of PAH  &   0.47, 1.12, 2.5 \\
		
		&  Minimum radiation field  &  0.10, 0.15, 0.20, 1.0 \\
		
		& Powerlaw slope  & 2.0, 3.0\\
  \hline
		
		AGN model   &  Average edge-on optical depth at 9.7 $\mu$m  &   7   \\	
		
		skirtor2016  &   inclination, i.e., viewing angle w.r.t. to the AGN axis  & 30,  70  \\ 	
		
		&     AGN fraction    &   0.0, 0.1, 0.2, 0.3, 0.4, 0.5, 0.8, 0.9  \\  
		\hline

		Galaxy radio synchrotron emission & FIR/radio correlation coefficient &  2.58 \\
	
		&  spectral index $^{a}$  &  0.8      \\
		
		\hline 
	\end{tabular}
	
	\caption{This table shows the model parameters and values set in the \enquote{pcigale.ini} configuration file. This \enquote{pcigale.ini} file was run to generate several models to fit our observed SED.  \\~\\
		NOTE. — \\ 
		For parameters and their corresponding values that were not listed here, we used the default values set by \texttt{CIGALE} V2022.0 \\ 
		a - For parameters relating to star-forming galaxies. } \label{table:CigaleParameters}
	\vspace{-4pt}	
\end{table*}

\bsp	
\label{lastpage}
\end{document}